\documentclass[12pt,preprint]{aastex}
\begin{document}

\newcommand{\dif}{\mathrm{d}}
\newcommand{\mclust}{{\em mclust}}
\newcommand{\optim}{``optim''}
\newcommand{\sparr}{{\em sparr}}
\newcommand{\spatstat}   {{\em spatstat}}
\newcommand{\mlpowerlaw}   {\texttt{ml\_powerlaw}}

\newcommand{\todo}[1]       {\noindent[{\bf \footnotesize #1}]} \newcommand{\tbr}[1]        {#1}
\newcommand{\tbd}           {\_\_\_\_  {\bf (TBD)}}     \newcommand{\revise}[1]     {{#1}}

\newcommand{\mystix} {MYStIX}
\newcommand{\Spitzer} {{\em Spitzer}}
\newcommand{\Herschel} {{\em Herschel}}
\newcommand{\Chandra} {{\em Chandra}}
\newcommand{\ACIS}    {{ACIS}}
\newcommand{\CIAO}    {{\em CIAO}}
\newcommand{\Sherpa}  {{\em Sherpa}}
\newcommand{\Chart}   {{\em ChaRT}}
\newcommand{\SAOTrace}{{\em SAOTrace}}
\newcommand{\DSnine}  {{\em DS9}}
\newcommand{\MARX}    {{\em MARX}}
\newcommand{\AEacro}  {{\em AE}}
\newcommand{\TARA}    {{\em TARA}}
\newcommand{\XSPEC}   {{\em XSPEC}}
\newcommand{\FTOOLS}  {{\em FTOOLS}}
\newcommand{\HEASOFT} {{\em HEASOFT}}
\newcommand{\IDL}     {{\em IDL}}
\newcommand{\Rlan}   {{\em R}}

\shorttitle{Stellar Subclusters in Star-Forming Regions}
\shortauthors{Kuhn et al.}
\slugcomment{Accepted, 13 March, 2014}

\title{The Spatial Structure of Young Stellar Clusters. I. Subclusters}  

\author{Michael A. Kuhn\altaffilmark{*}\altaffilmark{1}, Eric D. Feigelson\altaffilmark{1}, Konstantin V. Getman\altaffilmark{1}, Adrian J. Baddeley\altaffilmark{2}, Patrick S. Broos\altaffilmark{1}, Alison Sills\altaffilmark{3}, Matthew R. Bate\altaffilmark{4}, Matthew S. Povich\altaffilmark{5}, Kevin L. Luhman\altaffilmark{1}, Heather A. Busk\altaffilmark{1}, Tim Naylor\altaffilmark{4}, Robert R. King\altaffilmark{4}}

\altaffiltext{*} {mkuhn1@astro.psu.edu} 
\altaffiltext{1}{Department of Astronomy \& Astrophysics, Pennsylvania State University, 525 Davey Laboratory, University Park PA 16802}
\altaffiltext{2} {School of Mathematics and Statistics, University of Western Australia, 35 Stirling Highway, Crawley WA 6009, Australia}
\altaffiltext{3}{Department of Physics, McMaster University, 1280 Main Street West, Hamilton ON, L8S 4M1, Canada}
\altaffiltext{4}{Department of Physics and Astronomy, University of Exeter, Stocker Road, Exeter, Devon, EX4 4SB, UK}
\altaffiltext{5}{Department of Physics and Astronomy, California State Polytechnic University, 3801 West Temple Ave, Pomona, CA 91768}

\begin{abstract}

The clusters of young stars in massive star-forming regions show a wide range of sizes, morphologies, and numbers of stars. Their highly subclustered structures are revealed by the \mystix\ project's sample of 31,754 young stars in nearby sites of star formation (regions at distances $<$3.6~kpc that contain at least one O-type star.)
In 17 of the regions surveyed by \mystix, we identify subclusters of young stars using finite mixture models --- collections of isothermal ellipsoids that model individual subclusters. Maximum likelihood estimation is used to estimate the model parameters, and the Akaike Information Criterion is used to determine the number of subclusters. 
This procedure often successfully finds famous subclusters, such as the BN/KL complex behind the Orion Nebula Cluster and the KW-object complex in M~17.
A catalog of 142 subclusters is presented, with 1 to 20 subclusters per region. The subcluster core radius distribution for this sample is peaked at 0.17~pc with a standard deviation of 0.43~dex, and subcluster core radius is negatively correlated with gas/dust absorption of the stars --- a possible age effect. 
Based on the morphological arrangements of subclusters, we identify four classes of spatial structure: long chains of subclusters, clumpy structures, isolated clusters with a core-halo structure, and isolated clusters well fit by a single isothermal ellipsoid.

\end{abstract}


\section{Introduction \label{introduction_section}}

Massive star-forming regions (MSFRs) are a major, and perhaps the dominant, mode of star formation in the Galaxy \citep{Lada03,Fall09,Chandar11}. Most of the young stars in these massive complexes are clustered \citep{Clarke00,Lada03}, but a substantial, spatially distributed population of young stars also exists \citep{Evans09,Feigelson11}. Open questions on how young stellar clusters form and evolve in time in these harsh environments are outlined in the overview of the Massive Young Star-Forming Complex Study in Infrared and X-ray \citep[\mystix][]{overview}. This multifaceted project starts by characterizing the young stellar populations of 20 OB-dominated MSFRs at distances $d<3.6$~kpc.  The \mystix\  analysis provides a large sample of 31,754 young stellar members of these MSFRs \citep{mpcm}.  

The spatial distributions of young stars will show the imprint of cluster formation and evolution. For example, \citet{Elmegreen00} suggests that star formation occurs in a crossing time in a cloud with freely decaying turbulence, leaving behind stars that trace the subclustered structure of the natal cloud. However, \citet{Tan06} argue that gradual star formation in clouds with continually driven turbulence is necessary to explain smooth  stellar surface-density gradients in clusters like the Orion Nebula Cluster (ONC).
In addition, numerical simulations hint that dynamical processes of merging stellar subclusters may have a significant effect on the properties of the young clusters that are produced \citep[e.g.][]{McMillan07,Bate09a,Maschberger10}.
Observations of young stellar cluster sizes, densities, and morphologies will enable testing the theoretical models of star formation. However, the interpretation of spatial distributions faces several challenges.
\begin{enumerate}
\item 
Young stellar clusters, both real and simulated, form with a wide variety of intricate morphologies.
Therefore, summary statistics must be developed which capture the salient and astrophysically meaningful features of the distribution of star positions so that structure in one region may be compared quantitatively to another \citep[e.g.][]{Cartwright04,Allison09b,Gutermuth09}. For example, recent cluster-formation simulations produce stellar spatial distributions that resemble (to the eye) the stars in real star-forming regions \citep[e.g.][]{Bate09a,Bonnell11}, but robust comparison of structure should be based on calibrated statistical measures.
\item 
There may be ambiguities in determining which stars are clustered \citep{Bressert10,Gieles12,Pfalzner12}. Furthermore, molecular clouds and star formation show signs of fractal structure ranging from Galactic scales \citep{Elmegreen01} down to subcluster scales \citep{Schmeja11}. Were stellar distributions truly scale-invariant, the determination of cluster size and density would depend on the arbitrary choice of cluster boundaries. 
\item 
Different mechanisms of star cluster formation may produce clusters with similar properties.
For example, a young stellar cluster that has relaxed through two-body interactions may appear similar to a cluster that underwent rapid, violent relaxation induced by the merger of subclusters \citep{Allison09a} or gas removal \citep{Moeckel10}.
\end{enumerate}

A particularly salient features of the spatial distributions of young stars is the organization in clusters and subclusters\footnote{
The definitions of ``young stellar cluster'' in the star-formation literature are divergent \citep[for an alternative definition cf.][]{PortegiesZwart10}. Here, we use ``cluster'' in a statistical sense  (subclusters are the components of the finite mixture model in Section 3). 
}. 
\citet{Hillenbrand98} found that a \citet{King62} profile describes the radial surface density profile of the ONC, with $\rho_0=2-3\times 10^4$~M$_\odot$~pc$^{-3}$ and core radius $r_0=0.15-0.2$~pc. 
This profile has also been fit to NGC~6611, W~40, Tr~15, NGC~3603, the Arches cluster, h and $\chi$ Per, and young stellar clusters in the Large Magellanic Clouds \citep{Wang08, Kuhn10, Wang11, Sung04, Harfst10, Bragg05,Mackey03}. 
\citet{Pfalzner12} note that these clusters tend to have large ratios of half-mass radius to core radius, so the truncated King profile is a better model than the Plummer sphere. 

In addition to the main subclusters in regions, smaller groups of stars can be seen outside the main clusters, or even as subclusters within the main clusters. It is unclear whether a relaxed surface density profile is a good model for subclusters, in part because young stellar clusters are generally not expected to have had time to dynamically relax through two-body interactions and molecular gas may contribute significantly to the gravitational potential. Nevertheless, similar surface density profiles have been used by \citet{Smith11}, who use Plummer sphere subclusters in their simulations, and found by \citet{Maschberger10} in the radial density profile of simulated subcluster mergers. 

We use the young stellar samples from the \mystix\ project to analyze the subcluster properties in 17 MSFRs. This rich statistical sample of stars is particularly useful for investigation of spatial structure, since, prior to \mystix, the stellar populations in most of these regions were poorly characterized. \mystix\ Probable Complex Members include low-mass and high-mass stars and disk-bearing and disk-free stars \citep{mpcm}, and the multiwavelength X-ray/infrared observations ameliorate effects of high absorption from the natal cloud, infrared (IR) and optical nebulosity from ionized gas, and source crowding \citep{overview}. In particular \mystix\ greatly improves the detection of stars at the centers of dense subclusters, where IR-only studies often fare the worst \citep[e.g.][]{Bressert10}.  Although the \mystix\ samples are not ``complete,'' the identification of young stars is performed in a uniform way for the different regions which allows for comparative analysis of stellar populations in these different regions. 

Here, we analyze the projected spatial distributions of young stars from these censuses; analysis is done in parsec units to compare regions at different distances, and the distribution of star positions is analyzed using statistical theory for spatial point processes. This work (Paper I) discusses subclusters in these regions.
The subclusters have a variety of configurations that relate to different ways that star formation can occur and different stages in the evolution of a MSFR. 
Paper II will analyze the intrinsic stellar populations of individual subclusters cataloged here by correcting for incompleteness in each region. 
We recognize that the full spatial substructure of stars in a star-forming region cannot be reduced to discrete subclusters if fractal clustering is present; this will be investigated in Paper III using non-parametric, parametric, and stochastic methods for characterizing spatial structures. Mass segregation, the concentration of massive stars in the cores of many clusters will also be investigated.  Other papers in the \mystix\ project will use the subclusters derived here including study of the stellar ages \citep{Getman13a}, protoplanetary disk fractions, and spatial relationships between subclusters and their cloud environs.

This paper is organized as follows. In Section \ref{sample_section}, the \mystix\ census of young stars is described, and biases associated with the multiwavelength data are discussed. In Section~\ref{methodology_section} the statistical method for identifying and modeling subclusters is described. The cluster fitting results and catalogs are presented in Section \ref{results_section}.  The structure of star-forming complexes are examined in Section~5, and subcluster properties are investigated in Section~6. Section~7 is the discussion.  An Appendix gives code permitting the reader to recover and revise the subclusters obtained here using published \mystix\ source tables.   

\section{Stellar Samples: \mystix\ Probable Complex Members \label{sample_section}}

The regions included in this study, listed in Table \ref{target_table}, are nearby sites of high-mass star formation that are included in the \mystix\ sample \citep{overview}. Each of these regions has a population of hundreds or thousands of young stars in the \mystix\ sample. Some fields appear relatively simple with a single, dominant cluster, like the ONC or W~40; while others have complex structures, like the Rosette nebula which contains the dominant NGC~2244 cluster plus smaller, embedded clusters within the Rosette molecular cloud. The field of view must also be taken into consideration: for example, a single \Chandra\ exposure of the Orion Nebula captures only one major cluster on a $\sim$3~pc scale, while the mosaic of three \Chandra\ exposures of the more distant NGC~6357 complex captures three rich clusters on a $\sim$20~pc scale. The molecular clouds vary from region to region as well --- NGC~2362 has no molecular cloud left, others like Orion or NGC~1893 have blown a cavity in their molecular cloud around the star cluster, while others like NGC~2264 and DR~21 still have deeply embedded clusters. $V$-band extinction typically ranges from 2 to 10~mag, but the obscuration may be highly non-uniform with $A_V>100$ in the densest molecular cores and filaments. There may also be significant optical and IR nebulosity, and the polycyclic-aromatic-hydrocarbon (PAH) emission (seen with the  \Spitzer\ {\it Space Telescope}) can trace the interface between the ionized H~II region and the molecular cloud. 

These regions are covered in the near-IR by the UKIRT Infrared Deep Sky Survey \citep{nir} or Two-Micron All Sky Survey \citep{Skrutskie06}, in the mid-IR by \Spitzer\ IRAC observations \citep{mir}, and in the X-ray by the ACIS-I array on the \Chandra\ X-ray observatory in the 0.5-8.0~keV band \citep{xray,Townsley13}. 
Datasets are also obtained from projects by \citet[COUP;][]{Getman05}, \citet[GLIMPSE;][]{Benjamin03}, \citet{Kuhn10}, and \citet{Majewski07}.
The multiwavelength point-source catalogs are combined with a list of spectroscopic OB stars from the literature, and sources are classified probabilistically to obtain a list of MYStIX Probable Complex Members \citep[MPCM;][]{mpcm} Three types of sources make up the MPCMs: objects that are detected in the X-ray, objects with IR excess consistent with young stellar object (YSO) disk/envelope emission, and the spectroscopic OB stars. The MPCMs include sources with disks and/or envelopes and disk-free complex members, high-mass and low-mass members, and clustered and unclustered members. 

In the \mystix\ project, data reduction and classification are performed as uniformly as possible across the study to facilitate comparative analysis between different regions. The fields of view of the study are the intersections of the near-IR, mid-IR, and X-ray data. Nevertheless, differences among the observations (e.g.\ exposure times, coverage) and differences among the regions (e.g.\ distance, obscuration, nebulosity, and crowding) and biases in the selection of the different types of MPCMs are present. To improve spatial uniformity of the study, particularly variations in the X-ray sensitivity across a region,  our analysis here is restricted to a subset of MPCMs.  

Table \ref{target_table} lists the MYStIX regions analyzed here.  Column 3 gives the X-ray flux limit with uniform X-ray coverage.  Columns 4-7 give the number of MPCMs used for the spatial analysis, including the overlapping subsets selected by each criterion.

\begin{deluxetable}{lllrrrr}
\tablecaption{MYStIX Targets for Subcluster Analysis\label{target_table}}
\tablehead{
\colhead{} & \colhead{} & \colhead{} & \multicolumn{4}{c}{Number of Sources in Sample}\\
\cline{4-7}
\colhead{Region} & \colhead{Distance} & \colhead{$\log F_{X,\mathrm{limit}}$} & \colhead{Total} & \colhead{X-ray} &  \colhead{IR-excess} &  \colhead{OB}\\
\colhead{} & \colhead{(kpc)} & \colhead{(ph s$^{-1}$ cm$^{-2}$)} & \colhead{(stars)} & \colhead{(stars)} & \colhead{(stars)} & \colhead{(stars)}\\
\colhead{(1)} & \colhead{(2)} & \colhead{(3)} & \colhead{(4)} & \colhead{(5)} & \colhead{(6)} & \colhead{(7)}
}
\startdata
Orion Nebula & 0.414 & ~~~~~$-$6.6 & 1367 (90\%) & 1216 & 631 & 13\\
Flame Nebula & 0.414 & ~~~~~$-$6.2 & 342 (71\%) & 254 & 193 & 2\\
W~40 & 0.5 & ~~~~~$-$6.1 & 411 (96\%) & 174 & 309 & 3 \\
RCW~36 & 0.7 & ~~~~~$-$6.8 &307 (80\%) & 260 & 135 & 2\\
NGC~2264 & 0.914 & ~~~~~$-$6.1 &968 (83\%) & 599 & 555 & 7\\
Rosette & 1.33& ~~~~~$-$5.9 & 1195 (69\%) & 700 & 623 & 21\\
Lagoon & 1.3 & ~~~~~$-$6.05& 1251 (61\%) & 947 & 468 & 28\\
NGC~2362 & 1.48 & ~~~~~$-$5.95 &246 (50\%)& 207 & 49 & 12\\
DR~21 & 1.5 & ~~~~~$-$6.05 & 662  (67\%) & 199 & 507 & 1\\
RCW~38 & 1.7 & ~~~~~$-$6.2 & 495 (56\%) & 412 & 112 & 1\\
NGC~6334 & 1.7 & ~~~~~$-$5.9 & 987 (59\%) & 644 & 403 & 8\\
NGC~6357 & 1.7 & ~~~~~$-$6.0 & 1439 (64\%) & 1047 & 524 & 16\\
Eagle Nebula & 1.75 & ~~~~~$-$6.1 &1614 (63\%) & 1005 & 723 & 56\\
M~17 & 2.0& ~~~~~$-$6.5 & 1322 (57\%) & 1247  & 128 & 64\\
Carina Nebula & 2.3 &~~~~~$-$5.9 & 2790 (38\%) & 2043 & 815 & 134\\
Trifid Nebula & 2.7 & ~~~~~$-$6.0 &357 (67\%) & 227 & 174 & 2\\
NGC~1893 & 3.6& ~~~~~$-$6.65 & 854 (65\%) & 617 & 349 & 29 \\
\enddata
\tablecomments{Properties of star-forming regions and statistics of the MPCM sample used here. Column 1: Target region name. Column 2: Distance to region used in MYStIX analysis \citep{overview}. Column 3: Apparent $0.5-8.0$~keV X-ray photon flux limit imposed for spatial uniformity.  Column 4-7: Number of stars used in the analysis: total stars in the samples used here  (with percentage of all MPCMs), X-ray selected stars, \mystix\ InfraRed Excess (MIRES) stars, and published OB stars, respectively. See \citet{mpcm} and associated \mystix\ technical papers for details on the sample definitions and selection. }
\end{deluxetable}
\clearpage\clearpage

\subsection{X-ray Selected Members} \label{xray_section}

Young low-mass stars have strong X-ray emission compared to main-sequence stars, arising mostly from magnetic reconnection flares from the surface of pre-main sequence stars \citep{Gudel09,Feigelson10}. X-ray luminosities are typically $L_X\sim10^{-4}-10^{-3}L_\mathrm{bol}$ \citep{Preibisch05}, several orders of magnitude higher than X-ray emission from main-sequence stars like the Sun with $L_X\sim10^{-6}L_{\mathrm{bol}}$. Thus, X-ray observations can be used to discriminate young stars both disk-bearing and disk-free in star-forming regions from older Galactic field stars. Although most point sources detected in the \Chandra\ observation discussed here are young stars, some X-ray sources are non-member contaminants that include active galactic nuclei and foreground/background main-sequence stars \citep{Getman11}. Here, members and non-members are distinguished by Bayesian supervised learning using photometric, spectroscopic, and variability properties of the sources in our X-ray/IR data \citep{mpcm}. 

X-ray surveys are also effective at overcoming challenges to studying stars in the complex environments of MSFRs. Optical surveys are limited by absorption, which may exceed 100~mag in the $V$ band, but hard X-rays are effective at penetrating the interstellar medium. Optical and IR studies of stars are also limited by high background levels from nebular emission of the H~II region, which is largely absent in the X-ray. Often young stellar clusters are centered near the regions with the highest IR nebulosity, so the X-ray observations can be more effective for identifying low mass stars in areas with strong gas and dust emission near OB stars. In addition, the brightness contrast between T-Tauri and OB stars is not so dramatic in the X-ray as in optical and in IR bands \citep[e.g.][]{Cohen08}. Thus, low-mass stars can be detected in the X-ray, even when they are projected near OB stars in the dense young stellar cores. Finally, the low number of resolved field stars and the high resolution of the $Chandra$ mirrors reduce X-ray source confusion in the cluster centers to a low level for the \mystix\ MSFRs.

For X-ray selected complex members, source detection is limited by apparent photon flux in the $0.5-8.0$~keV band \citep{xray,Townsley13}. In the full MPCM list, sensitivity will vary over a considerable range in different parts of the field of view. This non-uniformity in sensitivity across a \mystix\ field is due to two effects: differing integration times at different locations in a \Chandra\ mosaic,  and telescope effects such as off-axis mirror vignetting and degradation of the point-spread function. Finally, comparison between \mystix\ regions must judiciously consider the different \Chandra\ exposure times and distances to the regions. In the present study, we compensate for the intra-region variations in sensitivity. Inter-region differences in sensitivity are treaded in the forthcoming Paper II (M. A. Kuhn et al., in preparation).  Readers are cautioned about comparing stellar populations in (sub)clusters of different \mystix\ regions from the analysis provided here.

The increased sensitivity to X-ray sources near the center of a \Chandra\ pointing and can produce artificially clustered sources; this has been called the ``egg-crate effect'' when many fields are combined into a mosaic. This problem can be seen clearly in the X-ray source distribution of the \Chandra\ Carina Complex Project \citep[][their Figure 4]{Townsley11} and is quantified by \citet[][their Figure 7]{Broos11}. Sensitivity is reduced by a factor of $\sim$4 in the outermost portions of a \Chandra\ field compared to on-axis portions. For our spatial analysis, we flatten these inhomogeneities in sensitivity by selecting 11,798 X-ray selected sources that have apparent $0.5-8.0$~keV band photon flux greater than the completeness limit for the observation, defined as the minimum apparent photon flux at which a young star has a nearly full chance of being detected anywhere in the field of view \citep{Feigelson05}. 

The completeness limit is calculated empirically with the assumption that the underlying X-ray photon flux distribution follows a power-law \citep[e.g.][]{Broos11}. For the \Chandra\ pointing in the ACIS-I mosaic that has the shortest exposure time, X-ray selected MPCMs are stratified by off-axis angle at 0.3, 5.1, 6.3, 7.5, 8.3, and 12.3 arcmin. Their apparent $0.5-8.0$~keV band photon flux distributions are examined to find a photon flux to truncate the sample such that the photon-flux distributions in all strata look similar. The imposed flux limits, $\log F_{X,\mathrm{limit}}$, are given by Column 3 of Table \ref{target_table}, and the number of X-ray selected MPCMs remaining in the flattened sample is given in Column 5. Typically, $\sim50$\% of the X-ray selected MPCMs \citep[][their Table 4]{mpcm} are included in the flattened sample\footnote{The young stars used in this study include MPCMs with $F_X>F_{X,\mathrm{limit}}$, IR-excess, or spectral classification as OB stars. If an X-ray selected MPCM is excluded from the ``flattened sample'', it may be included in the spatial analysis if it meets either of the other criteria.}. 

The uniform X-ray photon flux limit allows structures to be characterized independently of the $Chandra$ pointing history. But other spatial biases are present.  When variations in gas absorption are present, a fixed X-ray photon flux limit corresponds to a higher intrinsic luminosity limit in high-absorption regions.  Deeply embedded regions may this be deficient in X-ray MPCMs, but often this is compensated by the higher fraction of IR-excess sources.  Another bias may arise due to mass segregation.  X-ray luminosity is correlated to stellar mass \citep[e.g. XEST;][]{Gudel07}, so detection is not independent of mass, and MPCMs may more closely reflect the spatial distributions of the more massive population. A similar bias arises because X-ray observations are less sensitive to stars with circumstellar disks \citep[e.g.][]{Getman09}. Finally, the use of the spatial prior in the classification of MPCMs \citep{mpcm} will produce a bias against discovery of widely distributed MPCMs.

\subsection{IR-Excess Selected Members} \label{ire_section}

Dusty disks and/or infalling envelopes emit IR radiation above that produced by the young stellar photosphere. The spectral energy distributions (SEDs) may show an IR-excess in the $K$-band, but it is strongest in the \Spitzer\ mid-IR bands.  \citet{ised} catalog the \mystix\ InfraRed Excess Sources (MIRES) with SEDs consistent with disk-envelope-bearing young stars.  Povich et al.\ provide several filters to reduce likely background sources; non-member contaminants include reddened stellar photospheres, star-forming galaxies, active galactic nuclei, (post-)asymptotic giant branch stars, shock emission, unresolved knots of PAH emission, and background young stars.  We use here only MIRES stars that they designate $\mathrm{SED\_FLG}=0$ as likely young stars with protoplanetary disks. For the Orion Nebula and Carina Nebula fields, published lists of probable IR-excess members are used \citep{Megeath12,Povich11}.

IR-selected MPCMs are not biased by interstellar absorption and are detected down to very low masses. Therefore, the IR-excess selected MPCMs may be more effective than the X-ray selected MPCMs at revealing highly absorbed subclusters of young stars. The importance of the IR-only members to our multi-wavelength survey is illustrated by \citet{Getman12} where IR-excess stars dominate the stellar population in a triggered star-formation globule on the periphery of an H~II region.  

\Spitzer\ IRAC observations of MSFRs suffer from high, non-uniform nebulosity, primarily from polycyclic aromatic hydrocarbon emission lines in the 3.6, 5.8, and 8.0~$\micron$ bands, and from source confusion near the center of rich clusters \citep{mir}. In contrast, \Chandra's sensitivity is not affected by nebulosity it has several-times better on-axis spatial resolution than \Spitzer. These mid-IR effects may lead to severely decreased MIRES sensitivity in regions with the highest nebular contamination and stellar density. We do not attempt to correct these effects by removing weak sources, as we do for the X-ray selected MPCMs, because any attempt to apply a uniform flux limit across the field of view would leave only the few brightest stars.

The IR-excess selected MPCMs are also biased against widely distributed stars, because of the spatial filtering used to remove probable contaminants. This may enhance slight over-densities in source counts, and also interferes with attempts to characterize the distributed populations \citep{overview}.

\subsection{High-Mass Members from the Literature} \label{highmass_section}

Spectroscopic catalogs of OB-stars are available for many regions \citep[e.g.][and references therein]{Skiff09}, and all known OB stars in \mystix\ regions are included in our spatial analysis because of the large roll they play in clustered star formation. 
However, these lists may be incomplete due to limited spatial coverage by the spectrographic surveys. In addition, OB stars may hide in regions with high absorption and avoid detection in optical surveys \citep{Povich11}. Most of the OB stars have X-ray counterparts with luminosities comparable to or higher than the most X-ray luminous T-Tauri stars \citep[e.g.][]{Stelzer05}. The \mystix\ survey uncovers a moderate number of candidate new X-ray emitting OB stars (H. Busk et al., in preparation). 

\subsection{Summary of Targets} \label{summary_targets_section}

The statistical sample of stars used for the analysis of spatial distributions contains 16,608 MPCM out of 31,754 MPCMs in the entire \mystix\ project. The resulting samples range from a few hundred to a few thousand young stars for each \mystix\ region (Table \ref{target_table}, Column 4). The sample sizes reflect the intrinsic richnesses of the stellar populations, but are affected by ancillary effects such as region distance, obscuration, nebulosity, and observation exposure times. These observational selection effects will be treated in Paper~III; the present study is confined to the observed MPCM population. 

Nearly all regions have more X-ray selected stars than IR-excess stars. The major clusters are typically dominated by the X-ray selected sources, but peripheral, highly-obscured groupings of stars may be dominated by IR-excess sources. Even for W~40 and DR~21, 
where IR-excess sources outnumber X-ray sources in the whole region, the densest parts of the clusters are dominated by X-ray selected members.

\section{Modeling Spatial Distributions of Young Stars with Isothermal Ellipsoids \label{methodology_section}}

The star-forming regions investigated here include many famous star clusters, but the \mystix\ catalogs still reveal new structures in the spatial distributions of the young stellar members, which are complicated and multimodal. The structure is sometimes dominated by a single monolithic cluster, sometimes a dominant cluster with substructure, sometimes several clusters, and sometimes a dominant cluster with secondary subclusters. To model all of the spatial structures in an unbiased fashion, we need a statistical method that can find all of the clusters and subclusters in an objective fashion. 

A variety of methods exist to perform unsupervised cluster analysis including non-parametric methods (such as kernel density estimation, the $k$-nearest neighbor algorithm, and graph-based algorithms) and the parametric methods that we use for this study \citep[e.g.][]{Everitt11}.  The nonparametric methods require the choice of a critical value for the algorithm; for example,  the number of nearest neighbors \citep{RomanZuniga08}, threshold minimum spanning tree branch length \citep{Gutermuth09},  and a threshold surface density in a kernel density estimator  \citep{Feigelson11}.    Nevertheless, the ``No Free Lunch Theorem'' states that there is no general way to determine whether a particular cluster solution is better than another found by a different method in the multivariate unsupervised classification problem \citep{Wolpert97}.   The advantages and disadvantages of our parametric method are discussed in Section~3.6.

Our statistical procedure for identifying subclusters in MPCM stars is based on finite mixture models \citep{McLachlan00}. Here, we assume that the MPCM population is made up of subpopulations that can each be described by a parametric surface-density distribution, and the mixture model is the sum of these subcluster surface densities. Mixture models are commonly used in parametric modeling of point processes, though usually with Gaussian function (rather than an astrophysically motivated function) components \citep{McLachlan00, Fraley02}. Our procedure is based on several well-established steps: subcluster properties are obtained through maximum likelihood parameter estimation; the number of subclusters is determined by model selection; and subcluster assignments for young stars are determined using posterior probabilities from the fitted models \citep{Everitt11}. For example, these steps performed by the cluster-finding program \mclust\footnote{http://www.stat.washington.edu/mclust/} \citep{Fraley12}, which uses normal mixture models and performs model selection using the penalized likelihood Bayesian Information Criterion (BIC). Here, we use a subcluster model that has a roughly flat core and power-law wings, which is generally a better fit to young stellar clusters, and we perform model selection using the Akaike Information Criterion (AIC).

\begin{figure}
\centering
\includegraphics[angle=0.,width=6.5in]{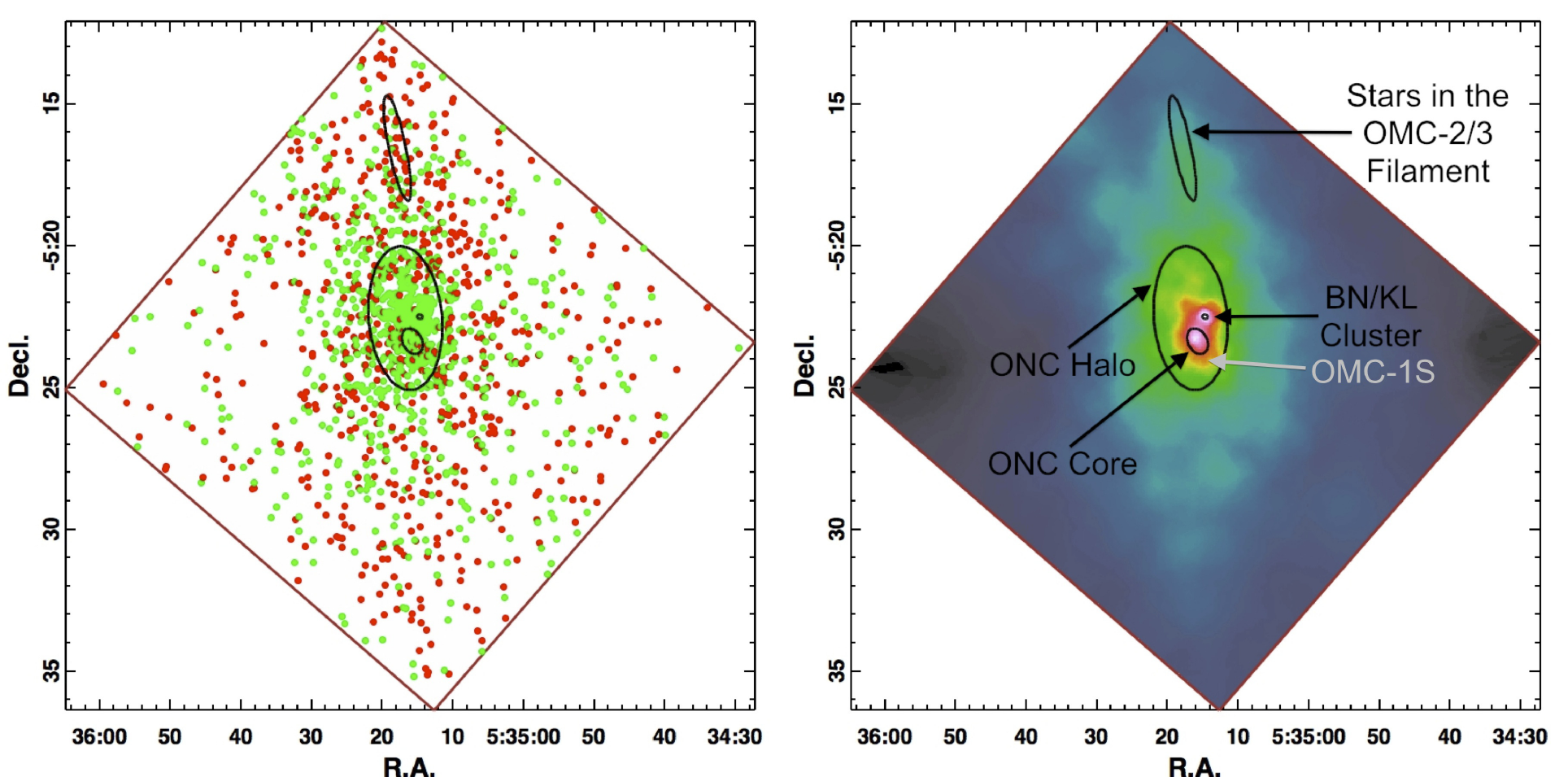}
\caption{Left: MPCMs in the Orion star-forming region (IR-excess selected stars are red points; X-ray selected stars without IR-excess detections are green points.) The clusters found by our statistical method are shown as black ellipses, indicating the cluster core region. Overlapping clusters are resolved, indicating the layered structure of the region. Right: The adaptively smoothed stellar surface density is shown, with the ellipses over-plotted. Subclusters from the literature that are identified in the finite mixture model are labeled (black arrows), including the BN/KL cluster \citep{Becklin67,Kleinmann67} and stars embedded in the southern end of the OMC-2/3 molecular filaments \citep{Megeath12}; the OMC-1S cluster is not resolved in the model (gray arrow). The ONC is modeled by an inner and outer ellipsoid, which may be interpreted a core-halo cluster structure. \label{orion_annotated_fig}
}
\end{figure}

Figure \ref{orion_annotated_fig} shows the projected spatial distribution of stars in the Orion Nebula region, in both individual stars and smoothed stellar surface density, with the clusters found from the finite mixture model algorithm indicated by black ellipses. In this field, this cluster-finding method identifies layered, overlapping clusters and clusters with different sizes and surface densities.  The statistical analysis is performed using software from the $R$ package \spatstat\footnote{http://www.spatstat.org/} \citep{Baddeley03}. This environment facilitates advanced statistical analysis of spatial point processes within irregularly shaped windows. 

\subsection{Constructing a Model} \label{construct_model_section}

In this study, we adopt a parametric model of young stellar clusters based on equilibrium configurations of self-gravitating stars in the form of isothermal ellipsoids\footnote{This distribution function is similar to King's profile, but it lacks the truncation radius where density goes to zero. The King profile's truncation radius is derived from the assumption of a tidal cutoff, which is unlikely to apply to young stellar clusters. For the datasets in our analysis, it is unlikely that model fitting would be able to constrain any truncation radius due to the limited field of view, the complex distributions of subclusters that overlap at large radii, and the presence of an unclustered component.} \citep{Chandrasekhar42}. The isothermal sphere distribution function may be used as the subcluster model. The isothermal sphere surface density profile can be approximated with the Hubble model
\begin{equation}\label{hubble_equation}
\Sigma(r)=\frac{\Sigma_0}{1+(r/r_c)^2},
\end{equation}
where $\Sigma(r)$ is the surface density of stars at a distance $r$ from the cluster center, which is parameterized by the coordinates of the cluster center, the central surface density $\Sigma_0$, and the core radius $r_c$.
This approximation is accurate to $<7$\% for $r<4r_c$ \citep{Binney08}. The central volume density $\rho_0$ is related to the central surface density by 
\begin{equation}\label{volume_density_equation}
\rho_0 = \Sigma_0 / 2 r_c,
\end{equation}
and the number of stars $N$ projected within a radius $r$ is 
\begin{equation}\label{integral_equation}
N(<r)=\pi r_c^2 \Sigma_0 \ln(1+r^2/r_c^2) .
\end{equation}
However, many young stellar clusters have elliptical shapes on the sky, for example \citet{Hillenbrand98} find that the ONC has elliptical contours of constant stellar surface density with ellipticity $\epsilon \simeq 0.30$. A more general surface density profile is produced by stretching the Hubble model to model ellipsoidal clusters. This fitting function introduces two new parameters, the ellipticity $\epsilon = (a-b)/a$ and the ellipse orientation $\phi$. The surface density for the isothermal ellipsoid becomes
\begin{equation}
\Sigma_\mathrm{ell}(\mathbf{r};\Sigma_0,r_0,r_c,\phi,\epsilon) = \Sigma_0 \left [1+ 
\left| \left( \begin{array}{cc}
(1-\epsilon)^{-1/2} & 0 \\
0 & (1-\epsilon)^{1/2} \\
\end{array} \right)
\hat{R}(\phi)(\mathbf{r}-\mathbf{r_0})
\right|^2\middle/r_c^2
\right]^{-1},
\end{equation}
where the matrix $\hat{R}(\phi)$ rotates the vector $\mathbf{r}-\mathbf{r_0}=(x-x_0, y-y_0)$ by angle $\phi$.
A single isothermal ellipsoid component has six parameters:  central right ascension $x_0$, central declination $y_0$, $\Sigma_0$, $r_c$, $\epsilon$, and $\phi$.

In addition to the subcluster components, a spatially flat unclustered component, $\Sigma_{\mathrm{U}} (r)=\Sigma_{\mathrm{U}}$, will be included to model non-clustered young stars in the region (and any remaining background contaminants). Distributed populations of young stars have been found in NGC~6334 and Carina where they may represent a previous generation of star formation \citep{Feigelson09,Feigelson11}.

The finite mixture model for a star-forming region will be the sum of the ellipsoid models for all $k$ subclusters (plus the unclustered component) multiplied by mixture coefficients, $a_i$, given by, 
\begin{equation}
\Sigma_\theta(\mathbf{r}) = \sum_{i=1}^{k+1}a_i\Sigma_{i}(\mathbf{r};\theta_i) = \sum_{i=1}^{k}a_i\Sigma_\mathrm{ell}(\mathbf{r};x_{0,i},y_{0,i},r_{c,i},\phi_i,\epsilon_i)+a_{k+1}\Sigma_{\mathrm{U}},
\end{equation}
where $\theta = \{a_1,x_{0,1},y_{0,1},r_{c,1},\phi_1,\epsilon_1,\ldots, a_k,x_{0,k},y_{0,k},r_{c,k},\phi_k,\epsilon_k,a_{k+1}\}$ denotes the model parameters. Our finite mixture model is only defined over the field of view, so it is integrable and can be used as a probability density function.

\subsection{Finding the Best-Fit Model} \label{fit_model_section}

We adopt the method of maximum likelihood estimation (MLE) that, following the principles laid out by \citet{Fisher22}, seeks the most likely model given the data. 
The log-likelihood function of the model parameters $\theta$ is given by the equation
\begin{equation}\label{likelihood}
\ln L(\theta;\mathbf{X})=\sum_{i=1}^{N} \ln \Sigma_{\theta}(\mathbf{r}_i)-\int_W \Sigma_{\theta}(\mathbf{r}^\prime)\dif \mathbf{r}^\prime,
\end{equation}
under the assumption of a Poisson point process $\mathbf{X}=\{\mathbf{r}_i\}_{i=1}^N$ containing $N$ points (stars), which are spatially distributed with the surface density of the finite mixture model $\Sigma_{\theta}$ in a window (field of view) $W$.
The integral in the second term is the expected number of stars within the field of view, and the model $\Sigma_\theta$ is normalized so that this term equals the total number of observed stars $N$. 

MLE of mixture models for cluster analysis is described by \citet[][Chpt. 6]{Everitt11}.  We seek the mode (highest value) of the likelihood function. 
This can be computationally challenging; for example, if 5 subclusters are present in a MYStIX field, then the parameter space to be searched to maximize the likelihood has $6 \times 5  = 30$ dimensions.  Commonly used computational methods, including the EM Algorithm \citep{McLachlan08} and Markov chain Monte Carlo (MCMC) procedures \citep{Brooks11}, will find good solutions but not necessarily the global maximum.  

Our problem is simpler than many problems where MCMC procedures are needed.  As we are seeking physical associations of stars in the two-dimensional (RA,Dec) subspace of a high-dimensional parameter space, any acceptable subcluster in the final model must be evident as a subcluster in the (RA,Dec) diagram.  We therefore start the MLE computation with a superset of possible clusters formed by visual examination of the MPCM spatial distribution and its adaptively smoothed surface-density map.  A Nelder-Mead optimization algorithm, implemented in \Rlan's function \optim\, is applied to iteratively find the global maximum likelihood solution.  Visual examination of a smoothed map of residuals between the data and the model shows that the MLE model solutions are quite satisfactory in explaining the MPCM spatial structure using isothermal ellipses.  

The \Rlan\ code for model fitting is presented in Appendix~A, using the NGC~6357 MSFR as an example. As is often the case for complex model fits, multiple fitting attempts are needed to find the best fitting model, with attention paid to individual parameters. 

\subsection{Model Selection} \label{model_selection_section}

The above steps require that the number of subclusters $k$ in the model be specified.  Model selection is the process by which $k$ is varied and the ``best'' value of $k$ is chosen that balances optimal fit to the data and some concept of model parsimony.  This requires that the maximum likelihood for each value of $k$ be ``penalized'' by some function of $k$.  The most common choices are the Bayesian Information Criterion (BIC) and the Akaike Information Criterion (AIC):
\begin{eqnarray}
BIC &=& -2 \ln L + (6k+1) \ln N \\
AIC &=& -2\ln L + 2(6k+1),
\end{eqnarray}
where $6k+1$ is the number of parameters for $k$ subclusters, and $N$ is the total number of stars in the MPCM sample.  The optimal $k$ value is chosen to minimize the BIC or AIC.   When models are found with AIC values very close to the best fit model, the final model is chosen to minimize maximum excursions in the smoothed residual maps.  The BIC places a stronger emphasis of model parsimony (simplicity) than the AIC, whereas the AIC sometimes overestimates the number of components.  The choice of AIC, BIC or other model selection approach is widely debated \citep{Lahiri01,Konishi08}. \citet{Burnham02} generally favor the AIC, while the BIC has more theoretical support \citep{Kass95}.  \citet[][sec 6.5.2]{Everitt11} discuss the issue in the context of mixture models for cluster analysis. 

Experimentation with MYStIX samples showed that the BIC model selection criterion produces final models with fewer subclusters, missing sparser clusters that are readily seen in visual inspection of the MPCM spatial distribution.  For a typical MYStIX field with $N \simeq 1000$, the addition of an additional subcluster is accepted to minimize the BIC only if it improves the log likelihood by $6 \times\ln 1000 \simeq 40$  in the likelihood.  However, a subcluster that improves the log likelihood by $\simeq 12$ will be acceptable to the AIC.  We therefore have chosen the AIC as our model selection criterion in order to capture a greater dynamic range of rich and poor clusters.  

This model selection has two further advantages.  First, it obviates the need to arbitrarily specify a critical parameter value in the clustering procedure, such as $k$ in $k$-nearest neighbors or a special branch length in the minimal spanning tree.  Second, it reduces features in the residual map is similar to model selection based on minimizing the ``final prediction error,'' an approach favorably presented by \citep{Rao01}. 

\subsection{Model Validation} \label{model_validation_section}

Maps of model residuals are valuable for both evaluating a model fits and identifying where new subclusters could be added to improve a model. The kernel smoothed ``raw'' residuals for point-process models are defined by \citep{Baddeley05,Baddeley08} and implemented by the ``diagnose.ppm'' function in \spatstat. There will be both positive and negative residuals, and the residuals over the entire region will integrate to zero. A better fit will have residuals with lower amplitude that are randomly distributed throughout the region without any coherent structures in the residuals. The residuals for \mystix\ clusters will be larger where the surface density is higher; however, there should be roughly equal numbers of positive and negative residuals that are not strongly correlated with the locations of the subclusters. 

To smooth the residual maps, we used kernels with bandwidths that are four times the median nearest-neighbor distance in a region, which produce residual maps for the different regions that have similar statistical balances between variance and bias.  Kernels that are too small produce lumpy structures due to stochastic fluctuations, while kernels that are too large blur distinct structures together.  The residual maps often show coherent positive residuals that cannot be eliminated by adding additional subclusters to the model.  These are typically curved or irregular structures that cannot be well modeled by the isothermal ellipsoids used here.

In addition to examination of the smoothed residual map, model goodness of fit may be determined using a $\chi^2$ test based on the number of counts in arbitrarily chosen spatial bins.  If the field of view is divided into $n$ regions $A_i$ with approximately equal number of counts, then the expected number of points in the $i$th regions is $\mathbb{E}(N_i)=\int_{A_{i}} \Sigma(u)\dif u$ for a surface-density model $\Sigma$. For a sufficiently large number of counts, we can test goodness of fit using Pearson's statistic
\begin{equation}
X^2=\sum_{i=1}^{n}\frac{\left[N_i-\mathbb{E}(N_i)\right]^2}{\mathbb{E}(N_i)}.
\end{equation}
For this test, the field of view is tessellated using adaptive polygonal cells that typically contain $\sim$20 points to ensure adequate counting statistics. We calculate p-values assuming $X^2$ is $\chi^2$ distributed with $n$ degrees of freedom.\footnote{The degrees of freedom of the problem may not be well-defined when the data are used for both model fitting and decisions on binning, affecting the interpretation of the $X^2$ statistics \citep{Feigelson12}.}  


\subsection{Subcluster Membership Assignment \label{subcluster_membership_section}}

The finite mixture modeling gives probabilities that any given star is part of a particular subcluster.  This is a ``soft classification'' where probabilities are calculated based on the relative contribution of the different components of the posterior model to the total surface density at the location of a star.  This stands in contrast to ``hard classification'' methods that places every object into a single ``best'' class.

However, approximate lists of stars belonging to each subcluster are very useful for astronomical study, so we generate hard classifications based on the mixture model soft classifier. One possibility for assigning stars to a particular subcluster (or to the unclustered component) is to assign them to the model component that is most probable; this strategy is used by the \mclust\ algorithm \citep{Fraley12}. However, to avoid including the most ambiguous stars in lists for specific subcusters, we additionally require that the probability for the assigned subcluster be better than 30\%. We further require that stars assigned to a subcluster be within an ellipse four times the size of the cluster core. (These decision rules play an analogous role to the choice of $k$ in $k$-nearest neighbor clustering algorithms.)

\subsection{Advantages and Disadvantages of the Finite Mixture Model \label{advantages_and_disadvantages_section}}

The choice of algorithm for cluster finding is related to the data that are used and the scientific goals of analysis. The MPCM data have variations in surface density of several orders of magnitude and complex, overlapping structures. Our algorithm resolves these structures into individual components, and constructs a model for a star-forming region based on a sum of simple astrophysically reasonable cluster shapes (isothermal ellipsoids) that can be used as a basis for studying subcluster ages, relationship to molecular material, physical conditions, and evolutionary state in later \mystix\ papers. These considerations led us to choose the parametric cluster-analysis methods.  However, a this approach may encounter a variety of challenges or disadvantages.

\begin{description}

\item[A model form must be assumed] It is not clear how well finite mixture models work when the component distribution models are poor descriptions of the data. Here we assume that real star clusters have an isothermal ellipsoid distribution function, which is clearly not always true. However, deviations from the isothermal structure can lead to interesting inferences about particular regions. Overall, the clean appearance of the residual maps and excellent global  $\chi^2$ values for the best-fit models (Table~4) show that the assumption of isothermal ellipsoids is satisfactory.

\item[Guidance on the model selection criteria is lacking] 
Statisticians have not settled on a universal penalty to maximum likelihood estimation for model selection. The AIC has a strong theoretical basis in information theory and thermodynamics, the BIC has a strong theoretical basis in Bayesian inference, and both are related to minimum-$\chi^2$ fitting under some circumstances.
Finite mixture model analysis using the BIC is less sensitive at identifying small clusters in fields of view that also contain rich clusters because the model selection criterion involves improvements to the global likelihood, and does not measure localized surface density enhancements. The AIC used here is more effective at finding the small clusters. But, the theory does not exist to interpret changes in the AIC in terms of statistical significance.

\item[Finding the best mixture model is computationally difficult] Fitting takes place in a high-dimensional parameter space, even though the point pattern data are two-dimensional. Additional problems may be encountered with infinite likelihoods, multimodality of the likelihood \citep{Kalai12}, and slow convergence of parameter estimators \citep{Chen95}.

\end{description}

Finite mixture models also have a number of advantages compared to other cluster-finding methods.

\begin{description}

\item[Estimators are based on well-accepted statistical procedures] Finite-mixture model fitting and selection rely on penalized likelihood techniques widely used across statistics.   In contrast, nonparametric methods, such as the friends-of-friends and the $k$-nearest neighbor algorithms, provide no mathematical guidance to estimate the number of clusters. They often exhibit weak performance in the presence of noisy or high-dynamic range structures \citep{Everitt11}.

\item[A critical scale or surface density is not assumed] Cluster size and central density are free parameters for each subcluster in the finite mixture models, so structures at vastly different spatial scales can be fit within a single region. This is in sharp contrast to most cluster algorithms, which rely on a single spatial threshold (e.g.\ critical distance or density) applied throughout the region. Given that it is an open area of research whether or not subclusters of young stars form with a uniform surface density, applying a single spatial threshold for cluster detection risks biasing the result with a biased assumption.  Decision rules that play the role of a critical scale are applied only when the stellar membership of a given subcluster needs to be defined.  

\item[Distributed populations are treated] The finite-mixture model does not insist that the clusters include every star, but rather allow relatively isolated stars to be assigned to a distributed population with uniform surface density. In this way, our approach is similar to density based clustering algorithms, rather than standard nonparametric clustering algorithms that require that every object lie in a cluster.

\item[Probabilistic cluster membership assignments are used] The finite-mixture model is a soft classification method, rather than a hard classification method. Each point has a membership probability associated with each cluster all of which sum to one, and often one probability is much greater than the others. Therefore, different probability thresholds can be used for different applications that require different levels of reliability. The hard classification methods do not include this information, and the ambiguous membership will be definitively assigned into a cluster.

\item[The models are flexible and effective] The parametric model chosen can be modified to suit the situation being modeled. Other methods may require incorrect assumptions (e.g.\ normal distributions) or lack any explicit assumptions.  An elliptical isothermal model permits a wide array of different morphologies. If the model is astrophysically reasonable, then the parametric clustering method is better optimized for detection structures than a nonparametric method.

\end{description}

\section{Results} \label{results_section}

\subsection{Surface Density Maps \label{maps_section}}

Surface density maps of young stars are created using an adaptive density estimator on projected star positions. The density estimator is based on the Voronoi tessellation \citep{Ogata03,Ogata04,Baddeley07}.  A randomly selected set of stars containing fraction $f$ of the stars is used to create a Voronoi tesselation of the region, which subdivides the field of view into cells that tend to be smaller in dense regions and larger in less dense regions. The rest of the stars are used to estimate the surface density in each cell. We choose a fraction $f=1/10$ and repeat this randomized procedure 100 times producing multiple estimates of surface density that are averaged to create the smoothed map.

The adaptively smoothed maps for each star-forming regions are presented in Figure~\ref{orion_map_fig}, with surface density indicated by color and brown contours showing increases by factors of 1.5. The maps show the surface density of the MPCMs from the ``flattened sample'' (Section~\ref{sample_section}).  These are $observed$ stellar surface densities not $intrinsic$ stellar surface densities that depend on the distances and telescope exposure times for each region.  The values therefore cannot be directly compared between regions.  

Surface densities vary up to a factor of $\sim$100 within a single regions, and many regions are multimodal with local maxima differing in surface density by more than an order of magnitude. An effective cluster finding algorithm must be able to identify these different subclusters despite the high contrast in surface density. 

\clearpage\clearpage

\begin{figure}
\centering
\includegraphics[angle=0.,width=3.0in]{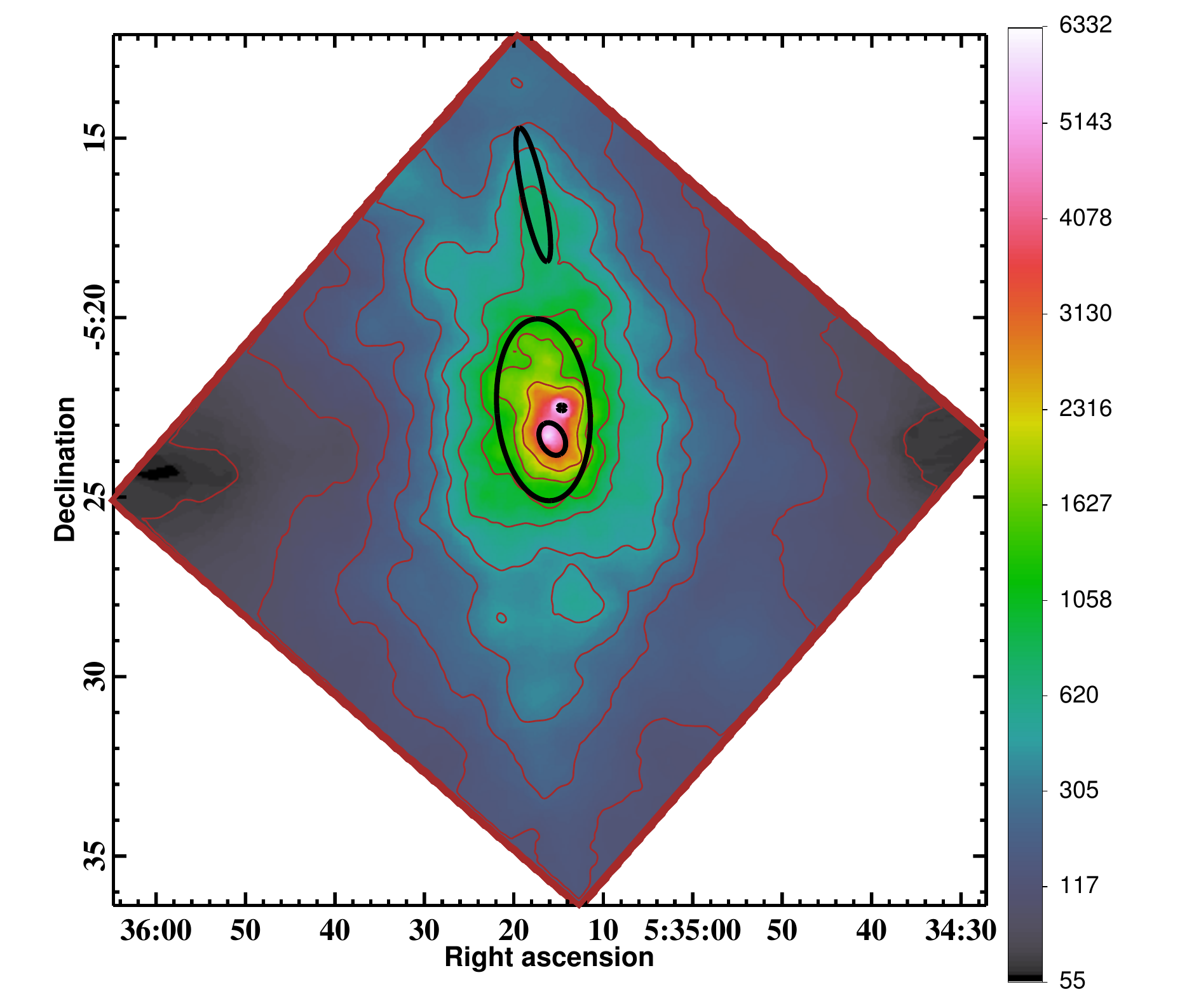}
\includegraphics[angle=0.,width=3.0in]{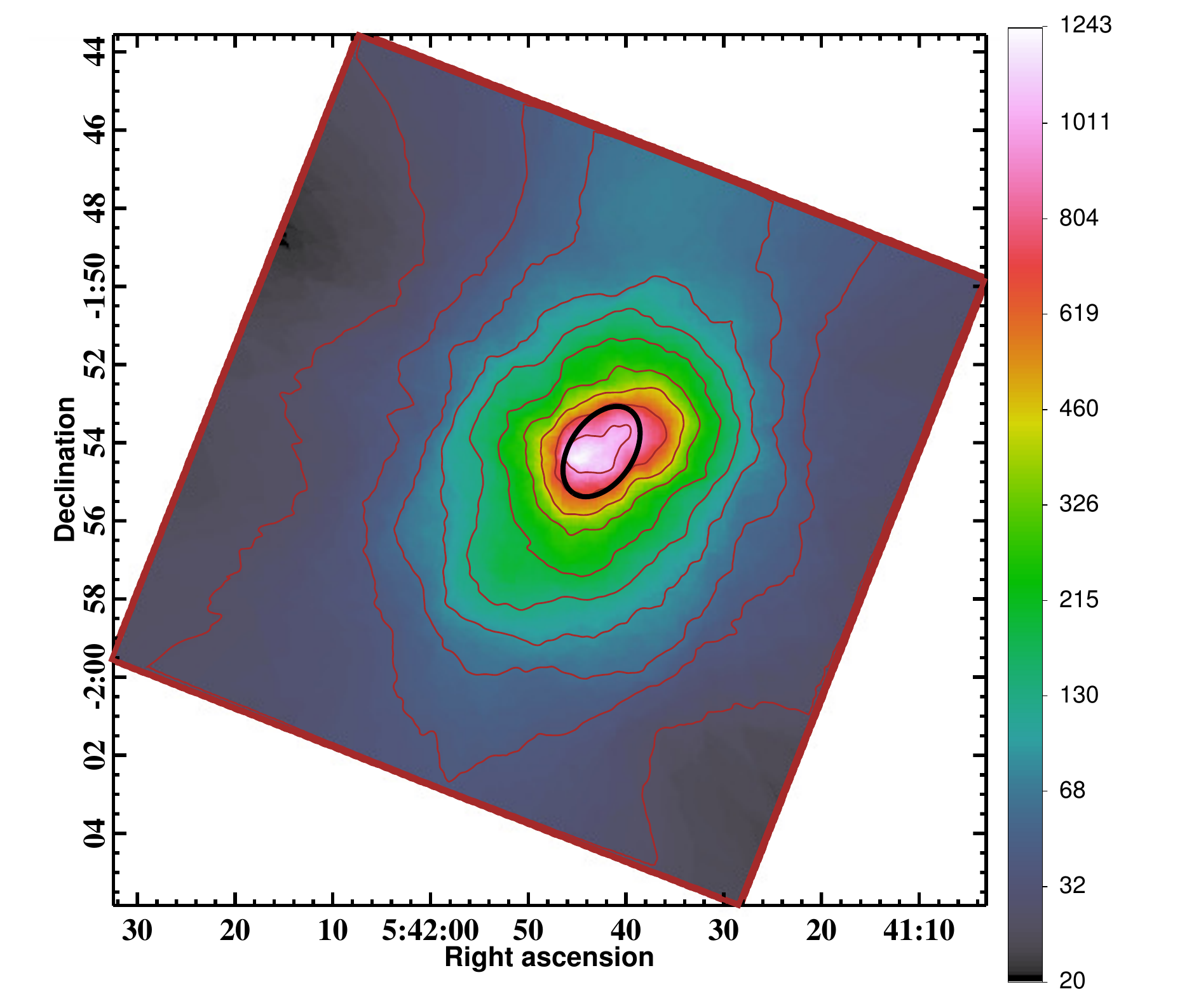}
\includegraphics[angle=0.,width=3.0in]{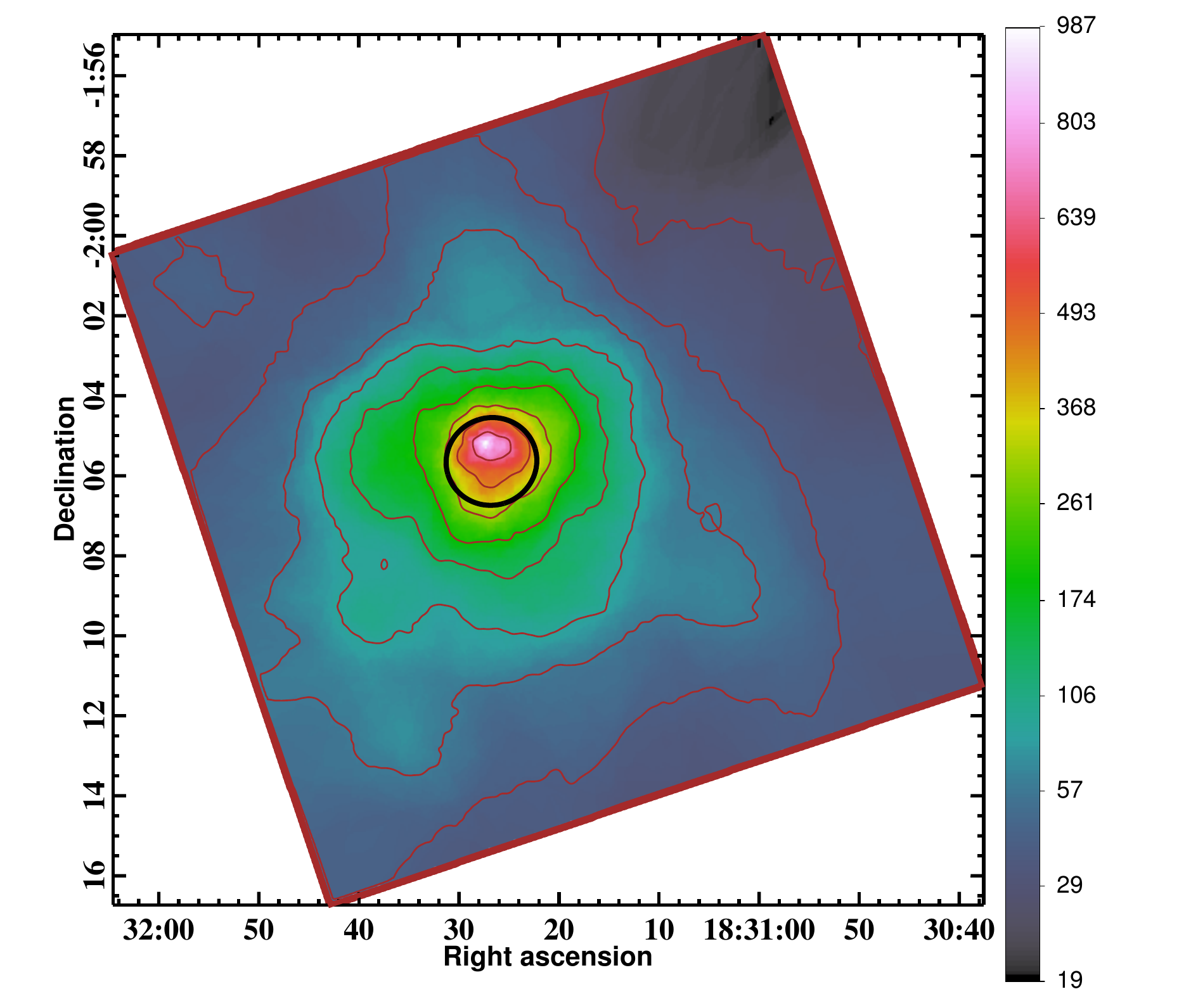}
\includegraphics[angle=0.,width=3.0in]{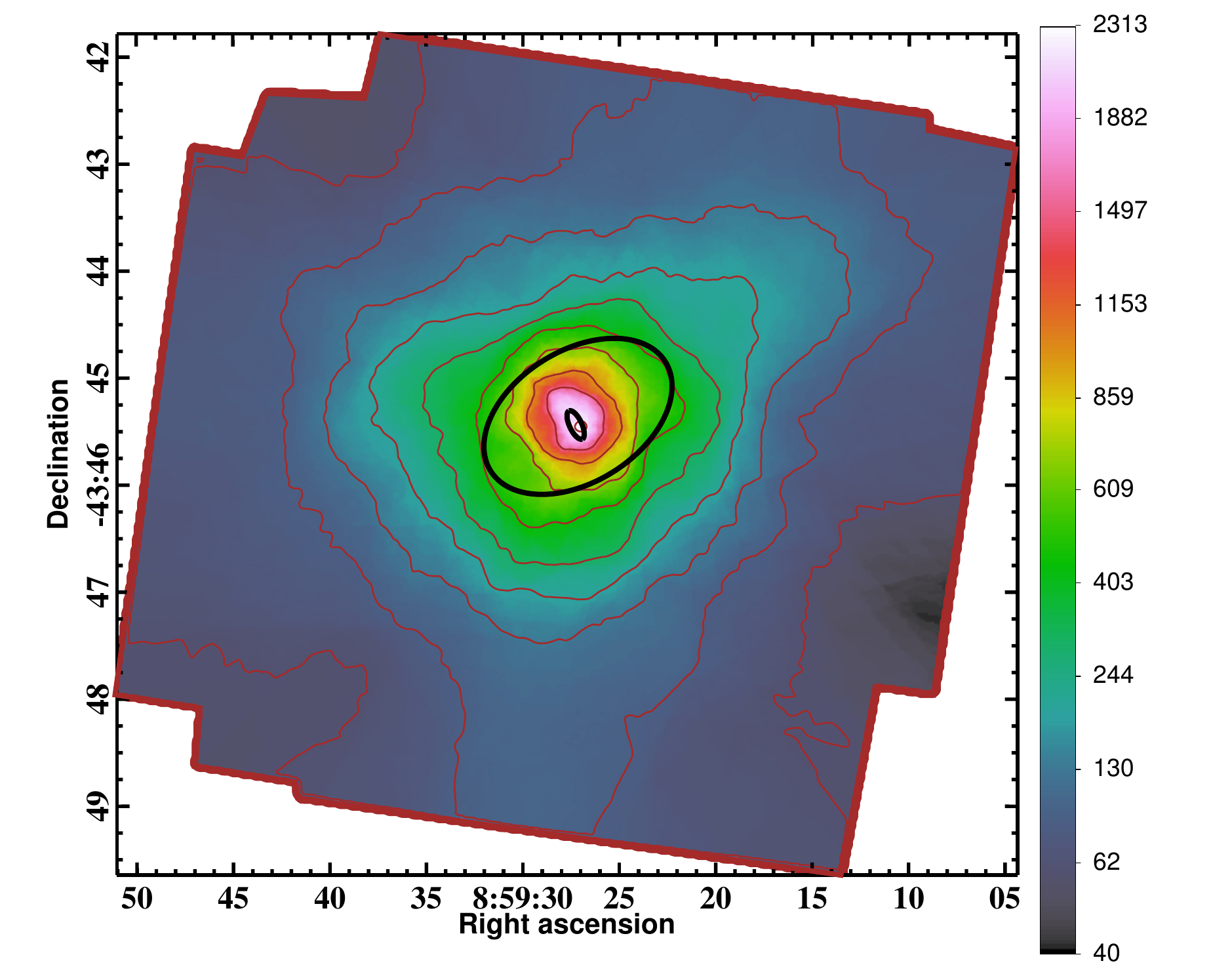}
\caption{The smoothed projected stellar surface density is shown, with a color bar in units of observed stars~pc$^{-2}$. Brown contours show increase in surface density by factors of 1.5. Black ellipses mark the core regions of the isothermal ellipsoid subclusters. Left to right and top to bottom: Orion, Flame, W 40, RCW 36.
\label{orion_map_fig}}
\end{figure}

\clearpage
\clearpage
\setcounter{figure}{1}
\begin{figure}
\centering
\includegraphics[angle=0.,width=5.0in]{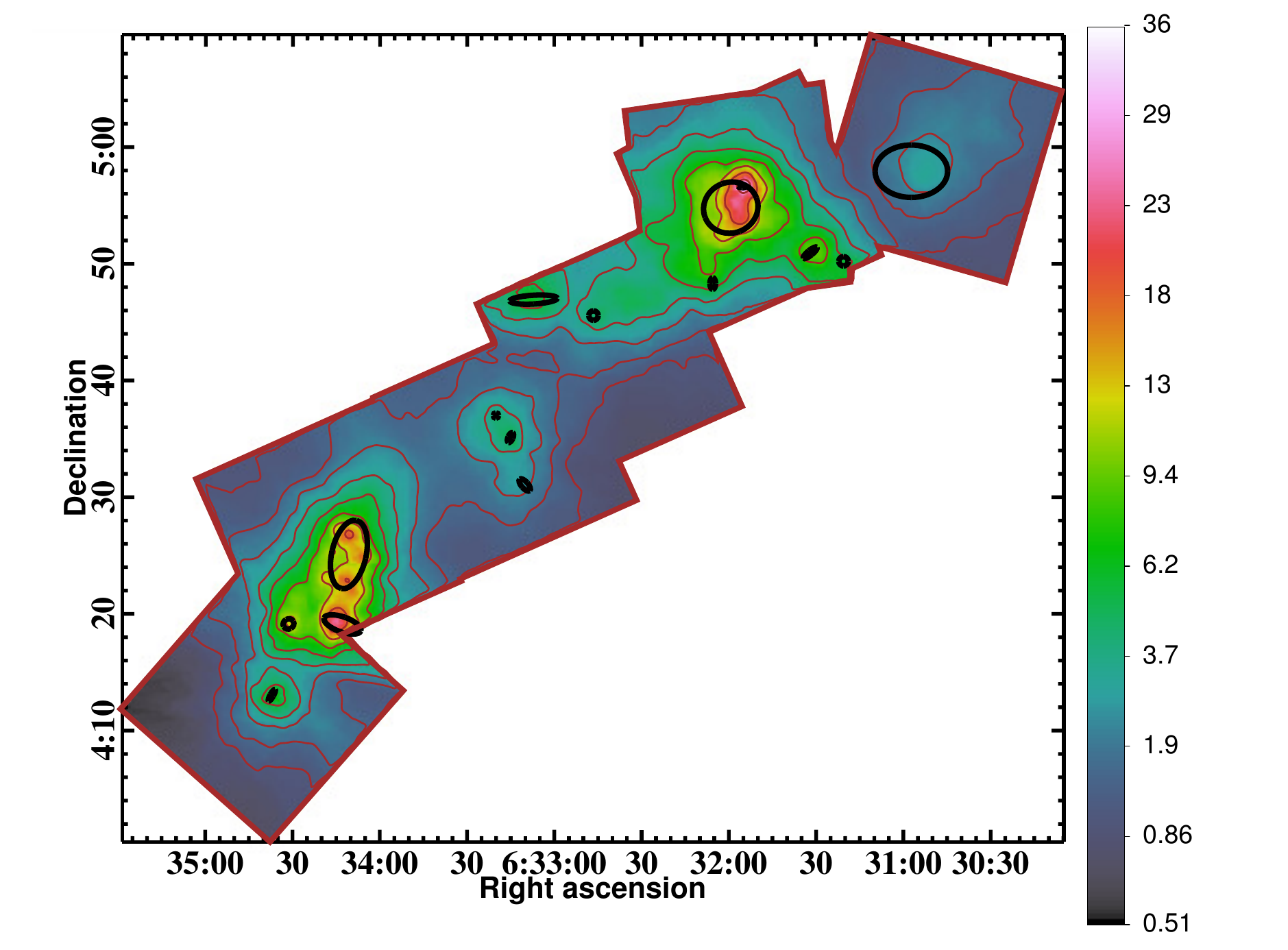}
\includegraphics[angle=0.,width=3.0in]{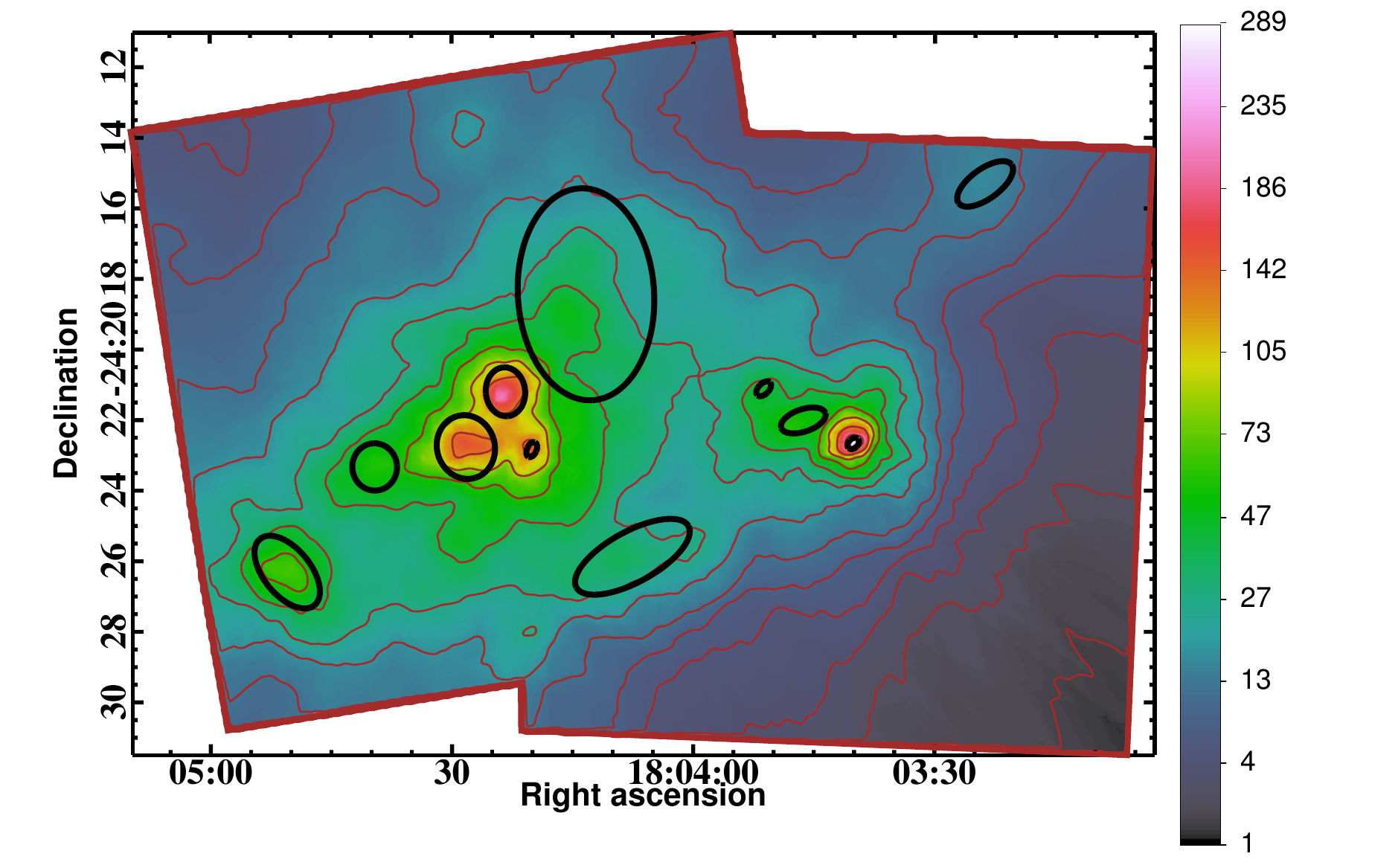}
\includegraphics[angle=0.,width=3.0in]{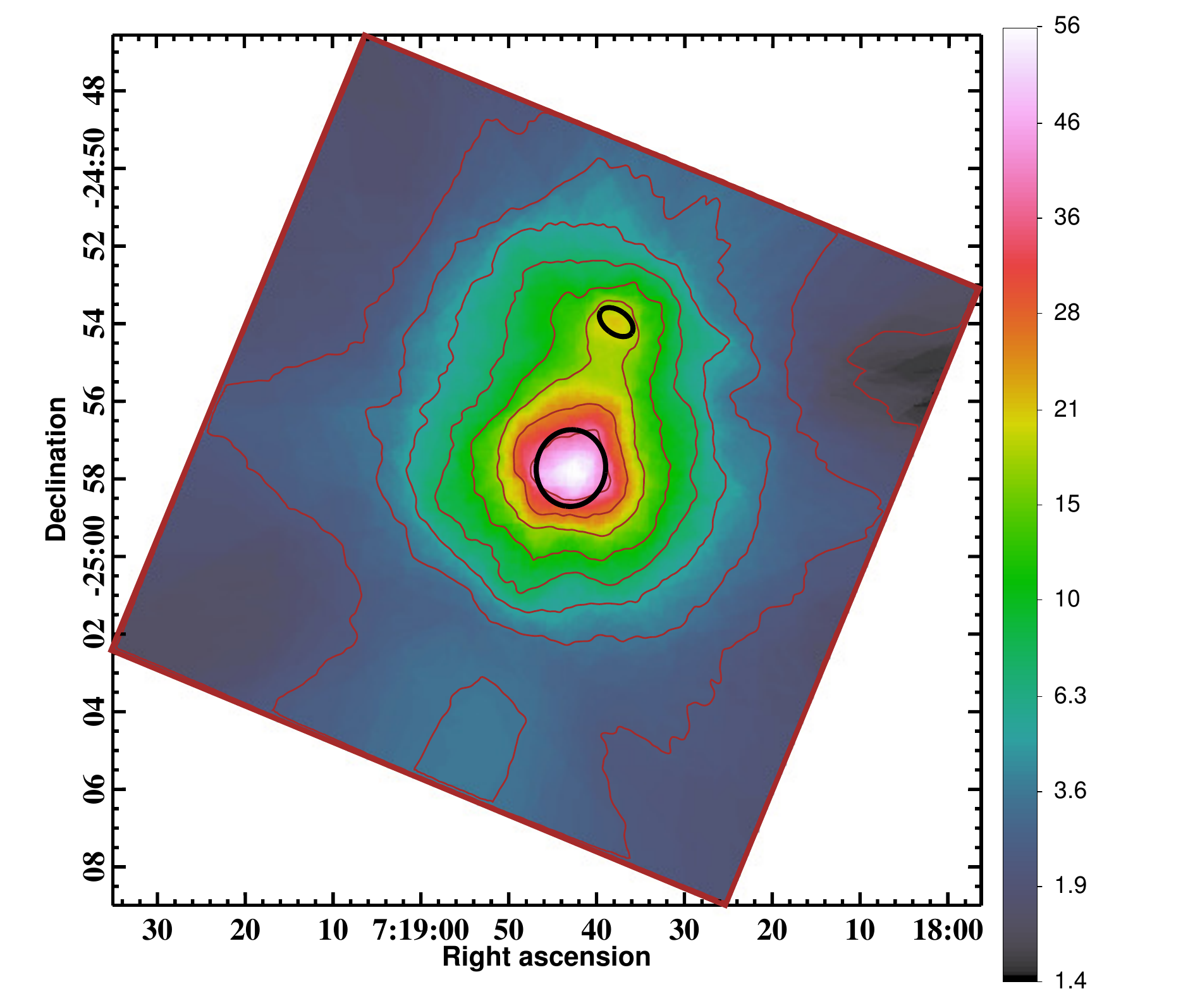}
\caption{Left to right and top to bottom: Rosette, Lagoon, NGC 2362.
\label{orion_map_fig}}
\end{figure}

\setcounter{figure}{1}
\begin{figure}
\centering
\includegraphics[angle=0.,width=3.0in]{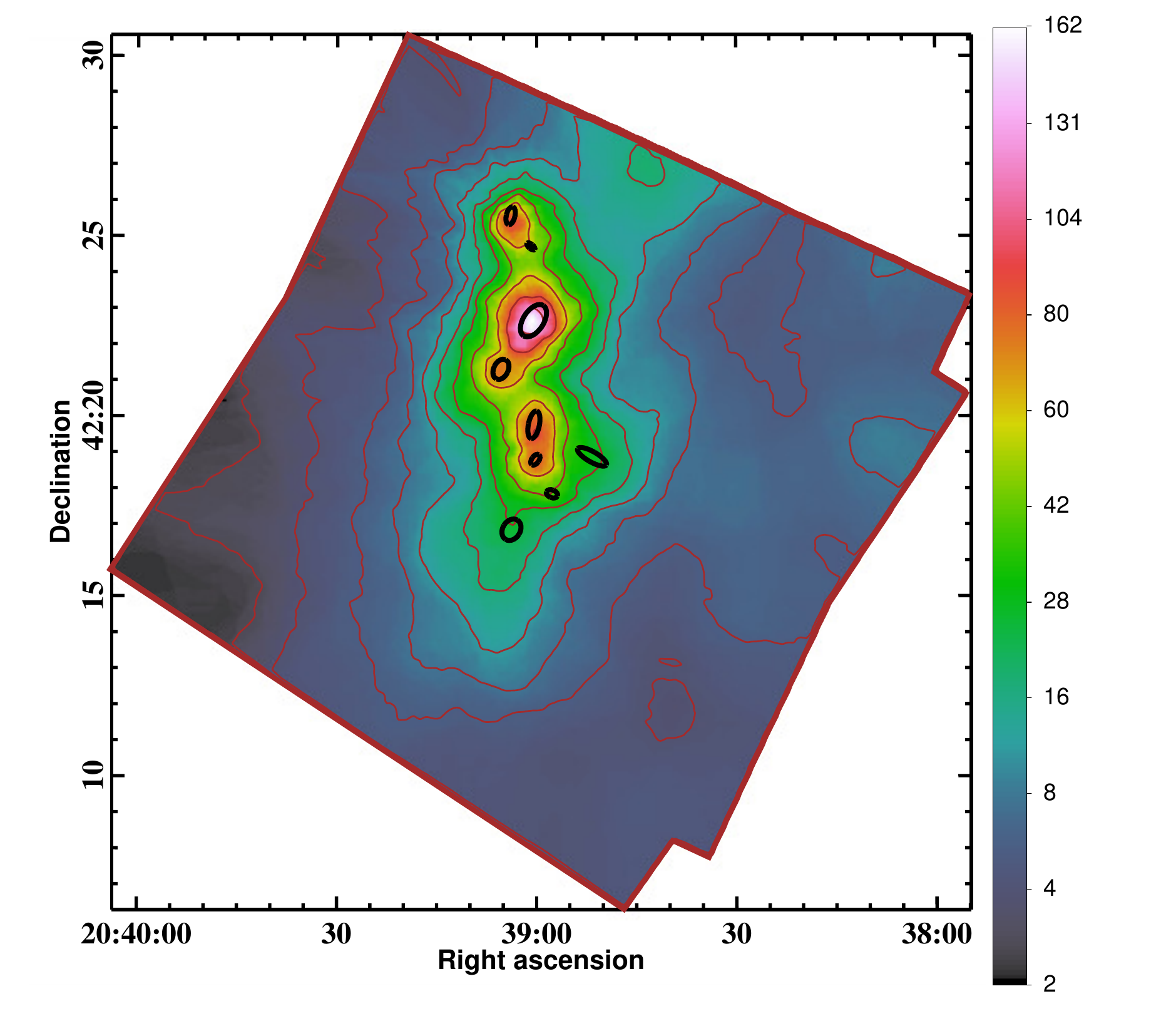}
\includegraphics[angle=0.,width=3.0in]{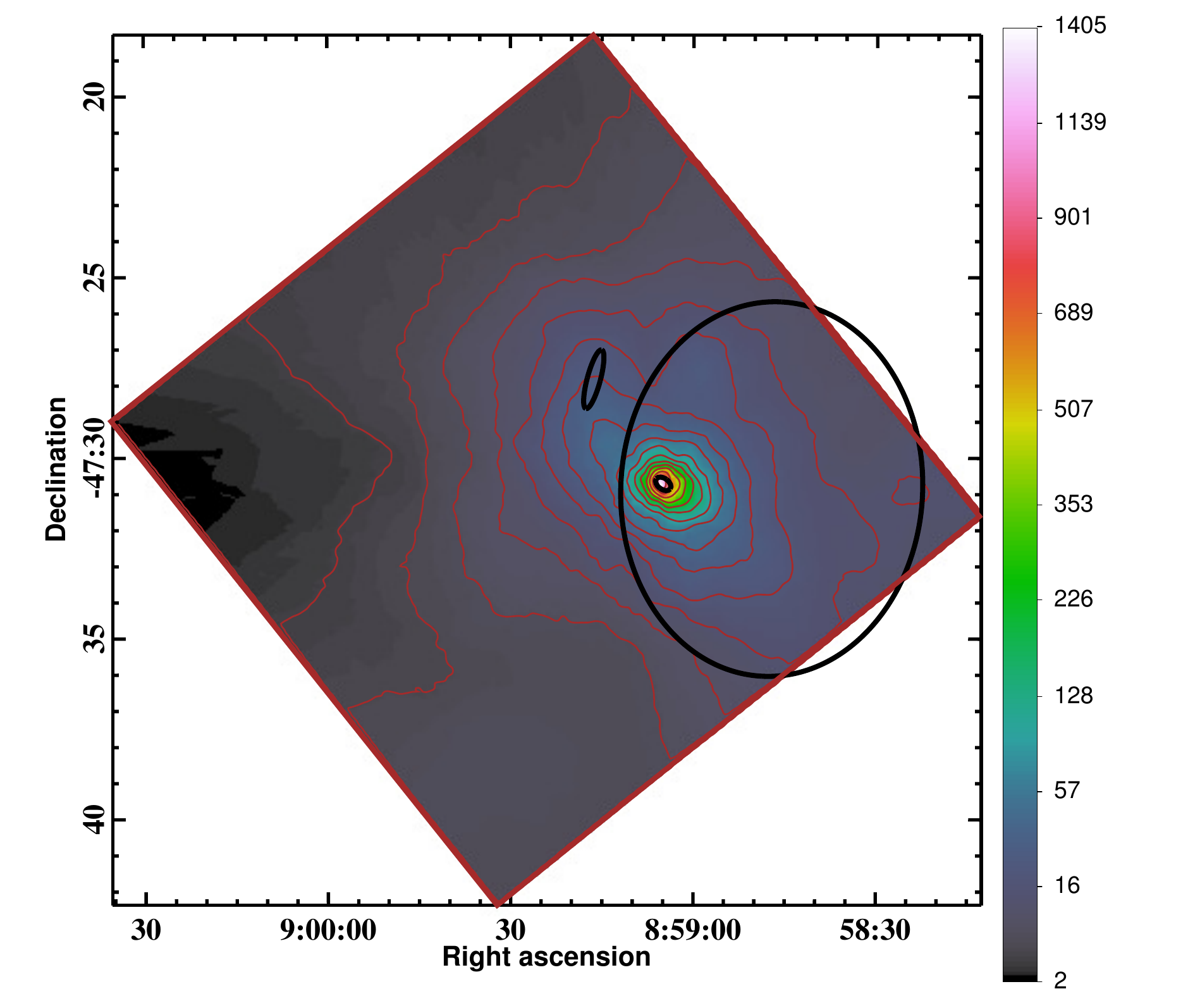}
\includegraphics[angle=0.,width=3.0in]{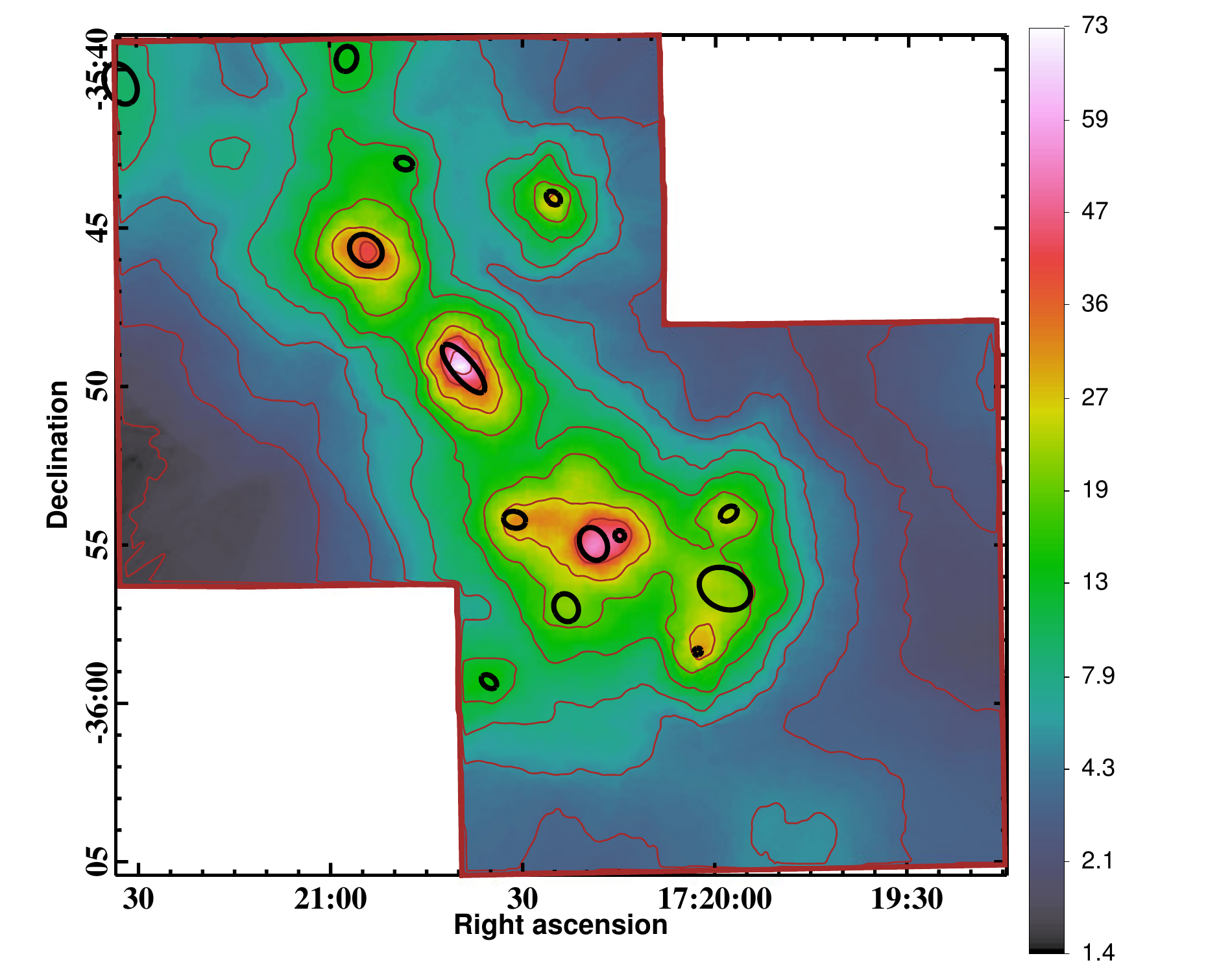}
\includegraphics[angle=0.,width=3.0in]{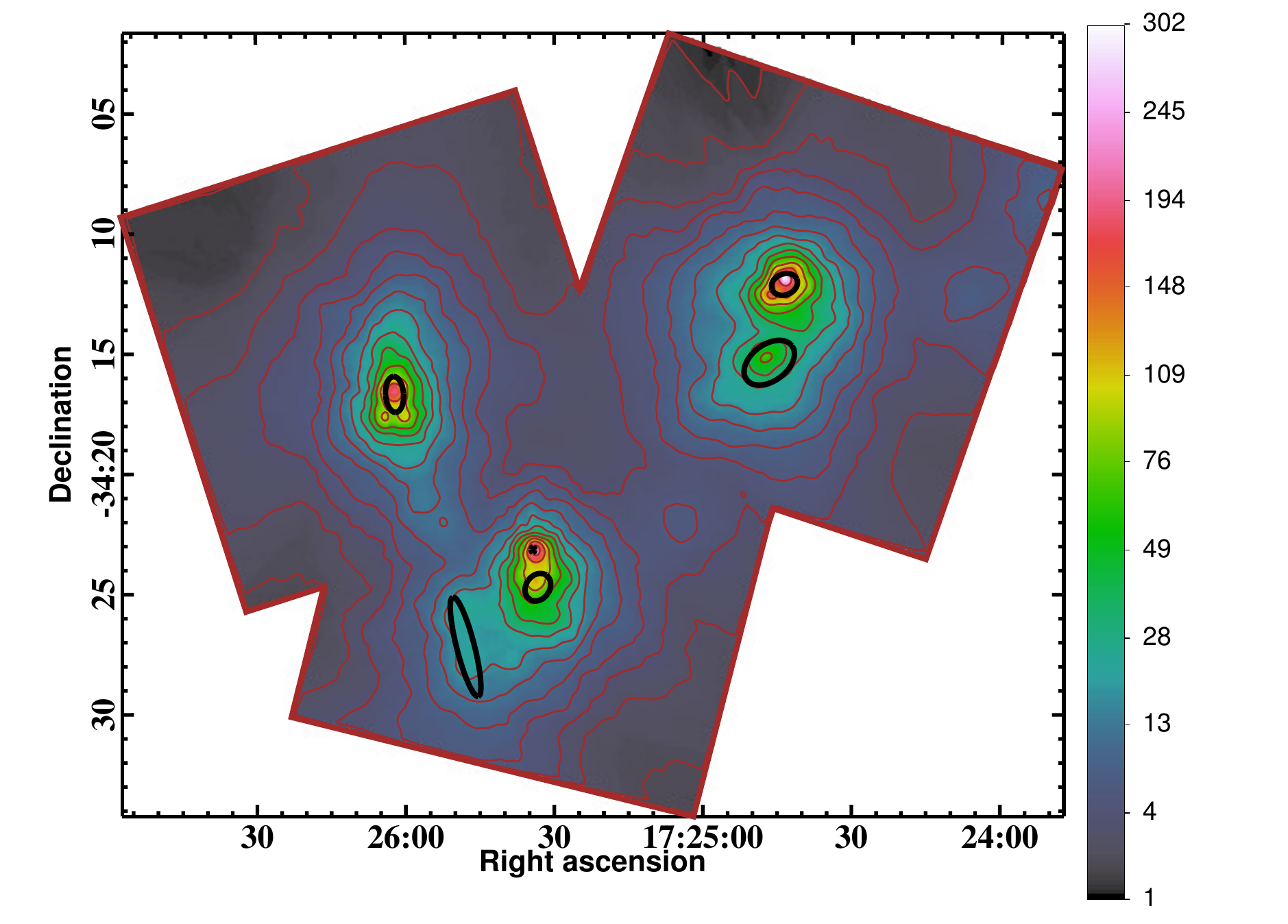}
\caption{Left to right and top to bottom: DR 21, RCW 38, NGC 6334, NGC 6357.
\label{orion_map_fig}}
\end{figure}

\setcounter{figure}{1}
\begin{figure}
\centering
\includegraphics[angle=0.,width=3.0in]{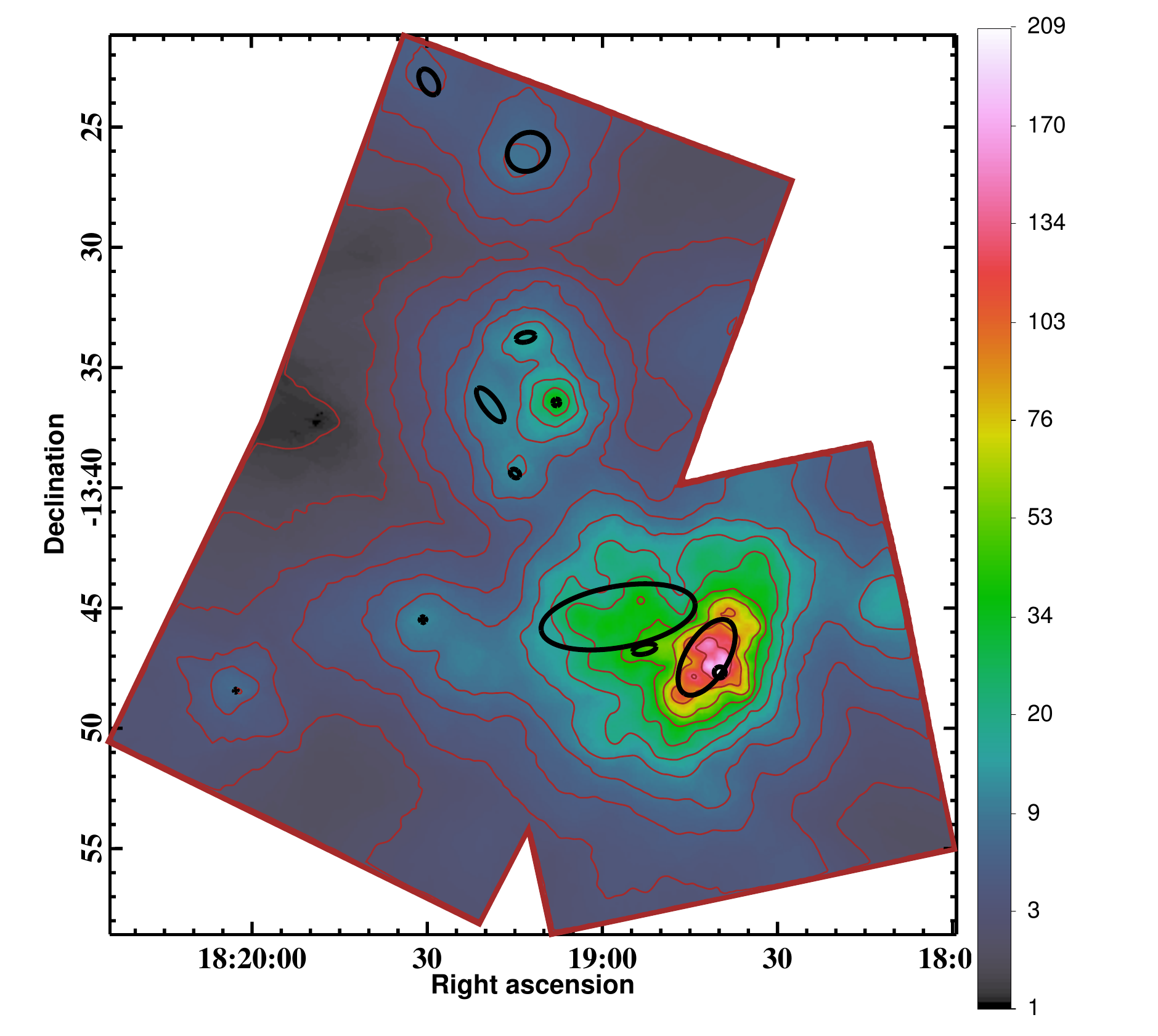}
\includegraphics[angle=0.,width=3.0in]{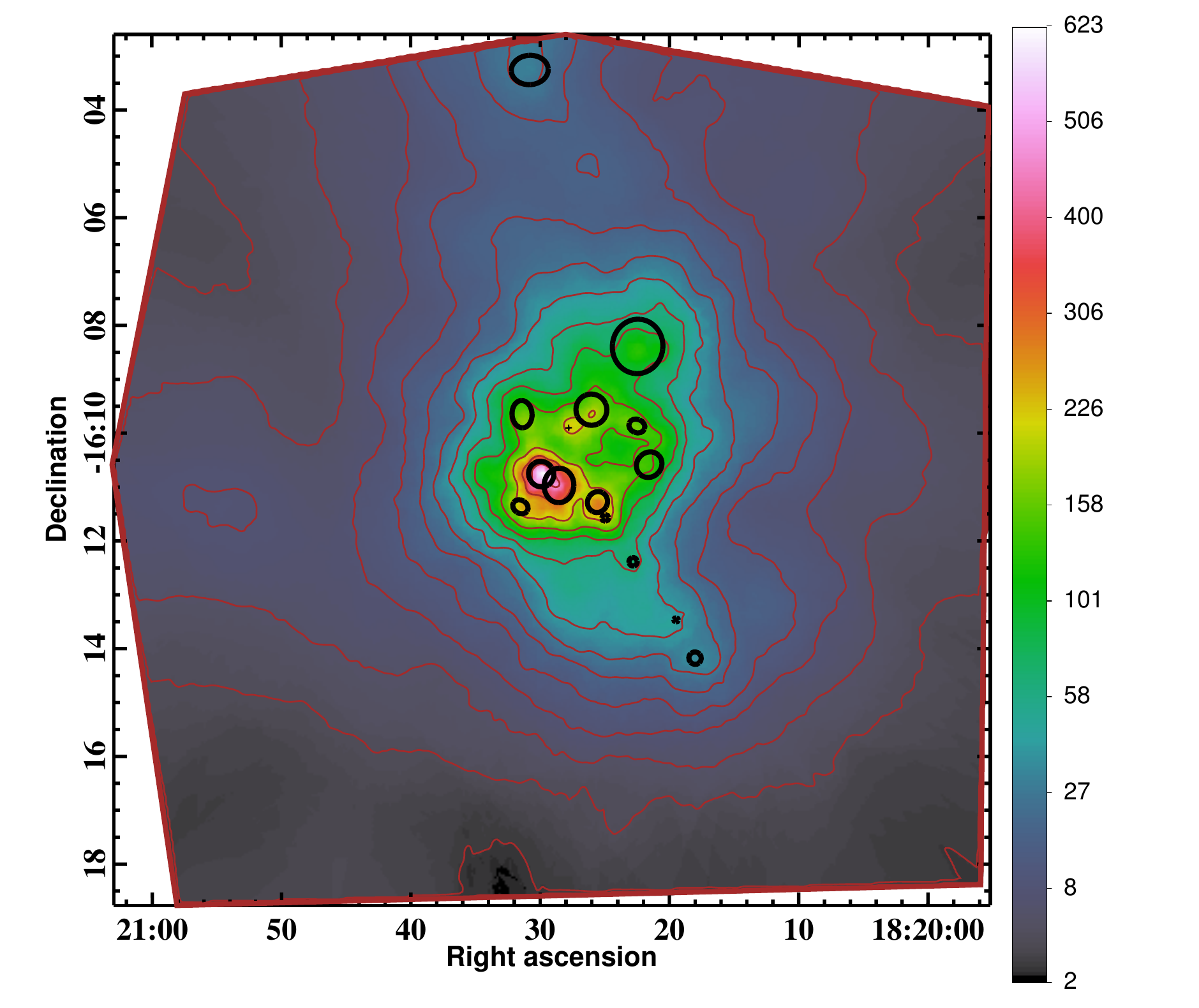}
\includegraphics[angle=0.,width=3.0in]{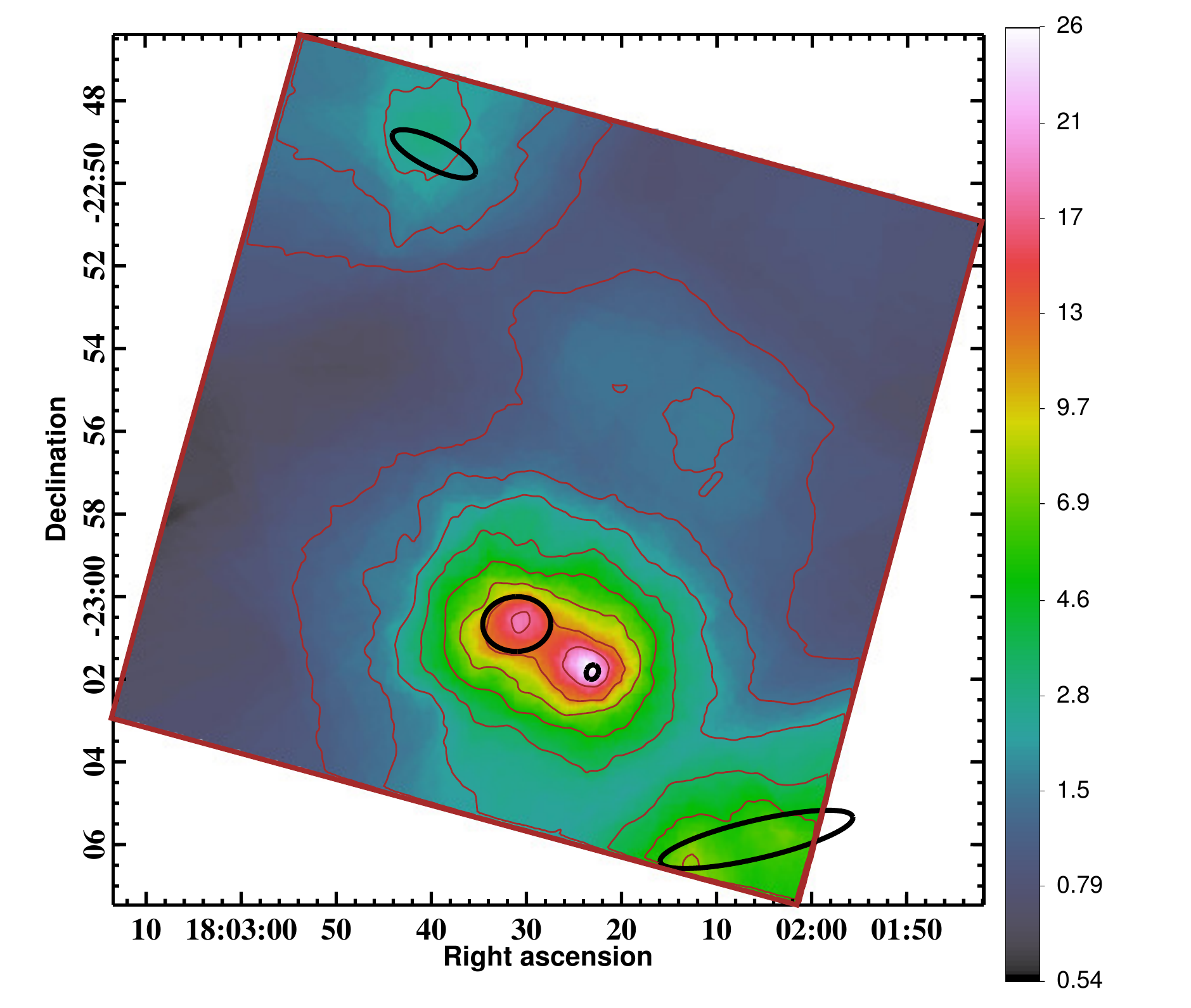}
\includegraphics[angle=0.,width=3.0in]{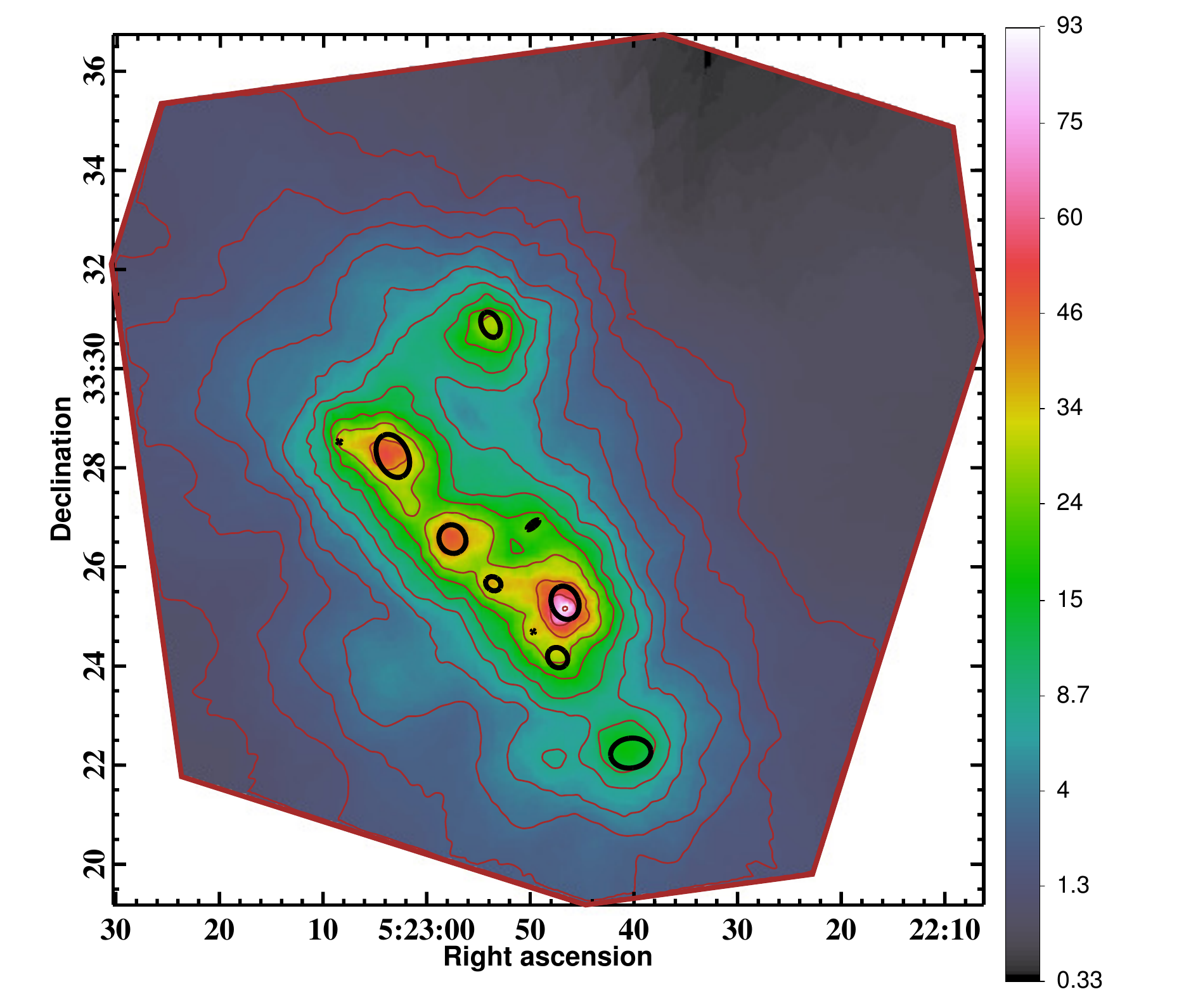}
\caption{Left to right and top to bottom: Eagle, M 17, Trifid, NGC 1893.
\label{orion_map_fig}}
\end{figure}

\setcounter{figure}{1}
\begin{figure}
\centering
\includegraphics[angle=0.,width=3.0in]{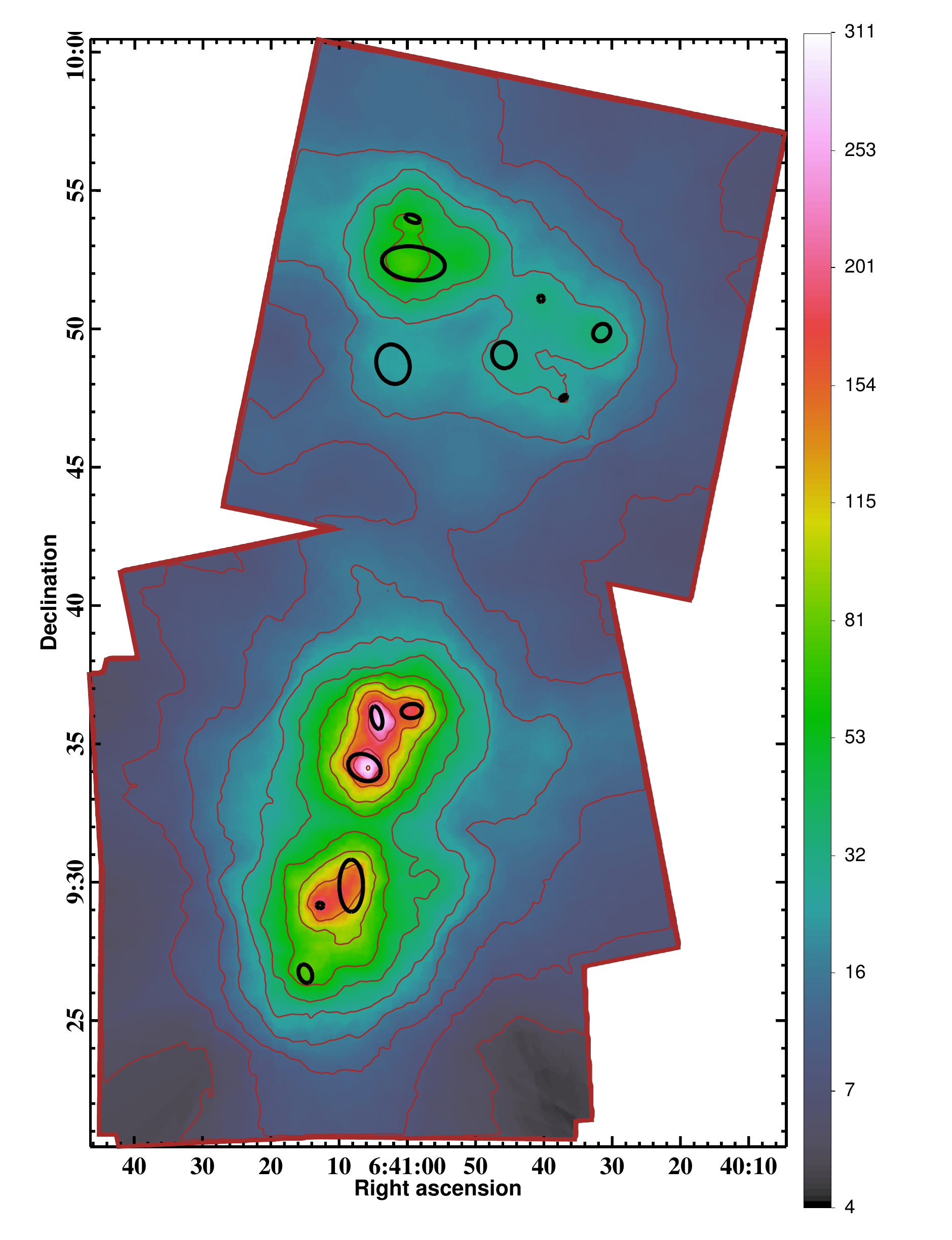}
\includegraphics[angle=0.,width=3.0in]{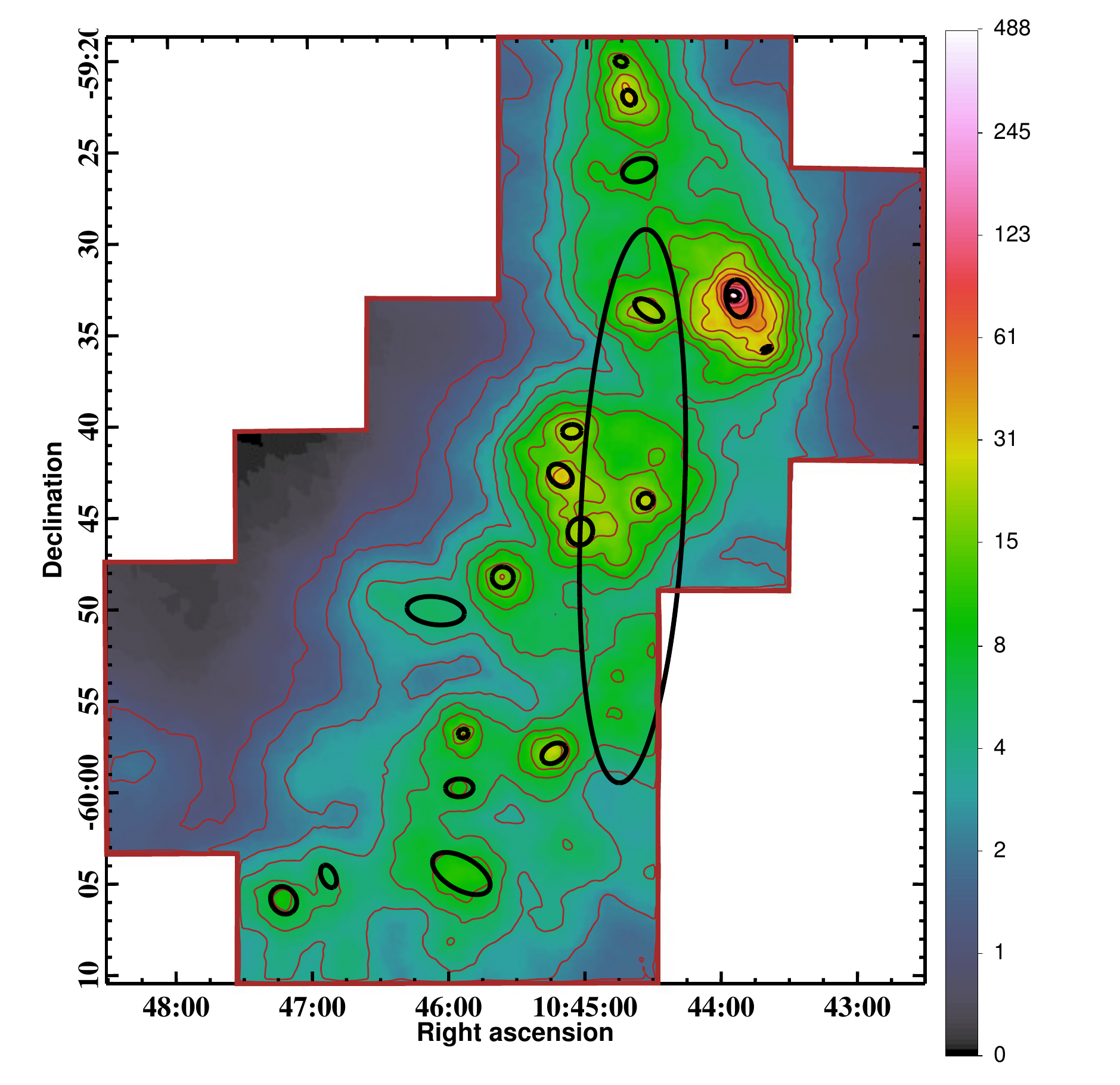}
\caption{Left to right: NGC 2264, Carina.
\label{orion_map_fig}}
\end{figure}

\clearpage\clearpage

\subsection{Subclusters \label{subclusters_section}}

Table \ref{ellipsoid_1_table} is a catalog of the 142 subclusters found in the best-fit models of 17  \mystix\ MSFRs. For each subcluster (letter designations are assigned from west to east) the table provides celestial coordinates of the subcluster center, semi-major and semi-minor axes of the core ellipse in parsecs, ellipticity, orientation, and total number of stars within an ellipse four times the size of the core --- all of which are obtained from model fitting (\S\ref{methodology_section}).  In Figure~\ref{orion_map_fig}, subclusters are shown overlaid on the smoothed surface density maps using a black ellipses to mark the core ellipse of each subcluster.

Some subclusters are dominated by X-ray selected sources, while others are dominated by IR-excess selected sources. For example, in the Eagle Nebula region, the main star cluster is dominated by the former while the embedded subclusters to the north and east are dominated by the later. Thus, the combined selection criteria of the \mystix\ survey help reveal a fuller picture of structures.  But observed star counts for individual subclusters can not be directly compared without correction for detection sensitivity effects.

\begin{figure}
\centering
\includegraphics[angle=0.,width=5.0in]{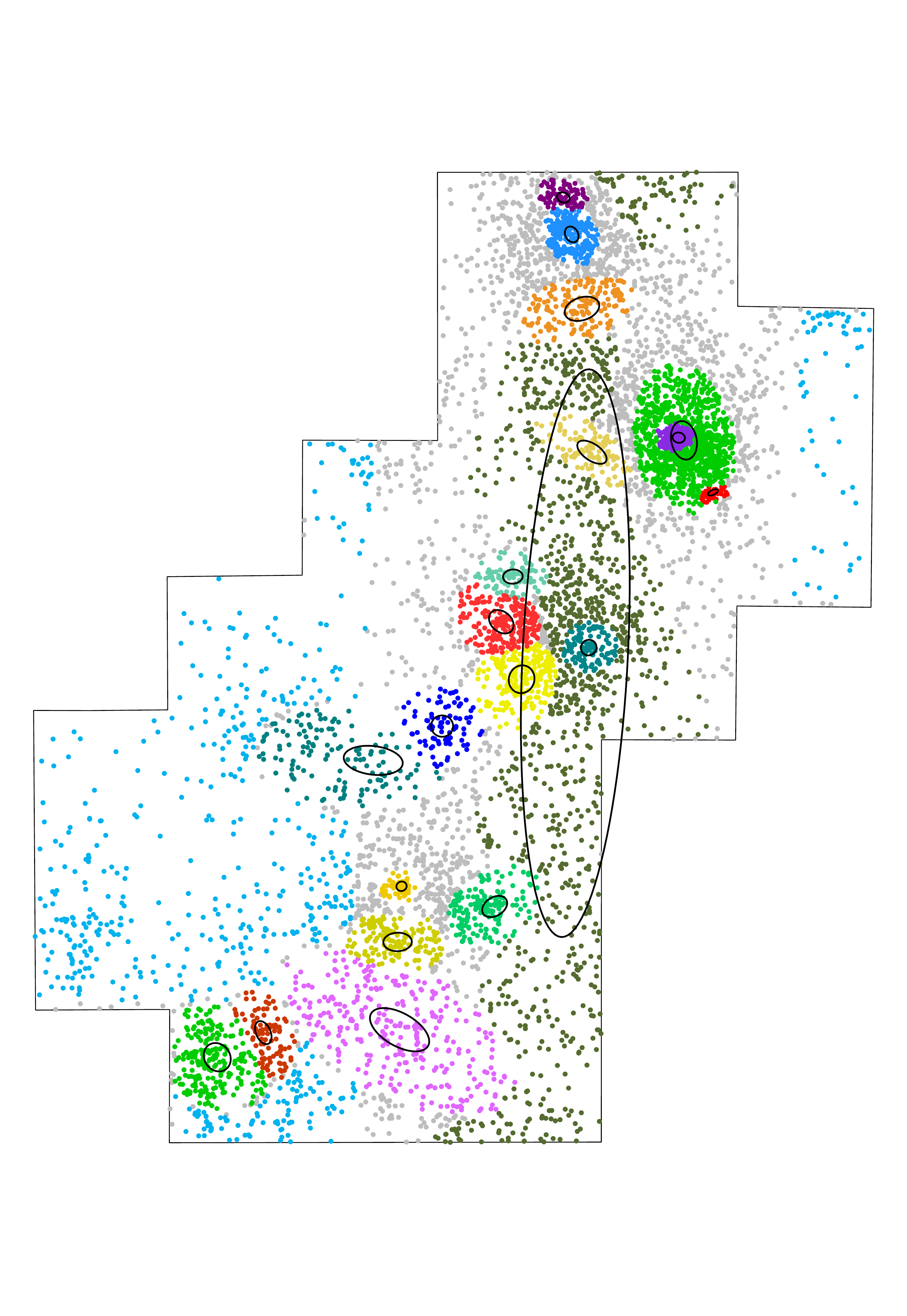}
\caption{
The MPCM stars in the Carina Nebula field of view are shown as colored points indicating subcluster assignments: 20 colors for 20 subclusters, light blue points for the non-clustered stellar population, and gray for the unassigned stars. Black ellipses represent the 20 subclusters. All MPCMs are shown (not only the statistical sample defined in Section~2.4), so the ``egg-create'' effect is evident. Diagrams of the other 16 regions are included in an online-only figure set.
\label{carina_membership_fig}
}
\end{figure}

The assignments of individual MPCM stars to subclusters (or to the unclustered young stellar component) is shown in Figure~\ref{carina_membership_fig} and listed in Table~\ref{subcluster_membership_table}. Table~\ref{subcluster_membership_table} also provides X-ray and near-infrared properties --- these are used in Table~\ref{ellipsoid_1_table} to calculate the median $J-H$ and median X-ray $ME$ values.  In low-density subclusters, the number of stars inferred from the model, $N_{4,\mathrm{obs}}$, may not precisely equal the number of stars assigned to that subcluster if there is spatial overlap with neighboring subclusters leading to ambiguities in assignments.

One can see some strong successes in the decomposition of stars in to isothermal ellipsoids, but also some limitations.  In the Orion region (Figure \ref{orion_annotated_fig}) the subclusters disentangle layered structures in the region, including the BN/KL cluster \citep{Becklin67,Kleinmann67}, the rich ONC, and stars in the north-south OMC-2/3 filament \citep{Megeath12}. Furthermore, two ellipsoids are needed to fit the ONC --- one corresponding to a dense ``core'' and the other corresponding to a less dense ``halo.'' This result agrees with \citet{Henney98}, who find that Orion proplyd and non-proplyd stars are strongly peaked around $\theta^1$~Ori C with statistically-significant excess above the surface densities expected from the King profile fit by \citet{Hillenbrand98}. On the other hand certain structures are not modeled: no distinction is made between the unabsorbed ONC members and the Orion Molecular Cloud stars \citep[OMC;][]{Feigelson05}, the OMC-1S subcluster \citep{Lada04} is not identified, and stars from the $\iota$~Ori cluster \citep{Alves12} may contribute to the subclusters and unclustered component. 

On occasion, the minimum AIC criterion requires a subcluster that appears very small with few stellar members; for example, subclusters B, E, F, J, K in Rosette and subcluster J in M~17 have fewer than 5 stars. This can be a consequence of poorly constrained core radii (see discussion in Section~6.1).  One of these sparse structures is Subcluster~B in M~17 which is associated with a well-studied embedded ultracompact H~II region \citet{Kleinmann73}.  Nevertheless, these subclusters often correspond to clumps of stars that appear real to visual inspection. On the other hand, occasionally model components contain large numbers of stars that do not appear to be discrete subclusters. This includes subcluster~G in Carina and subcluster E in NGC~6357. 

\citet[][their Figures 7--22]{Getman13a} show subclusters superimposed on 500~$\micron$ {\it Herschel}-SPIRE maps of the molecular clouds in the star-forming region. Some of the sparser structures are coincident with molecular cloud cores, providing corroborating evidence that they are real embedded subclusters. Some examples include, NGC~2264 (Subclusters~G, I, J, K, L, and M), Rosette (Subclusters~J, M, N, O), Lagoon (Subclusters~B, K), DR~21 (Subclusters~H, F, E, D, C), NGC~6334 (Subclusters~M, L, J, E, D), Eagle (Subclusters~E, G, H, L), and NGC~1893 (Subclusters~J and G).

\clearpage\clearpage
\begin{deluxetable}{lcccccrrcc}
\tablecaption{MYStIX Subcluster Properties\label{ellipsoid_1_table}}
\tabletypesize{\small}
\tablewidth{0pt}
\tablehead{
\colhead{Subcluster} & \colhead{$\alpha$} & \colhead{$\delta$} & \colhead{$r_{c,\mathrm{major}}$} & \colhead{$r_{c,\mathrm{minor}}$}& \colhead{$\epsilon$} & \colhead{$\phi$} & \colhead{$N_{4,\mathrm{obs}}$} & \colhead{$J-H$} & \colhead{$ME$} \\
\colhead{} & \colhead{(J2000)} & \colhead{(J2000)} & \colhead{(pc)} & \colhead{(pc)} & \colhead{} & \colhead{(deg)}  & \colhead{(stars)} & \colhead{(mag)} & \colhead{(keV)} \\
\colhead{(1)} & \colhead{(2)} & \colhead{(3)} & \colhead{(4)} & \colhead{(5)} & \colhead{(6)} & \colhead{(7)} & \colhead{(8)} & \colhead{(9)} & \colhead{(10)}
}
\startdata
Orion A &  05 35 14.6  &-05 22 31 & 0.01 & 0.01 & 0.23 & 85 & 18 & 0.96 & 3.6 \\
Orion B &  05 35 15.7  &-05 23 23 & 0.06 & 0.04 & 0.30 & 28 & 73 & 0.87 & 1.6 \\
Orion C &  05 35 16.7  &-05 22 34 & 0.31 & 0.16 & 0.49 & 5 & 834 & 1.05 & 1.6 \\
Orion D &  05 35 17.8  &-05 16 35 & 0.23 & 0.04 & 0.84 & 12 & 48 & 1.17 & 1.4 \\
\cline{1-10}
Flame A &  05 41 42.5  &-01 54 14 & 0.15 & 0.10 & 0.37 & 146 & 219 & 1.79 & 2.8 \\
\cline{1-10}
W 40 A &  18 31 26.7  & -02 05 39 & 0.17 & 0.16 & 0.04 & 107 & 187 & 2.10 & 2.5 \\
\cline{1-10}
RCW36 A &  08 59 27.1  &-43 45 22 & 0.20 & 0.13 & 0.36 & 122 & 196 & 1.63 & 2.3\\
RCW36 B &  08 59 27.3  &-43 45 26 & 0.03 & 0.01 & 0.61 & 25 & 24 & 2.14 & 2.8 \\
\cline{1-10}
NGC 2264 A &  06 40 31.5  &+09 49 52 & 0.09 & 0.08 & 0.14 & 136 & 11 & 0.83 & 1.1 \\
NGC 2264 B &  06 40 37.1  &+09 47 31 & 0.04 & 0.02 & 0.44 & 124 & 5 & 0.60 & 1.1 \\
NGC 2264 C &  06 40 40.4  &+09 51 06 & 0.03 & 0.03 & 0.11 & 8 & 6 & 0.73 & 1.1 \\
NGC 2264 D &  06 40 45.8  &+09 49 03 & 0.13 & 0.11 & 0.11 & 194 & 9 & 0.65 & 1.1 \\
NGC 2264 E &  06 40 59.1  &+09 52 22 & 0.30 & 0.16 & 0.47 & 83 & 54 & 0.63 & 1.1 \\
NGC 2264 F &  06 40 59.2  &+09 53 59 & 0.07 & 0.03 & 0.54 & 69 & 12 & 0.61 & 1.0 \\
NGC 2264 G &  06 40 59.4 & +09 36 12 & 0.10 & 0.07 & 0.31 & 96 & 31 & 2.11 & 2.3 \\
NGC 2264 H &  06 41 02.1 & +09 48 44 & 0.19 & 0.16 & 0.19 & 19 & 15 & 0.65 & 1.0 \\
NGC 2264 I &  06 41 04.5  &+09 35 57 & 0.11 & 0.05 & 0.55 & 14 & 31 & 1.16 & 1.6 \\
NGC 2264 J &  06 41 06.3  &+09 34 09 & 0.16 & 0.12 & 0.25 & 63 & 62 & 1.22 & 1.7 \\
NGC 2264 K &  06 41 08.2 & +09 29 53 & 0.25 & 0.11 & 0.55 & 0 & 71 & 0.75 & 1.3 \\
NGC 2264 L &  06 41 12.8  &+09 29 10 & 0.03 & 0.03 & 0.13 & 83 & 16 & 1.84 & 3.5 \\
NGC 2264 M &  06 41 14.9 & +09 26 42 & 0.09 & 0.06 & 0.32 & 20 & 20 & 0.73 & 1.2 \\
\cline{1-10}
Rosette A &  06 30 57.1  &+04 57 57 & 1.20 & 0.87 & 0.28 & 90 & 58 & 0.79 & 1.4 \\
Rosette B &  06 31 20.5  &+04 50 14 & 0.16 & 0.16 & 0.00 & 0 & 4 & 0.73 & 1.4 \\
Rosette C &  06 31 32.0  &+04 50 58 & 0.29 & 0.09 & 0.71 & 129 & 7 & 0.77 & 1.4 \\
Rosette D &  06 31 55.4  &+04 56 39 & 0.14 & 0.04 & 0.72 & 87 & 11 & 1.17 & 1.4 \\
Rosette E &  06 31 59.3  &+04 54 50 & 0.91 & 0.83 & 0.08 & 111 & 244 & 0.76 & 1.3 \\
Rosette F &  06 32 05.5  &+04 48 20 & 0.20 & 0.09 & 0.51 & 0 & 4 & 0.83 & 1.4 \\
Rosette G &  06 32 46.5 & +04 45 35 & 0.16 & 0.16 & 0.00 & 0 & 2 & 0.76 & 1.0 \\
Rosette H &  06 33 07.2 & +04 46 57 & 0.79 & 0.15 & 0.82 & 93 & 34 & 0.84 & 1.3 \\
Rosette I &  06 33 10.2  &+04 31 04 & 0.25 & 0.11 & 0.56 & 42 & 5 & 0.94 & 1.6 \\
Rosette J &  06 33 15.1  &+04 35 08 & 0.18 & 0.09 & 0.53 & 156 & 4 & 2.18 & 2.4 \\
Rosette K &  06 33 20.1 & +04 37 01 & 0.09 & 0.09 & 0.00 & 90 & 2 & 2.00 & 2.6 \\
Rosette L &  06 34 10.7  &+04 25 06 & 1.16 & 0.57 & 0.51 & 167 & 158 & 1.10 & 1.5 \\
Rosette M &  06 34 12.9 & +04 19 07 & 0.65 & 0.25 & 0.61 & 71 & 81 & 2.21 & 2.3 \\
Rosette N &  06 34 31.4  &+04 19 09 & 0.18 & 0.16 & 0.08 & 119 & 10 & 1.62 & 1.9 \\
Rosette O &  06 34 37.1  &+04 13 02 & 0.22 & 0.05 & 0.75 & 151 & 9 & 1.76 & 2.2 \\
\cline{1-10}
Lagoon A &  18 03 23.8  &-24 15 19 & 0.35 & 0.16 & 0.55 & 126 & 30 & 0.84 & 1.4 \\
Lagoon B &  18 03 40.1  &-24 22 40 & 0.07 & 0.05 & 0.28 & 134 & 34 & 1.22 & 1.8 \\
Lagoon C &  18 03 46.3  &-24 22 01 & 0.25 & 0.12 & 0.52 & 109 & 27 & 0.85 & 1.4 \\
Lagoon D &  18 03 51.3  &-24 21 08 & 0.10 & 0.06 & 0.41 & 135 & 9 & 0.78 & 1.3 \\
Lagoon E &  18 04 07.6  &-24 25 53 & 0.68 & 0.27 & 0.60 & 119 & 62 & 0.83 & 1.3 \\
Lagoon F &  18 04 13.3  &-24 18 27 & 1.14 & 0.73 & 0.36 & 3 & 243 & 0.80 & 1.3 \\
Lagoon G &  18 04 20.1 & -24 22 51 & 0.09 & 0.05 & 0.40 & 156 & 17 & 0.83 & 1.3 \\
Lagoon H &  18 04 23.3 & -24 21 13 & 0.26 & 0.21 & 0.20 & 6 & 81 & 0.79 & 1.3 \\
Lagoon I &  18 04 28.3  &-24 22 46 & 0.34 & 0.31 & 0.10 & 17 & 126 & 0.83 & 1.3 \\
Lagoon J &  18 04 39.6  &-24 23 20 & 0.25 & 0.24 & 0.05 & 4 & 47 & 0.87 & 1.3 \\
Lagoon K &  18 04 50.5 & -24 26 19 & 0.46 & 0.25 & 0.45 & 39 & 86 & 1.03 & 1.4 \\
\cline{1-10}
NGC 2362 A &  07 18 37.8  &-24 53 58 & 0.21 & 0.13 & 0.38 & 54 & 20 & 0.65 & 1.2 \\
NGC 2362 B &  07 18 42.9  &-24 57 44 & 0.43 & 0.38 & 0.10 & 172 & 126 & 0.62 & 1.09 \\
\cline{1-10}
DR 21 A &  20 38 51.7 & +42 18 52 & 0.20 & 0.06 & 0.69 & 60 & 12 & 1.97 & 2.3 \\
DR 21 B &  20 38 57.7 & +42 17 51 & 0.07 & 0.05 & 0.34 & 248 & 5 & 2.04 & 2.8 \\
DR 21 C &  20 39 00.2 & +42 18 47 & 0.07 & 0.04 & 0.41 & 144 & 11 & 3.00 & 3.5 \\
DR 21 D &  20 39 00.4 & +42 19 46 & 0.17 & 0.07 & 0.61 & 168 & 22 & 2.92 & 4.0 \\
DR 21 E &  20 39 00.5 & +42 22 39 & 0.22 & 0.12 & 0.44 & 147 & 75 & 2.36 & 3.7 \\
DR 21 F &  20 39 00.9  &+42 24 43 & 0.06 & 0.03 & 0.55 & 50 & 5 & 2.55 & 4.0 \\
DR 21 G &  20 39 03.8 & +42 16 51 & 0.14 & 0.11 & 0.22 & 149 & 11 & 2.51 & 3.3 \\
DR 21 H &  20 39 03.9 & +42 25 34 & 0.11 & 0.05 & 0.53 & 167 & 22 & 2.96 & 3.3 \\
DR 21 I &  20 39 05.4  &+42 21 18 & 0.13 & 0.09 & 0.32 & 154 & 15 & 2.66 & 3.0 \\
\cline{1-10}
RCW 38 A &  08 58 47.2 & -47 30 52 & 2.57 & 2.07 & 0.19 & 177 & 314 & 1.36 & 2.2 \\
RCW 38 B &  08 59 05.0 & -47 30 43 & 0.12 & 0.08 & 0.36 & 54 & 117 & 1.06 & 2.6 \\
RCW 38 C &  08 59 16.4 & -47 27 50 & 0.42 & 0.09 & 0.78 & 165 & 15 & 1.62 & 2.5 \\
\cline{1-10}
NGC 6334 A &  17 19 57.9 & -35 54 02 & 0.15 & 0.10 & 0.31 & 126 & 17 & 1.44 & 2.1 \\
NGC 6334 B &  17 19 58.5 & -35 56 24 & 0.42 & 0.31 & 0.25 & 66 & 42 & 1.44 & 2.0 \\
NGC 6334 C &  17 20 02.7 & -35 58 23 & 0.06 & 0.05 & 0.15 & 107 & 11 & 1.10 & 1.8 \\
NGC 6334 D &  17 20 14.9 & -35 54 43 & 0.08 & 0.07 & 0.09 & 35 & 10 & 1.88 & 2.8 \\
NGC 6334 E &  17 20 19.0 & -35 54 59 & 0.27 & 0.21 & 0.24 & 28 & 44 & 1.88 & 3.1 \\
NGC 6334 F &  17 20 23.2 & -35 57 00 & 0.22 & 0.19 & 0.17 & 25 & 18 & 1.28 & 1.8 \\
NGC 6334 G &  17 20 25.2&  -35 44 04 & 0.13 & 0.09 & 0.26 & 49 & 21 & 1.61 & 2.5 \\
NGC 6334 H &  17 20 31.2 & -35 54 14 & 0.18 & 0.13 & 0.26 & 77 & 23 & 1.38 & 1.8 \\
NGC 6334 I &  17 20 35.3  &-35 59 20 & 0.14 & 0.09 & 0.32 & 52 & 12 & 1.01 & 1.6 \\
NGC 6334 J &  17 20 39.2  &-35 49 27 & 0.48 & 0.17 & 0.65 & 40 & 75 & 2.27 & 3.2 \\
NGC 6334 K &  17 20 48.4 & -35 42 59 & 0.14 & 0.09 & 0.31 & 75 & 9 & 3.00 & 3.1 \\
NGC 6334 L &  17 20 54.4  &-35 45 43 & 0.27 & 0.23 & 0.16 & 50 & 50 & 2.44 & 3.2 \\
NGC 6334 M &  17 20 57.3 & -35 39 41 & 0.20 & 0.16 & 0.20 & 167 & 18 & 2.03 & 2.7 \\
NGC 6334 N &  17 21 32.5  &-35 40 27 & 0.34 & 0.24 & 0.30 & 28 & 25 & 2.20 & 1.6 \\
\cline{1-10}
NGC 6357 A &  17 24 43.7  &-34 12 07 & 0.28 & 0.22 & 0.22 & 113 & 151 & 1.26 & 1.9 \\
NGC 6357 B &  17 24 46.7  &-34 15 23 & 0.58 & 0.37 & 0.36 & 127 & 133 & 1.30 & 1.9 \\
NGC 6357 C &  17 25 33.3  &-34 24 43 & 0.30 & 0.24 & 0.19 & 145 & 128 & 1.29 & 1.9 \\
NGC 6357 D &  17 25 34.3  &-34 23 10 & 0.06 & 0.04 & 0.38 & 176 & 27 & 1.27 & 2.0 \\
NGC 6357 E &  17 25 47.9  &-34 27 12 & 1.06 & 0.18 & 0.83 & 14 & 60 & 1.33 & 1.9 \\
NGC 6357 F &  17 26 02.2  &-34 16 42 & 0.37 & 0.19 & 0.50 & 4 & 162 & 1.41 & 2.1 \\
\cline{1-10}
Eagle A &  18 18 39.9 & -13 47 41 & 0.12 & 0.12 & 0.01 & 139 & 46 & 0.88 & 1.4 \\
Eagle B &  18 18 42.2 & -13 47 03 & 0.90 & 0.45 & 0.50 & 148 & 451 & 0.98 & 1.5 \\
Eagle C &  18 18 52.8 & -13 46 43 & 0.25 & 0.10 & 0.60 & 103 & 20 & 0.96 & 1.4 \\
Eagle D &  18 18 57.3 & -13 45 23 & 1.65 & 0.65 & 0.61 & 100 & 275 & 1.10 & 1.6 \\
Eagle E &  18 19 07.9 & -13 36 28 & 0.08 & 0.07 & 0.07 & 13 & 20 & 2.40 & 2.6 \\
Eagle F &  18 19 12.8 & -13 26 03 & 0.45 & 0.40 & 0.13 & 123 & 37 & 1.61 & 2.3 \\
Eagle G &  18 19 13.1&  -13 33 45 & 0.20 & 0.10 & 0.51 & 102 & 20 & 2.59 & 3.7 \\
Eagle H &  18 19 15.0 & -13 39 25 & 0.11 & 0.08 & 0.26 & 51 & 10 & 1.57 & 1.6 \\
Eagle I &  18 19 19.2  &-13 36 34 & 0.44 & 0.16 & 0.64 & 37 & 18 & 1.64 & 2.4 \\
Eagle J &  18 19 29.7  &-13 23 08 & 0.30 & 0.18 & 0.41 & 30 & 12 & 1.51 & 2.1 \\
Eagle K &  18 19 30.7 & -13 45 30 & 0.06 & 0.06 & 0.04 & 50 & 6 & 1.42 & 2.0 \\
Eagle L &  18 20 02.8  &-13 48 26 & 0.04 & 0.02 & 0.31 & 143 & 7 & 2.01 & 3.0 \\
\cline{1-10}
M 17 A &  18 20 18.0  &-16 14 11 & 0.07 & 0.07 & 0.03 & 41 & 10 & 1.92 & 3.2 \\
M 17 B &  18 20 19.5  &-16 13 28 & 0.03 & 0.02 & 0.10 & 80 & 5 & 2.49 & 2.9 \\
M 17 C &  18 20 21.6  &-16 10 35 & 0.14 & 0.13 & 0.09 & 135 & 31 & 1.70 & 2.5 \\
M 17 D &  18 20 22.5  &-16 08 23 & 0.29 & 0.27 & 0.07 & 173 & 119 & 1.53 & 2.4 \\
M 17 E &  18 20 22.5  &-16 09 52 & 0.09 & 0.07 & 0.18 & 68 & 13 & 1.56 & 2.5 \\
M 17 F &  18 20 22.8  &-16 12 23 & 0.05 & 0.04 & 0.03 & 15 & 7 & 1.70 & 4.4 \\
M 17 G &  18 20 25.0 & -16 11 35 & 0.04 & 0.04 & 0.05 & 94 & 6 & 1.55 & 3.9 \\
M 17 H &  18 20 25.6 & -16 11 17 & 0.11 & 0.11 & 0.07 & 151 & 31 & 1.25 & 2.2 \\
M 17 I &  18 20 26.0  &-16 09 34 & 0.17 & 0.17 & 0.02 & 24 & 48 & 1.45 & 2.1 \\
M 17 J &  18 20 27.8  &-16 09 54 & 0.01 & 0.01 & 0.02 & 137 & 4 & 1.56 & 2.0 \\
M 17 K &  18 20 28.5 & -16 10 58 & 0.19 & 0.16 & 0.14 & 177 & 78 & 1.41 & 2.2 \\
M 17 L &  18 20 29.9  &-16 10 46 & 0.14 & 0.13 & 0.05 & 34 & 95 & 1.82 & 2.8 \\
M 17 M &  18 20 30.8 & -16 03 16 & 0.20 & 0.16 & 0.21 & 94 & 31 & 1.83 & 2.5 \\
M 17 N &  18 20 31.4  &-16 09 39 & 0.15 & 0.11 & 0.27 & 4 & 32 & 1.33 & 2.1 \\
M 17 O &  18 20 31.5  &-16 11 22 & 0.09 & 0.07 & 0.18 & 57 & 26 & 1.42 & 2.4 \\
\cline{1-10}
Carina A &  10 43 41.9  &-59 35 49 & 0.19 & 0.09 & 0.55 & 115 & 11 & 1.03 & 1.6 \\
Carina B &  10 43 54.2  &-59 33 02 & 0.70 & 0.45 & 0.35 & 191 & 193 & 0.97 & 1.5 \\
Carina C &  10 43 56.4  &-59 32 54 & 0.22 & 0.18 & 0.17 & 80 & 106 & 0.94 & 1.4 \\
Carina D &  10 44 32.9  &-59 33 42 & 0.60 & 0.30 & 0.50 & 56 & 46 & 0.90 & 1.4 \\
Carina E &  10 44 34.0  &-59 44 08 & 0.28 & 0.27 & 0.05 & 137 & 36 & 0.84 & 1.4 \\
Carina F &  10 44 37.4  &-59 26 03 & 0.64 & 0.40 & 0.36 & 108 & 38 & 0.94 & 1.5 \\
Carina G &  10 44 39.7 & -59 44 26 & 10.15 & 1.88 & 0.81 & 177 & 741 & 0.91 & 1.5 \\
Carina H &  10 44 41.8 & -59 22 05 & 0.30 & 0.23 & 0.23 & 209 & 49 & 0.83 & 1.3 \\
Carina I &  10 44 45.3  &-59 20 07 & 0.23 & 0.18 & 0.24 & 70 & 18 & 0.80 & 1.3 \\
Carina J &  10 45 02.4  &-59 45 50 & 0.50 & 0.46 & 0.09 & 163 & 65 & 0.96 & 1.5 \\
Carina K &  10 45 06.2 & -59 40 21 & 0.35 & 0.25 & 0.28 & 94 & 36 & 0.89 & 1.5 \\
Carina L &  10 45 11.1  &-59 42 46 & 0.49 & 0.36 & 0.27 & 229 & 58 & 0.91 & 1.5 \\
Carina M &  10 45 13.7 & -59 57 58 & 0.48 & 0.33 & 0.32 & 122 & 52 & 0.95 & 1.6 \\
Carina N &  10 45 36.1  &-59 48 20 & 0.39 & 0.38 & 0.03 & 84 & 37 & 1.51 & 2.1 \\
Carina O &  10 45 53.4  &-59 56 53 & 0.18 & 0.17 & 0.09 & 110 & 25 & 1.15 & 1.8 \\
Carina P &  10 45 54.4  &-60 04 32 & 1.17 & 0.59 & 0.50 & 60 & 118 & 0.87 & 1.5 \\
Carina Q &  10 45 55.2 & -59 59 51 & 0.51 & 0.33 & 0.35 & 92 & 27 & 0.87 & 1.4 \\
Carina R &  10 46 05.4 & -59 50 09 & 1.06 & 0.51 & 0.52 & 83 & 46 & 0.99 & 1.6 \\
Carina S &  10 46 52.7 & -60 04 40 & 0.44 & 0.26 & 0.41 & 24 & 24 & 0.90 & 1.5 \\
Carina T &  10 47 12.5 & -60 05 58 & 0.53 & 0.46 & 0.14 & 215 & 48 & 0.92 & 1.5 \\
\cline{1-10}
Trifid A &  18 02 05.8  &-23 05 53 & 1.88 & 0.36 & 0.81 & 103 & 87 & 1.53 & 1.3 \\
Trifid B &  18 02 23.1  &-23 01 50 & 0.14 & 0.11 & 0.20 & 158 & 25 & 0.92 & 1.4 \\
Trifid C &  18 02 31.0  &-23 00 40 & 0.65 & 0.52 & 0.20 & 92 & 82 & 0.87 & 1.3 \\
Trifid D &  18 02 39.7  &-22 49 18 & 0.87 & 0.28 & 0.68 & 64 & 27 & 1.27 & 1.4 \\
\cline{1-10}
NGC 1893 A &  05 22 40.3  &+33 22 15 & 0.43 & 0.32 & 0.26 & 98 & 46 & 0.81 & 1.5 \\
NGC 1893 B &  05 22 46.6  &+33 25 17 & 0.36 & 0.29 & 0.21 & 22 & 110 & 0.81 & 1.5 \\
NGC 1893 C &  05 22 47.4  &+33 24 11 & 0.23 & 0.20 & 0.14 & 44 & 20 & 0.85 & 1.5 \\
NGC 1893 D &  05 22 49.7  &+33 24 42 & 0.03 & 0.03 & 0.09 & 13 & 5 & 0.75 & 1.4 \\
NGC 1893 E &  05 22 49.7  &+33 26 51 & 0.15 & 0.05 & 0.67 & 129 & 9 & 0.80 & 1.4 \\
NGC 1893 F &  05 22 53.6  &+33 25 40 & 0.17 & 0.14 & 0.16 & 58 & 18 & 0.81 & 1.4 \\
NGC 1893 G &  05 22 53.8 & +33 30 54 & 0.28 & 0.19 & 0.33 & 25 & 46 & 0.89 & 1.6 \\
NGC 1893 H &  05 22 57.5 & +33 26 34 & 0.31 & 0.28 & 0.10 & 28 & 60 & 0.82 & 1.4 \\
NGC 1893 I &  05 23 03.3  &+33 28 15 & 0.48 & 0.32 & 0.33 & 25 & 114 & 0.83 & 1.5 \\
NGC 1893 J &  05 23 08.5  &+33 28 32 & 0.04 & 0.03 & 0.37 & 82 & 11 & 0.97 & 1.7 \\
\enddata
\tablecomments{The ellipsoid components from the selected finite-mixture model. Column 1: The component name, labeled from east to west. Columns 2-3: Celestial coordinates (J2000) for the ellipsoid center. Column 4-5: Semi-major and minor axes of the ellipsoid core region in parsec units. Column 6: Ellipticity. Column 7: Orientation angle of the ellipse in degrees east from north. Column 8: Number of stars estimated by integrating the model component out to four times the size of the core. Column 9: Median $J-H$ color index as a measure of absorption.  Column 10: Median X-ray median energy as a measure of absorption.}
\end{deluxetable}
\clearpage\clearpage

\clearpage\clearpage
\begin{deluxetable}{lrrccclllll}
\tablecaption{Membership of MYStIX Subclusters\label{subcluster_membership_table}}
\tabletypesize{\tiny}
\tablehead{\colhead{Designation} & \colhead{R.A. (J2000)} &\colhead{Dec. (J2000)} &\colhead{X-ray} & \colhead{IR-excess} & \colhead{OB} & \colhead{ME} & \colhead{$J-H$} & \colhead{$\log L_{X,\mathrm{tc}}$}  & \colhead{MSFR} & \colhead{Subcluster}\\
\colhead{} & \colhead{deg} & \colhead{deg} & \colhead{} &\colhead{} & \colhead{} & \colhead{(keV)} & \colhead{(mag)} & \colhead{(erg s$^{-1}$)} & \colhead{}& \colhead{}\\
\colhead{(1)} & \colhead{(2)} & \colhead{(3)} & \colhead{(4)} & \colhead{(5)} & \colhead{(6)} &  \colhead{(7)} & \colhead{(8)} &  \colhead{(9)} & \colhead{(10)}& \colhead{(11)}}
\startdata
182025.46-160943.4 &275.1060920 &-16.1620560 &1 &0 &0  &2.10 & 1.38 & 30.78 &M 17  &    I\\
182025.49-161213.8 &275.1062390 &-16.2038430 &1 &0 &1  &3.70 &-9.99 & 30.66 &M 17  &    U\\
182025.49-161053.8 &275.1062440 &-16.1816170 &0 &0 &0  &2.28 & 1.31 & 31.65 &M 17  &    H\\
182025.52-161016.0 &275.1063340 &-16.1711260 &1 &0 &0  &1.98 & 0.96 & 30.44 &M 17  &    X\\
182025.52-160906.8 &275.1063380 &-16.1518940 &1 &0 &0  &1.55 & 1.15 & 29.79 &M 17  &    I\\
182025.52-160558.5 &275.1063600 &-16.0996060 &1 &0 &0  &3.70 &-9.99 & 31.04 &M 17  &    U\\
182025.53-161004.6 &275.1063810 &-16.1679530 &0 &0 &0  &1.76 & 1.11 & 30.79 &M 17  &    I\\
182025.53-161157.7 &275.1063840 &-16.1993820 &1 &0 &0  &3.31 & 3.18 & 31.08 &M 17  &    X\\
182025.55-160931.2 &275.1064880 &-16.1586940 &0 &0 &0  &1.93 & 1.24 & 30.66 &M 17  &    I\\
182025.56-161124.7 &275.1065330 &-16.1901970 &0 &0 &0  &3.26 &-9.99 & 30.48 &M 17  &    H\\
\enddata
\tablecomments{
The source properties and subcluster assignments for all MPCMs from \citet{mpcm} within the spatial analysis fields of view.  This table thus contains stars that are not in the ``flux flattened'' sample used to define the subclusters. Column 1: \mystix\ MPCM designation \citep{mpcm}. Columns 2--3: MPCM celestial coordinates. Columns 4--6: Binary variables indicating inclusion (1) or exclusion (0) in the flux flattened X-ray sample, IR-excess selection, or OB selection. Column 7: X-ray median energy in the 0.5--8.0~keV band, when sufficient counts are available. Column 8: $J-H$ color index. Column 9: Inferred absorption corrected X-ray luminosities in the 0.5--8.0~keV band, when available.  Column 10: The name of the \mystix\ MSFR. Column 11: Subcluster assignment; ``A''--``T'' indicates the assigned subcluster, ``U'' indicates the unclustered stellar population, and ``X'' indicates uncertain assignment.
This table is available in its entirety in the electronic edition of the journal. A portion is shown here for guidance regarding its form and content.}
\end{deluxetable}
\clearpage\clearpage

\clearpage\clearpage

\setcounter{figure}{2}
\begin{figure}
\centering
Orion, Flame, W 40, RCW 36\\
\includegraphics[width=0.2\textwidth]{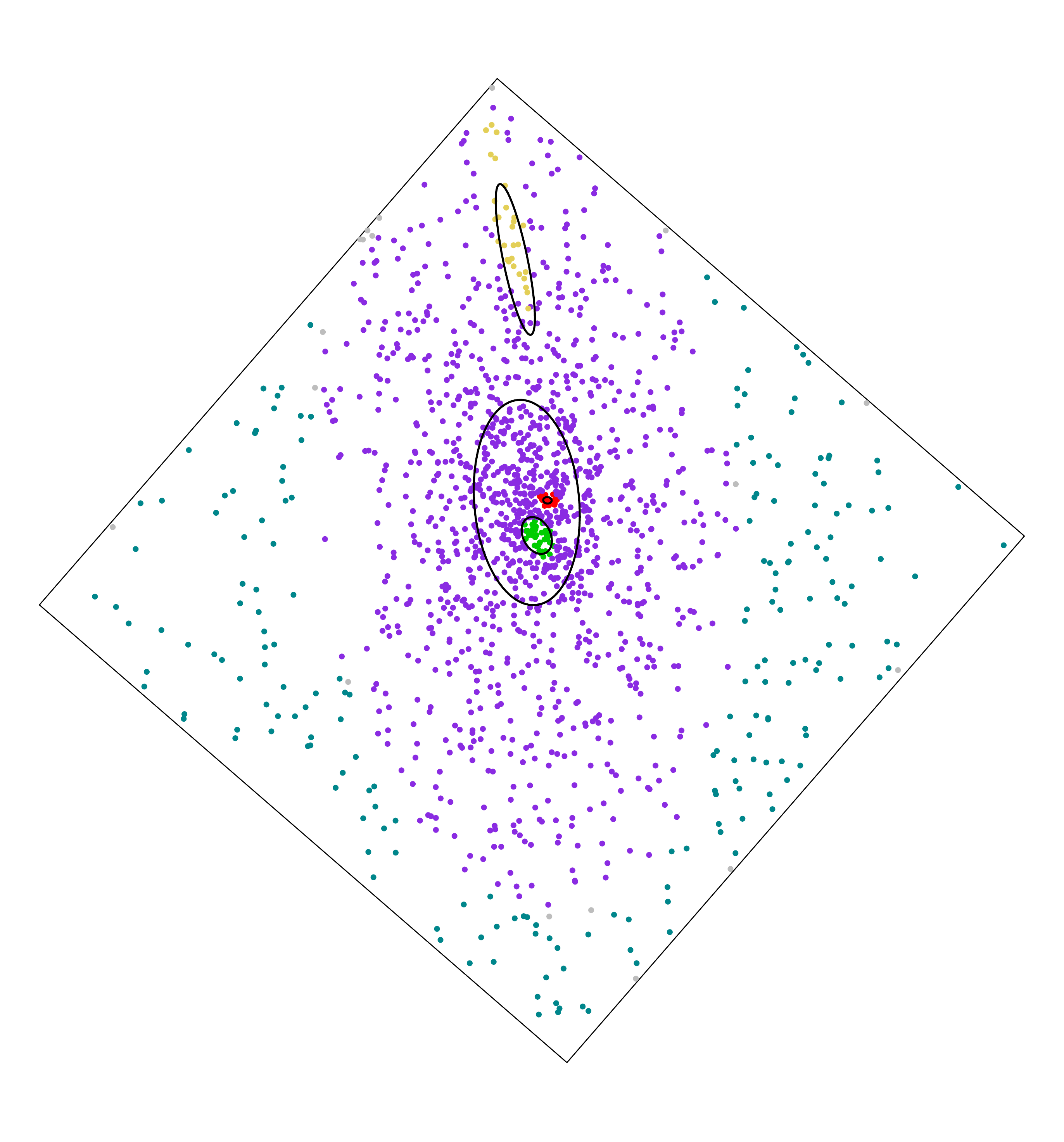}
\includegraphics[width=0.2\textwidth]{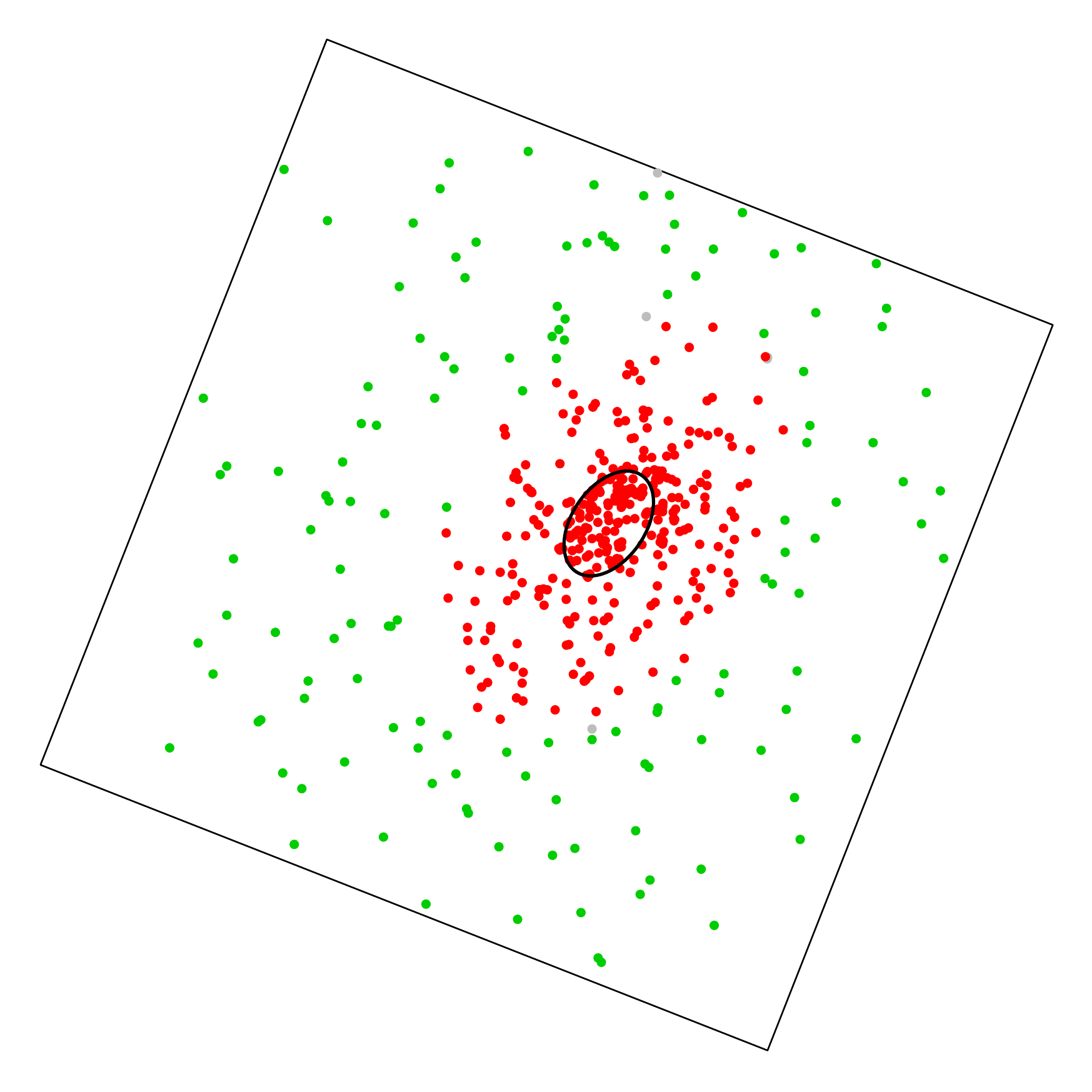}
\includegraphics[width=0.2\textwidth]{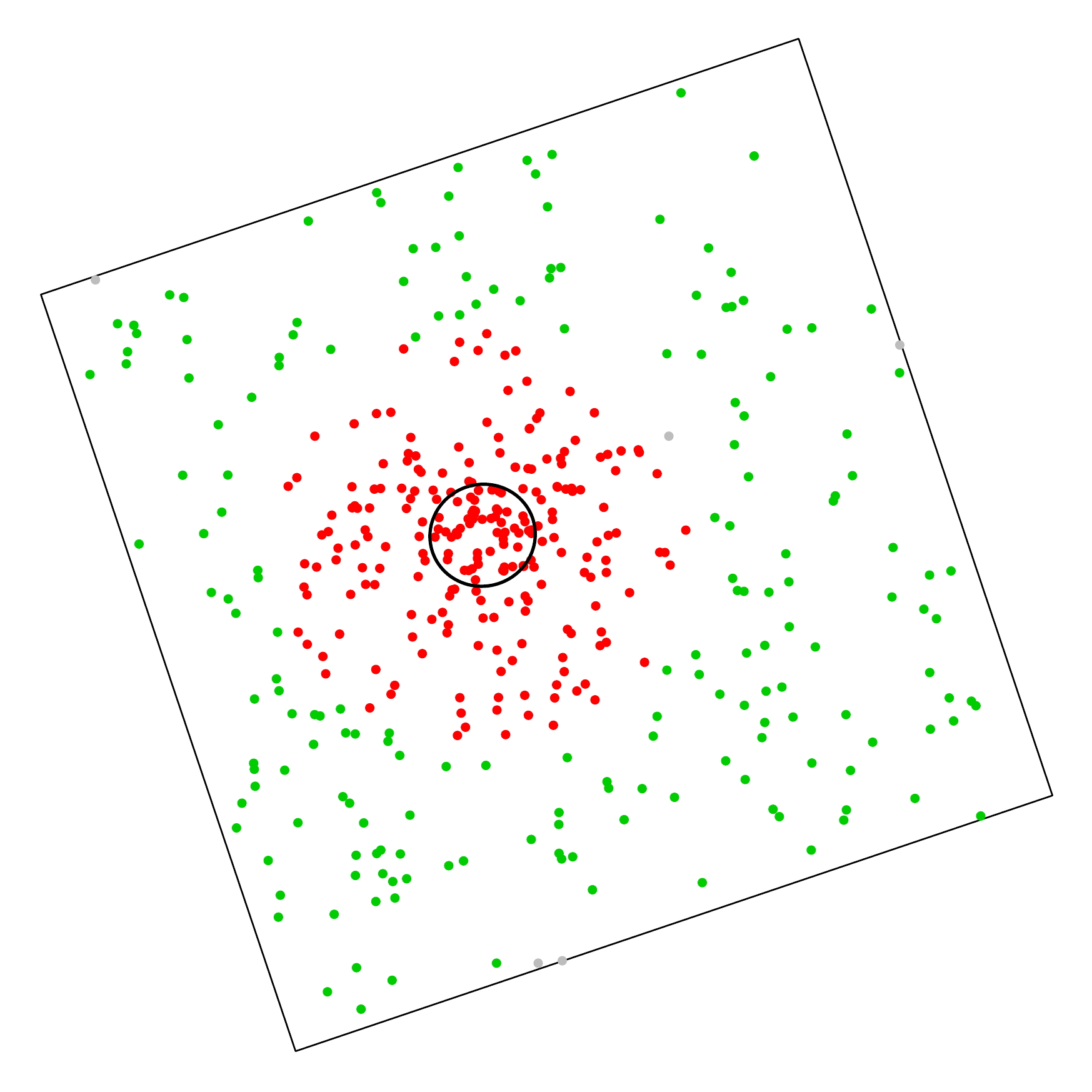}
\includegraphics[width=0.2\textwidth]{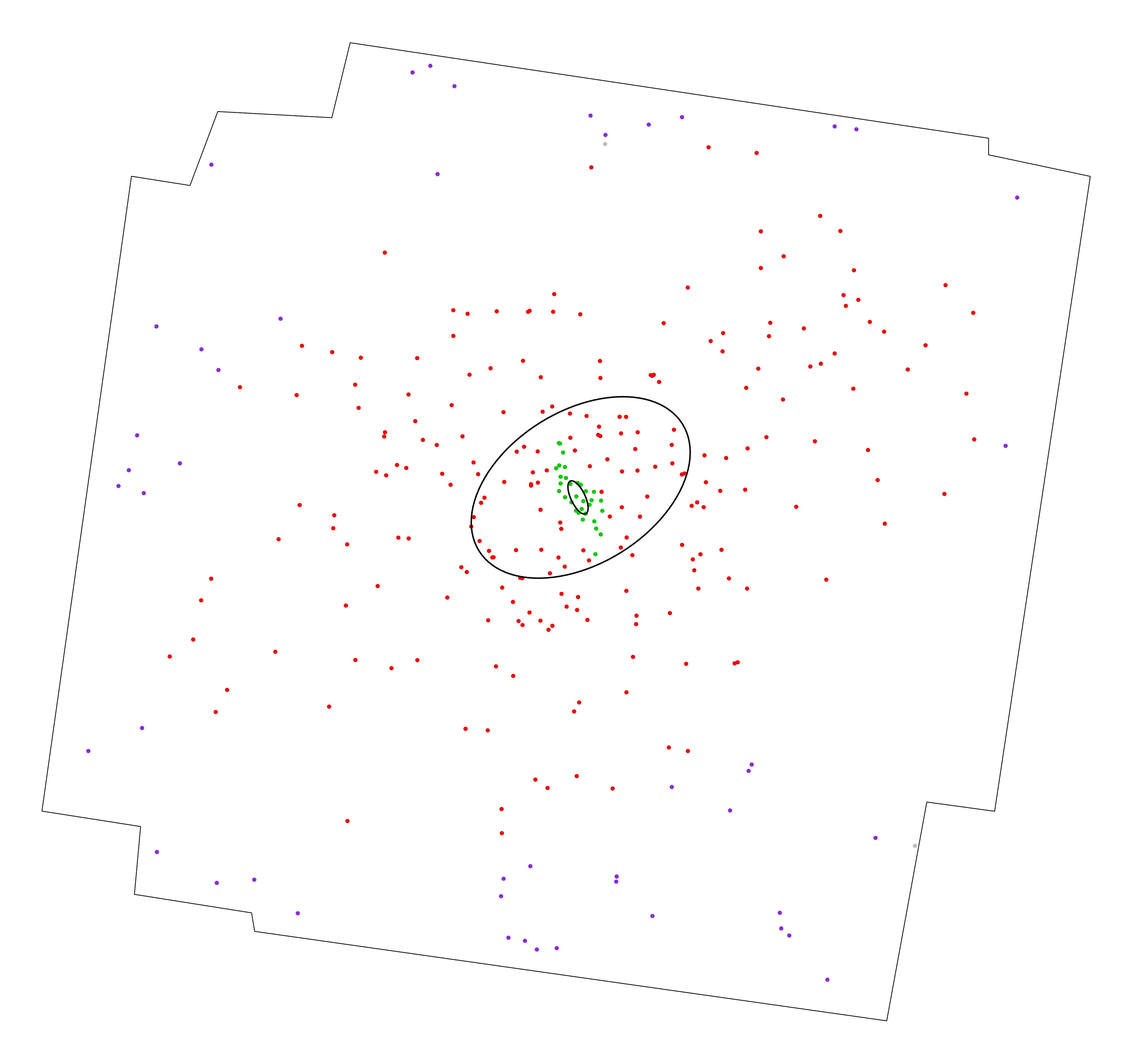}\\
Rosette, Lagoon, NGC 2362, DR 21\\
\includegraphics[width=0.2\textwidth]{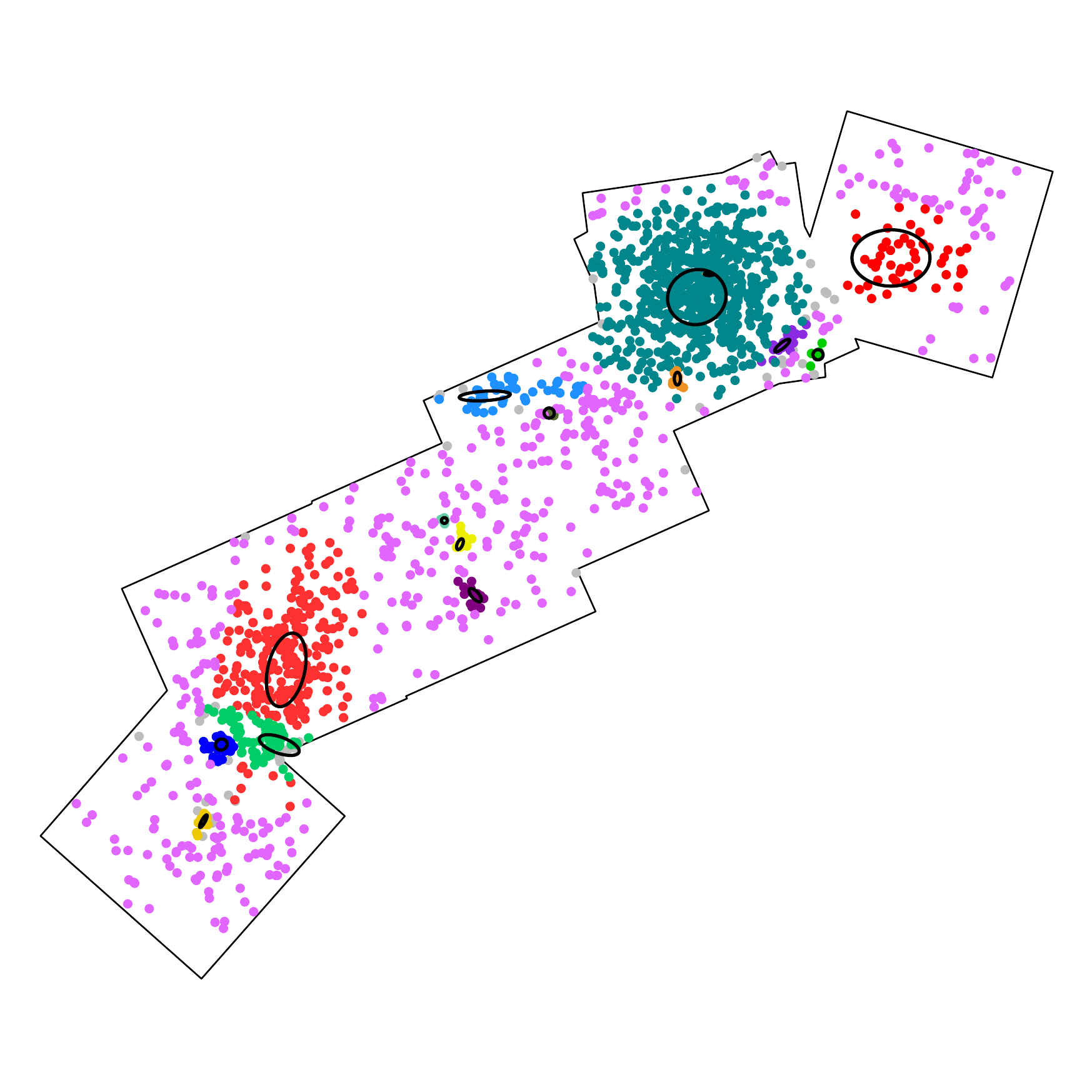}
\includegraphics[width=0.2\textwidth]{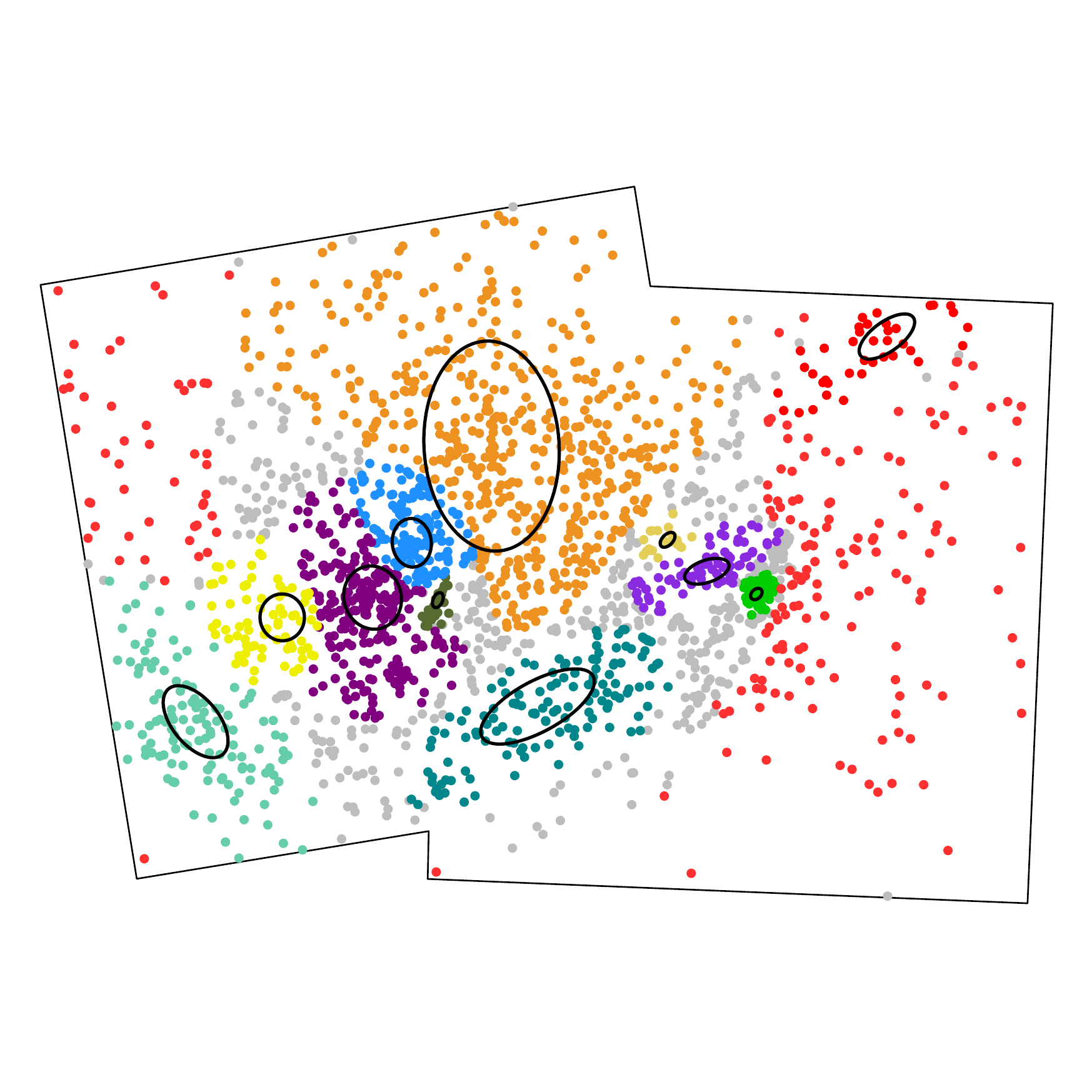}
\includegraphics[width=0.2\textwidth]{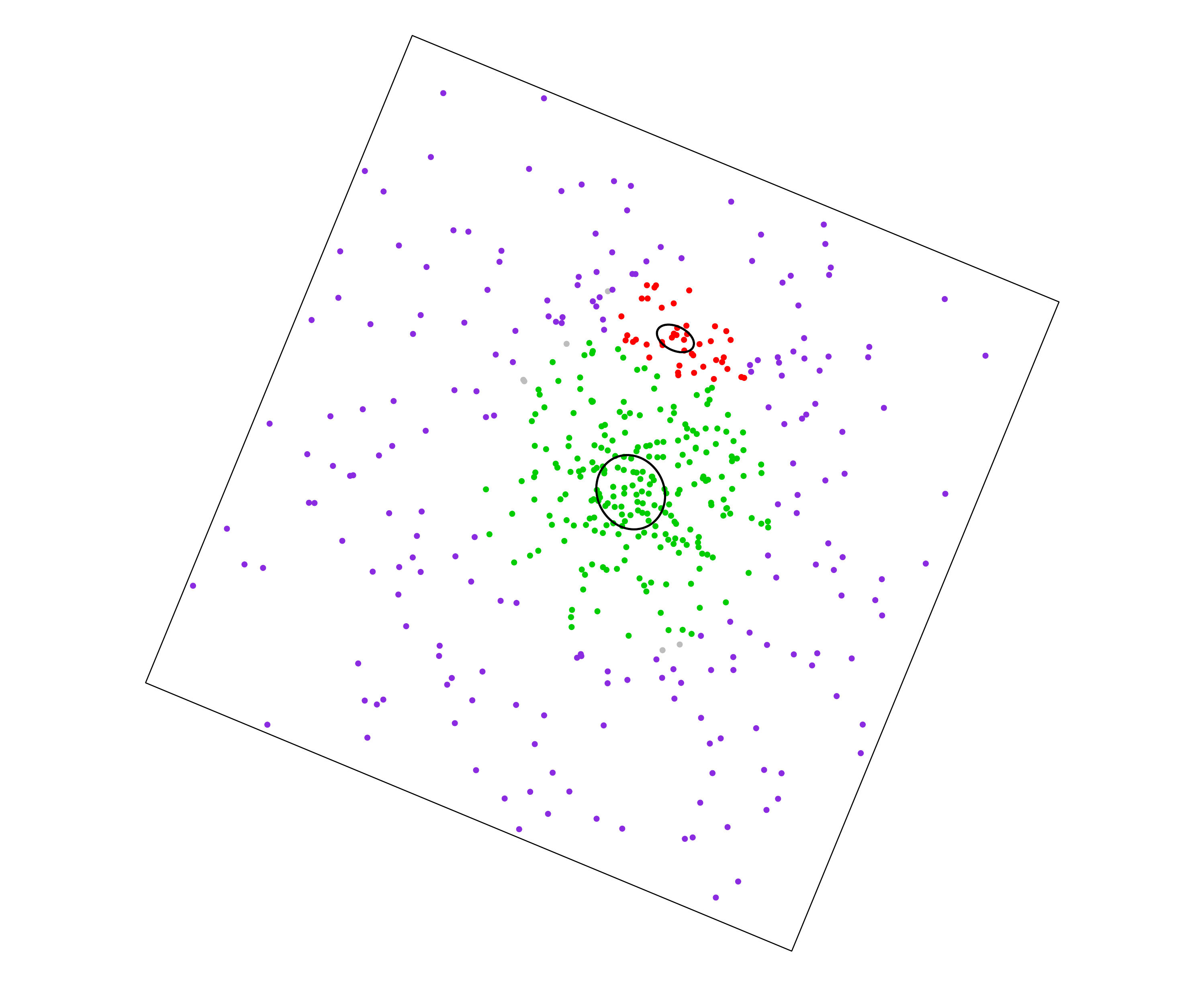}
\includegraphics[width=0.2\textwidth]{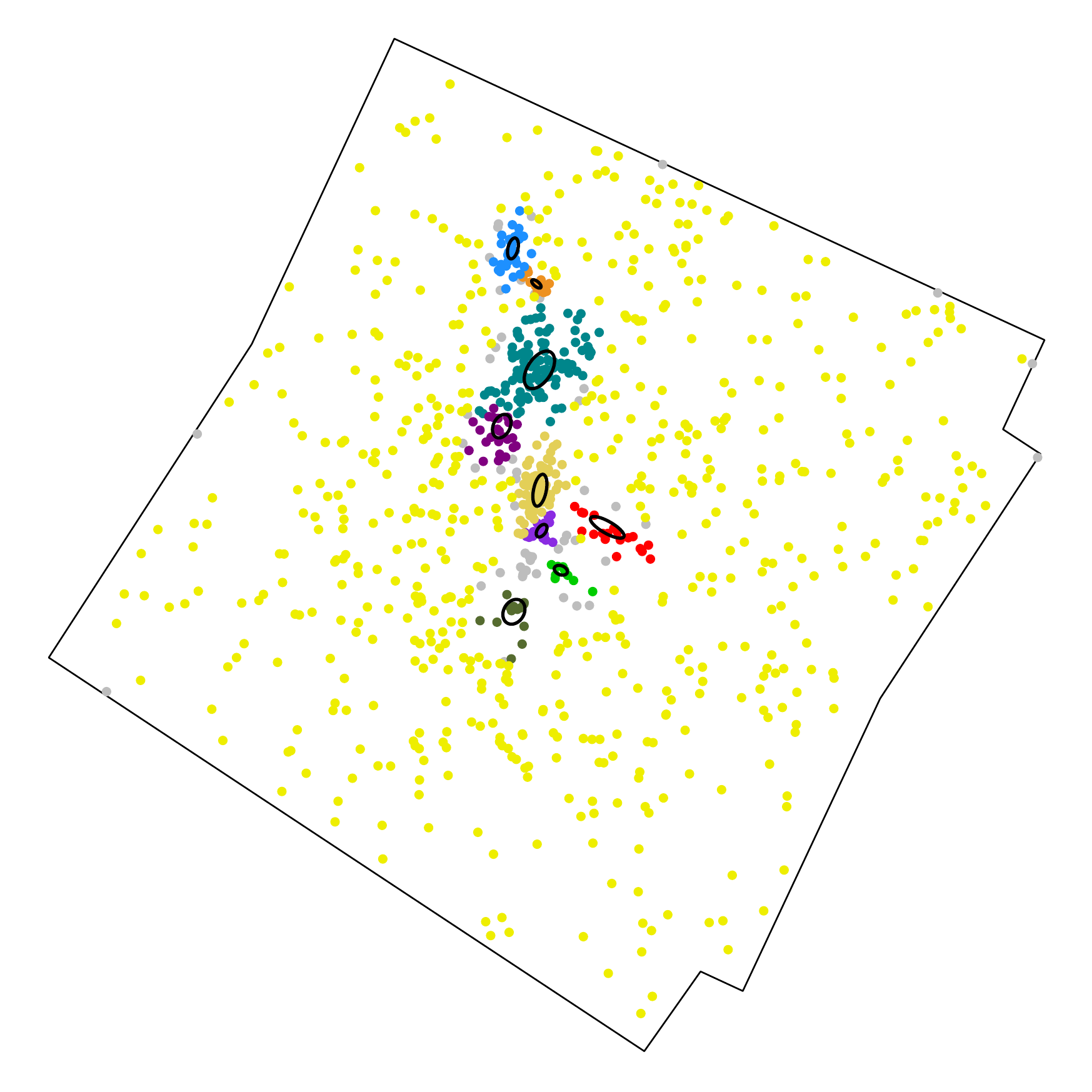}\\
RCW 38, NGC 6334, NGC 6357, Eagle\\
\includegraphics[width=0.2\textwidth]{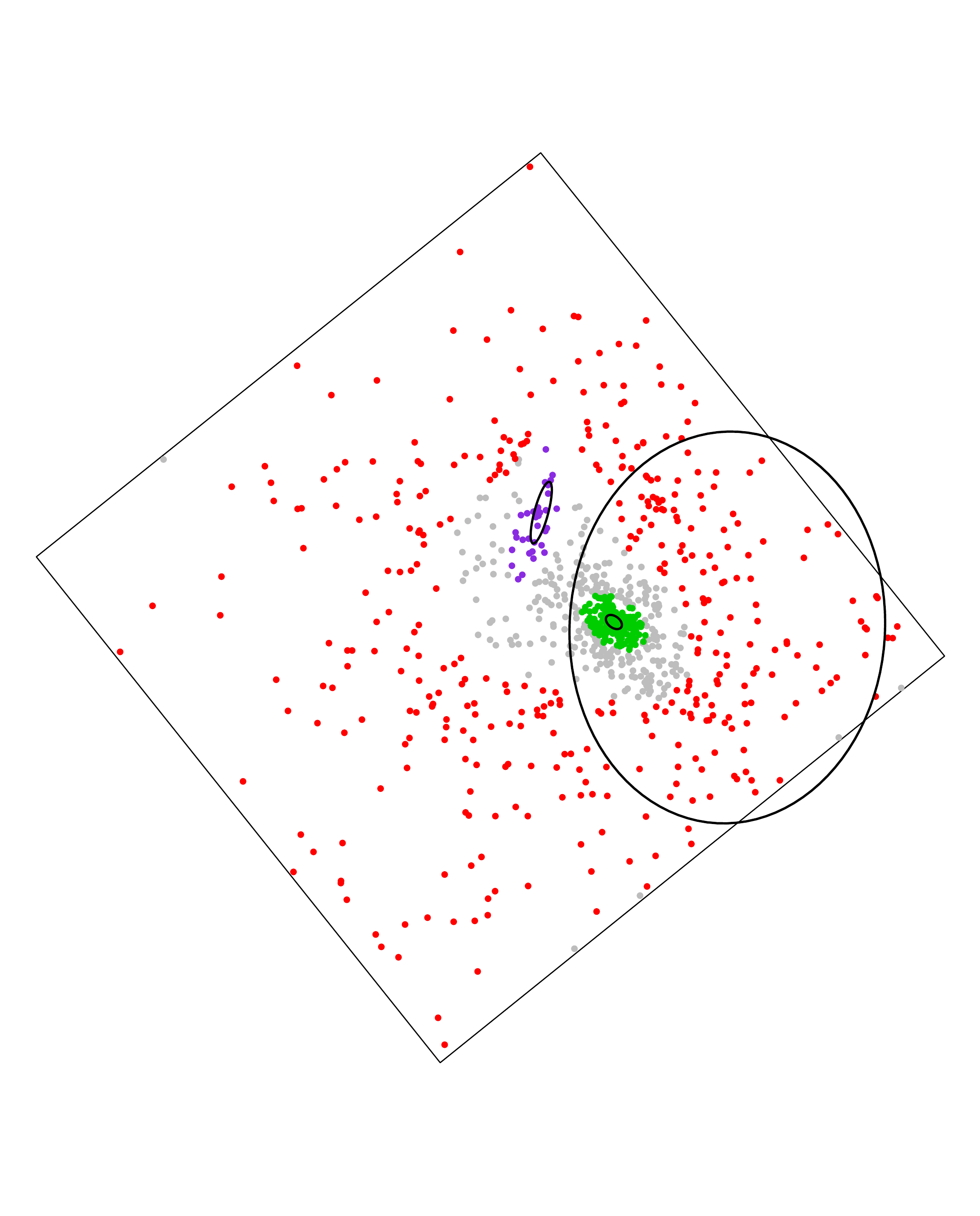}
\includegraphics[width=0.2\textwidth]{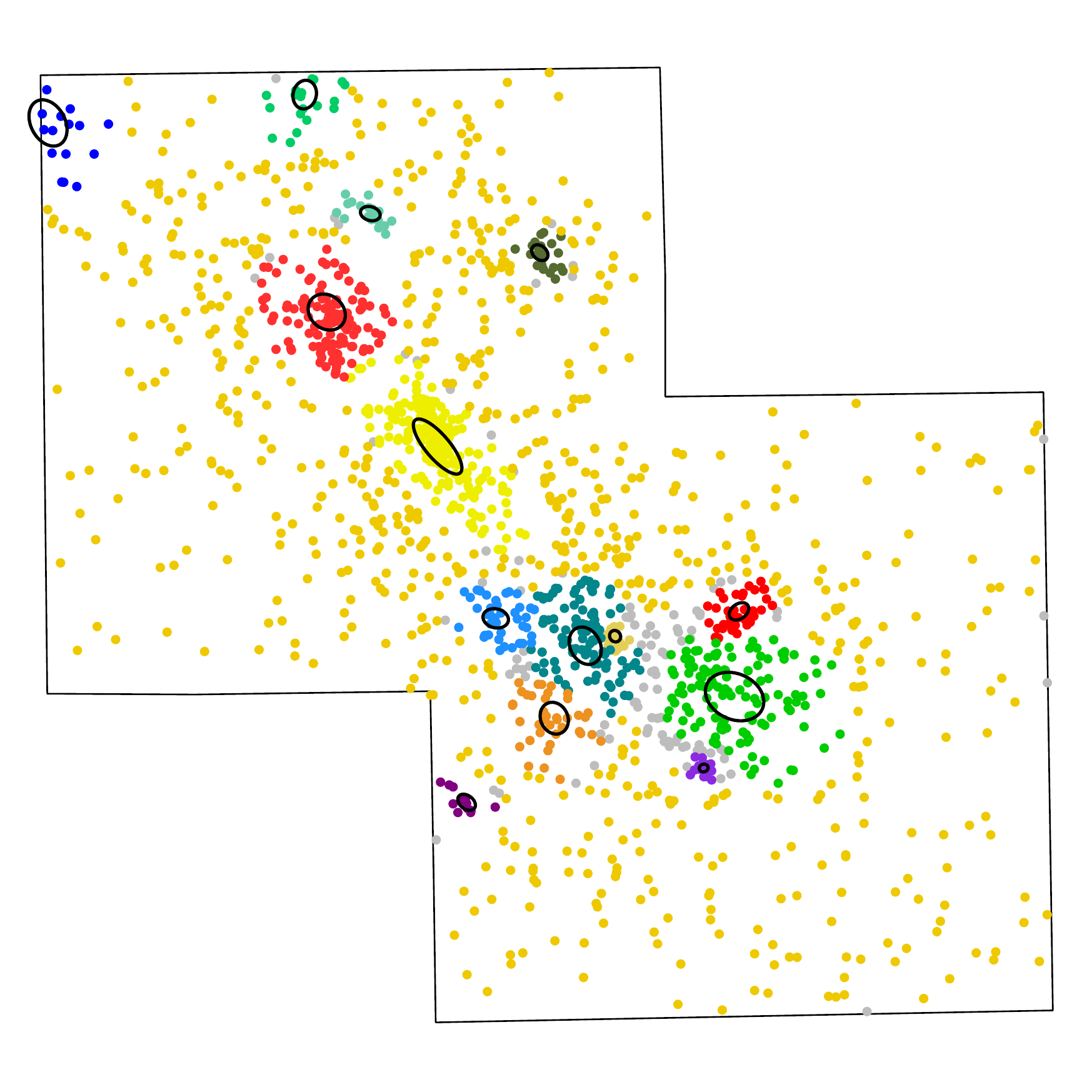}
\includegraphics[width=0.2\textwidth]{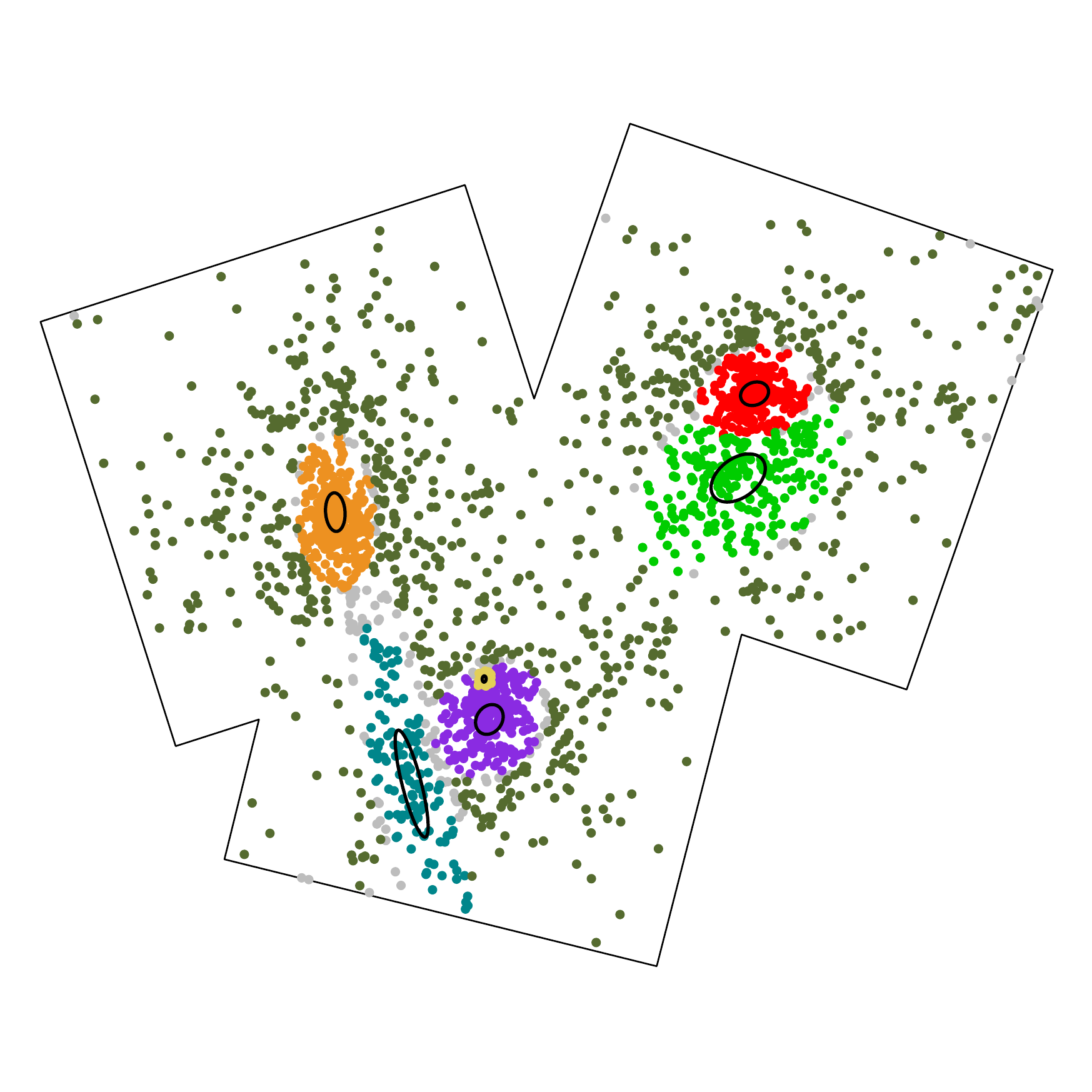}
\includegraphics[width=0.2\textwidth]{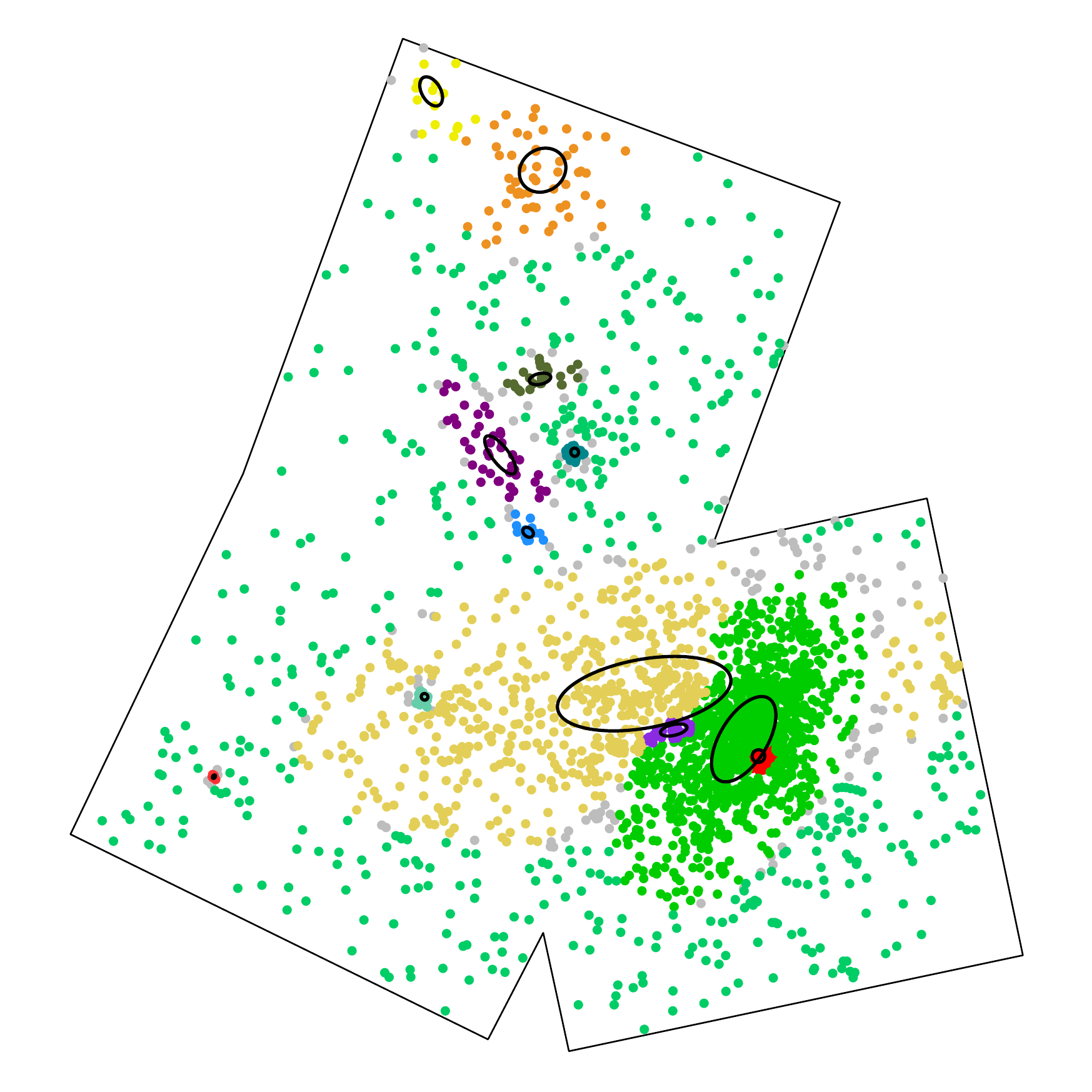}\\
M 17, Trifid, NGC 1893, NGC 2264\\
\includegraphics[width=0.2\textwidth]{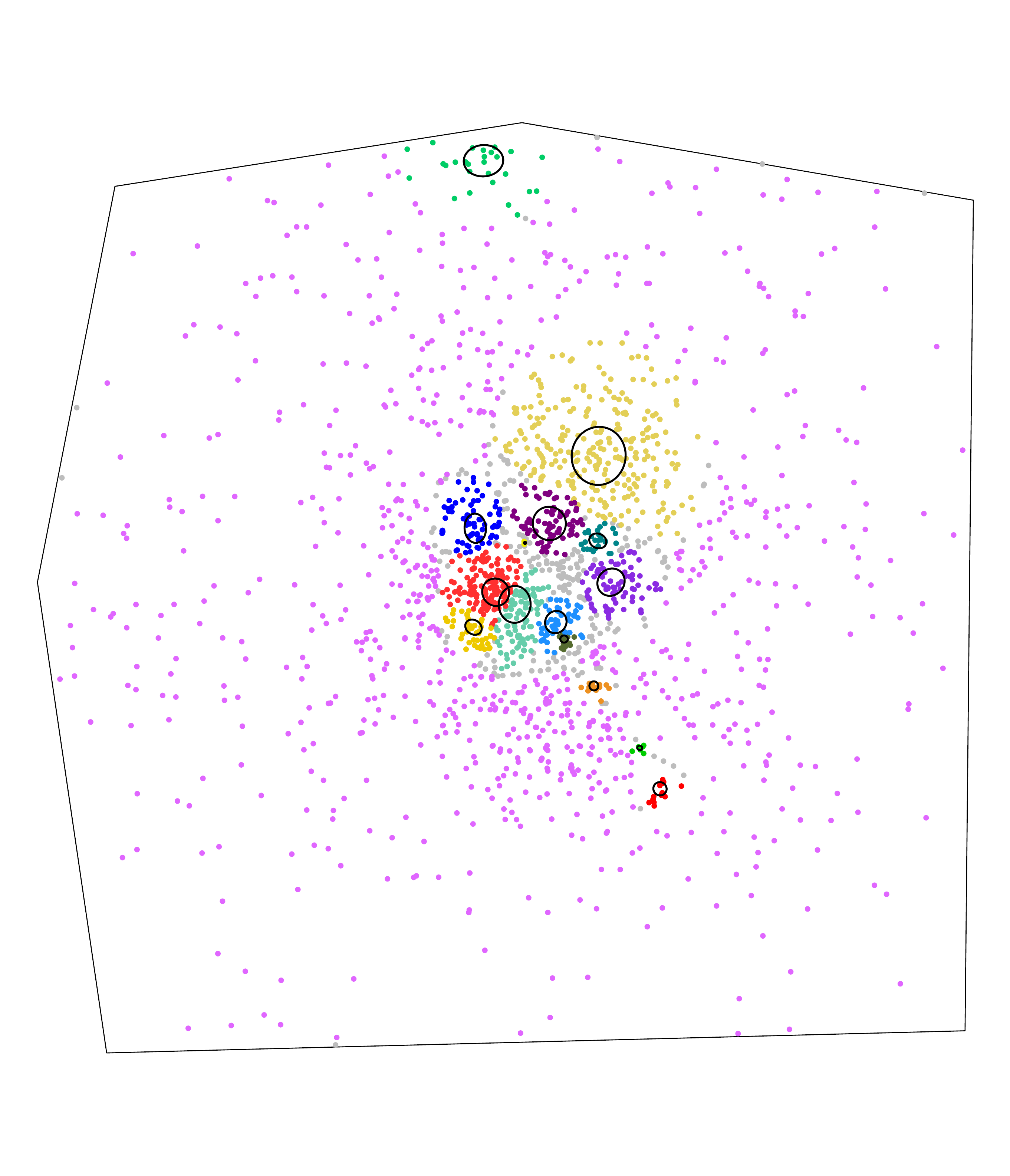}
\includegraphics[width=0.2\textwidth]{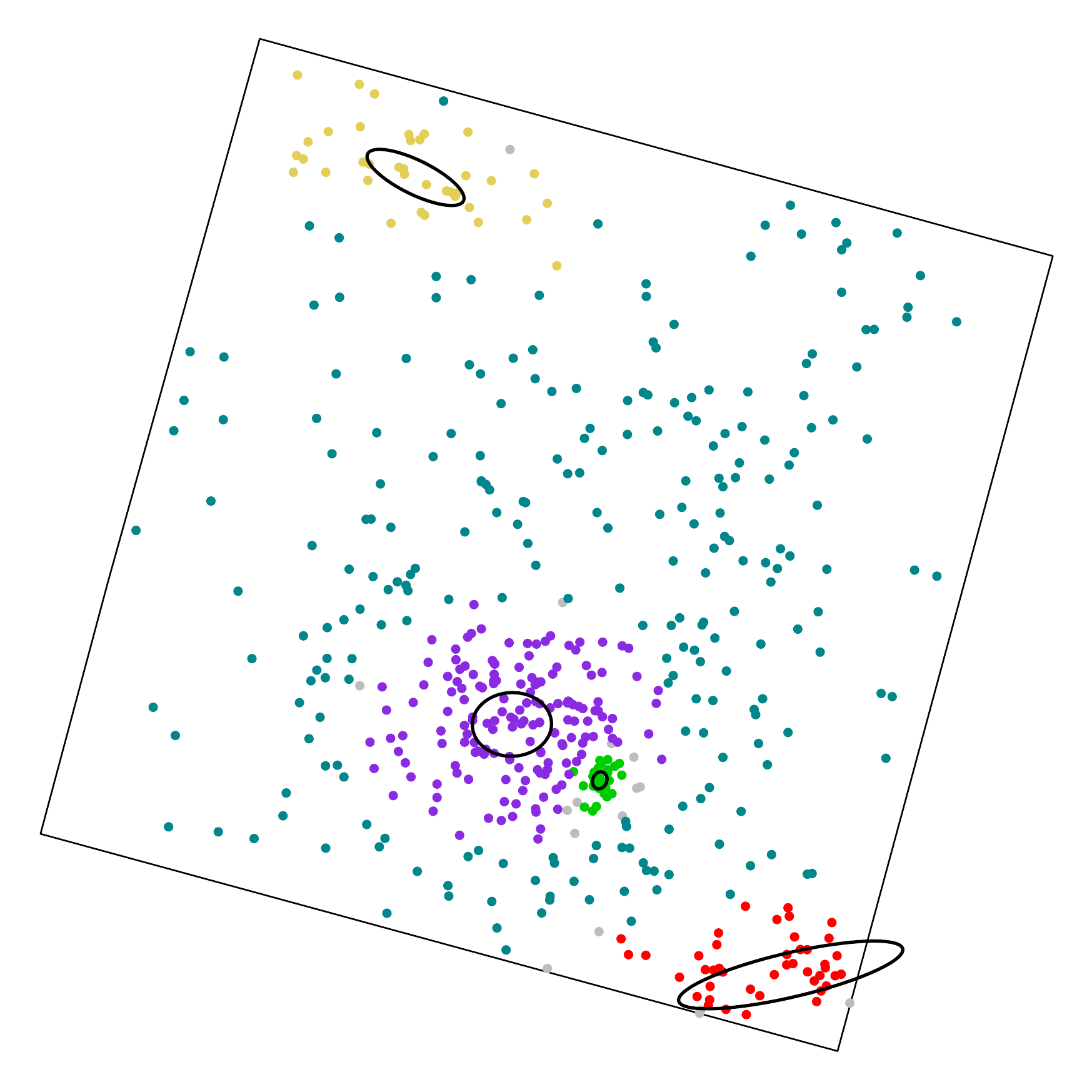}
\includegraphics[width=0.2\textwidth]{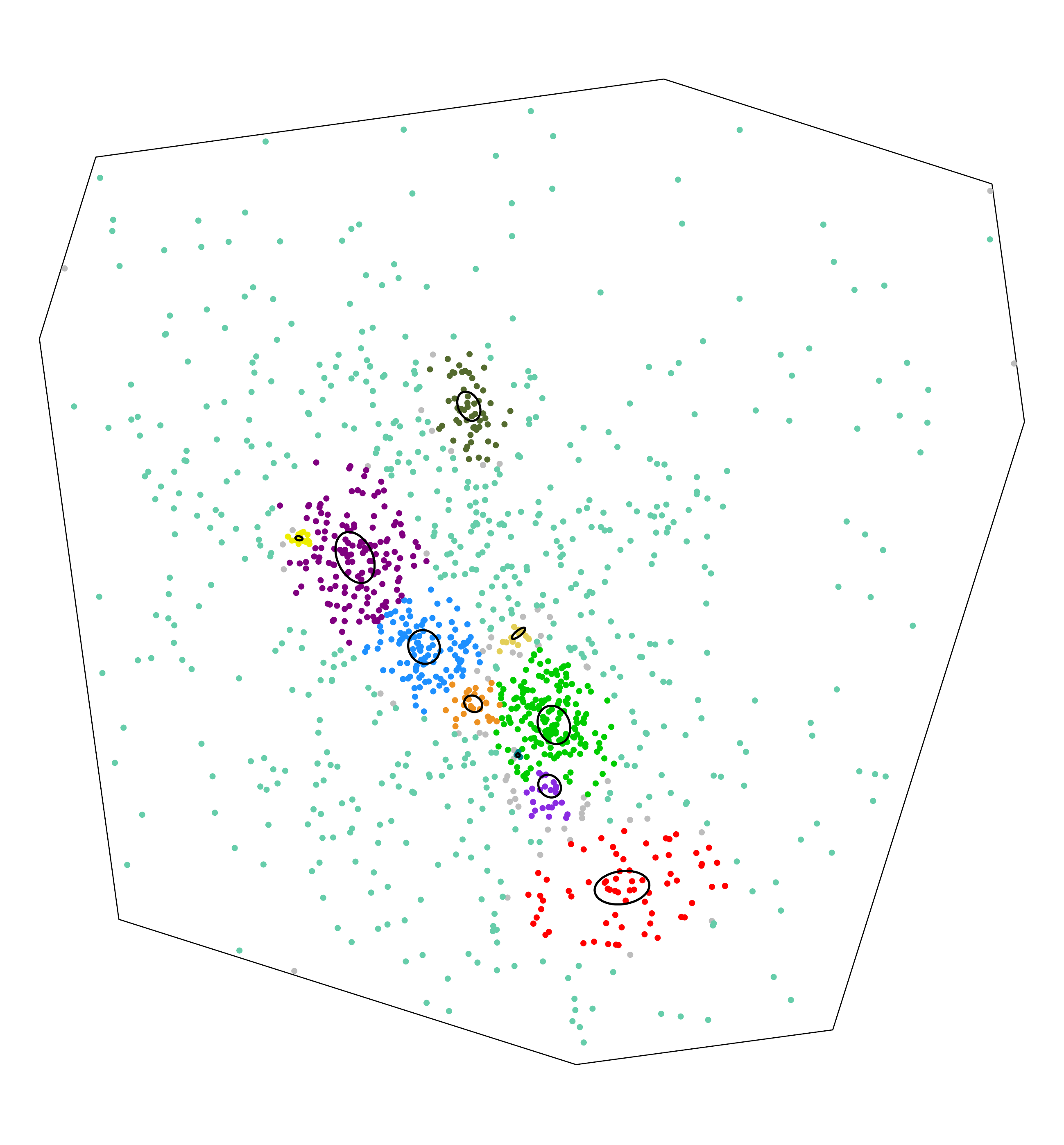}
\includegraphics[width=0.2\textwidth]{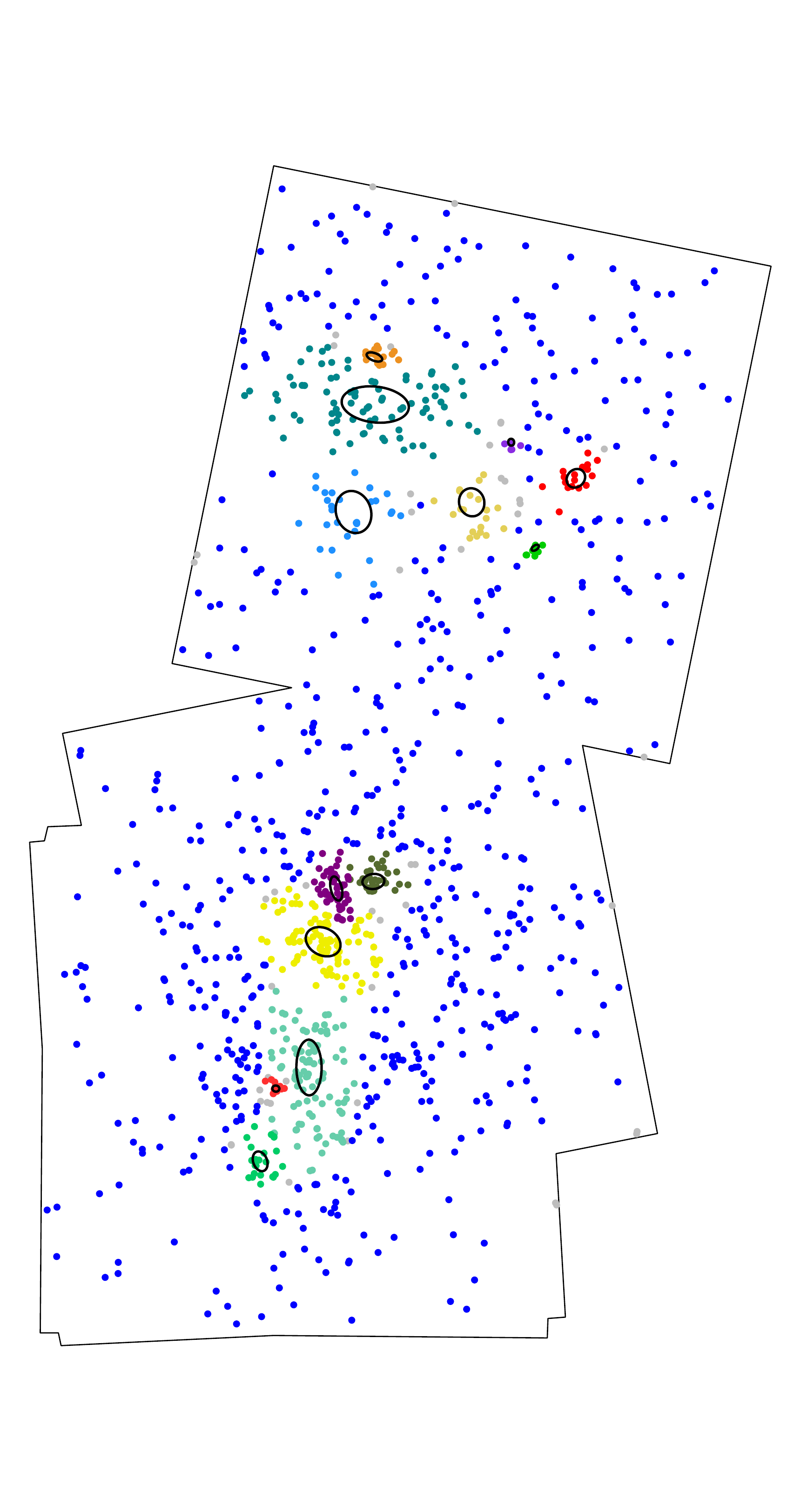}
\caption{electronic figure set}
\end{figure}

\clearpage\clearpage

\subsection{Global Fits \label{regional_models_section}}

The properties of the best-fit model for each region are provided in Table \ref{best_fit_models_tab}, including the number of subclusters, the observed surface density of the unclustered young stars, the minimum and maximum of the fitting error residuals and the kernel size used to calculate these residuals, and a $\chi^2$ goodness-of-fit statistic.

The number of subclusters per region ranges from 1 (Flame and W~40) to 20 (Carina). The variation in this number is an indication of the vast diversity in spatial structure. However, this number also relates to the size of the field of view. The fractal distribution of star formation throughout the Galaxy makes it difficult to choose a sample field of view that adequately characterizes a region. Often clusters are located near each other and it is unclear whether they should be treated separately or as subclusters, as is the case with the three rich clusters in NGC~6357 and the Tr~14-15-16 clusters in the Greater Carina Nebula. Selection effects in our survey of young stars also influence the number of inferred subclusters. 

The goodness-of-fit statistics for the tessellated observed vs.\ model MPCM counts in Table~\ref{best_fit_models_tab} suggest that, in most cases, the finite mixture models are neither over-fitted ($\chi^2\ll\mathrm{d.o.f.}$) nor poorly fit ($\chi^2\gg\mathrm{d.o.f.}$). 
Poor fits are an indication that structures are not simple isothermal ellipsoids, which is not surprising in elaborate, complex structures seen in regions like Rosette, DR~21, NGC~6357, and Carina.

Figure \ref{resid_fig} (left) shows a map of the smoothed fitting residuals for the Carina region (residual maps for the other regions are provided by the electronic version of this article) that were smoothed by a 0.59~pc Gaussian kernel. Red and blue colors indicate positive and negative residuals in units of stars~pc$^{-2}$, respectively. For Carina the maximum residual is located south-west of Tr~14, between Subclusters A and B -- this indicates an asymmetry to Tr~14 due to an excess of stars at the edge of the Northern Cloud in this star-formation complex. Otherwise, the positive and negative residuals are low and more-or-less evenly distributed. 

In all the \mystix\ regions minimum/maximum error residuals for the fits are typically $<$10\% of the peak stellar surface density in a cluster. Salient features of residual maps in select regions are described here: 
\begin{description}
\item[Orion Nebula] The minimum residual is located south of the ONC cluster, and may indicate some deviation from a true isothermal ellipsoid shape. There is also a large negative patch east of the ONC, which was noted as a void by \citet{Feigelson05}. 
\item[Flame Nebula] There is a large positive residual to the north-west of the cluster, which corresponds to a group of highly absorbed stars. 
\item[W~40] The residuals are nearly zero in the cluster core, but there is a positive residual to the south-east, composed mostly of IR-excess selected stars, with a population too small to be accepted as a subcluster by the AIC statistic. 
\item[Rosette Nebula] The residuals are higher amplitude than in other \mystix\ regions, and may indicate that our finite mixture model fit is inadequate for this complex region. The $\chi^2$ value for this fit is also poor.
\item[Lagoon Nebula] Residuals are low indicating an overall good fit. However, there is a long positive residual stretching from the west to the south of NGC~6530.  This residual coincides with the edge of the molecular cloud and a bright rimmed cloud; it may represent an area of triggered star formation. 
\item[DR~21] The large blue residual along the chain of subclusters in DR~21 indicates problems with this fit.  This is also implied by the poor $\chi^2$ statistic for this region. These results suggest that the deeply embedded clusters in DR~21 are distributed along the molecular filament and not yet fully collected into isothermal ellipsoids.  
\item[RCW~38] The small residuals suggest a good match between the data and the ``core-halo'' model, especially in the center of the cluster. The maximum residual is located north of the cluster and does not appear to be associated with any identified subcluster. 
\item[NGC~6357] A strong negative residual is located in the center of the southern complex of stars \citep[G353.1+0.6;][]{Townsley13}. The best AIC is obtained when three ellipsoids were used to fit this complex, although three distinct components are not obvious from a visual examination of the asymmetrical G353.1+0.6 cluster.
\item[Eagle Nebula] The main cluster in the Eagle Nebula is composed of two overlapping subclusters (A and B). There are large positive residuals at the edges of the cluster; nevertheless, different configurations of the two subclusters do not produce lower residuals, so the structure can not be fully represented with isothermal ellipsoids. 
\end{description}

In the right panels of Figure~\ref{resid_fig}, the surface density obtained from the nonparametric adaptive smoothing (x-axis) is plotted against the surface density obtained from the parametric finite mixture mode (y-axis) where each point corresponds to a pixel in the residual map. These generally agree to within 0.1~dex. The local maxima from the surface density maps are typically somewhat higher in the models than in the smoothed maps;  this is an artificial effect of the smoothing algorithm. For Carina, the agreement is good at surface densities above $\sim$3~stars~pc$^{-2}$. Deviations at lower surface densities may reveal differences between the modeling of unclustered young stars with a constant surface density and the true surface-density variations of the unclustered stars.

\clearpage\clearpage
\begin{deluxetable}{lrcrrrrrr}
\tablecaption{Best-Fit Models\label{model_table} \label{resid_fig}}
\tabletypesize{\small}
\tablewidth{0pt}
\tablehead{\colhead{} & \colhead{} & \colhead{} & \multicolumn{3}{c}{Residual Map} & \multicolumn{3}{c}{Goodness of Fit}\\
\cline{4-6}\cline{7-9}
\colhead{Region} & \colhead{$N_\mathrm{subclus}$} & \colhead{$\Sigma_\mathrm{U}$} &  \colhead{kernel} &\colhead{min} & \colhead{max} & \colhead{$\chi^2$} & \colhead{d.o.f.} & \colhead{$P_{\chi^2}$}\\
\colhead{} & \colhead{} & \colhead{(stars pc$^{-2}$)} & \colhead{(pc)} & \colhead{(stars pc$^{-2}$)}& \colhead{(stars pc$^{-2}$)} & \colhead{} & \colhead{} & \colhead{}\\
\colhead{(1)} & \colhead{(2)} & \colhead{(3)} & \colhead{(4)} & \colhead{(5)} & \colhead{(6)} & \colhead{(7)} & \colhead{(8)} & \colhead{(9)}}
\startdata
Orion & 4 & 		30  & 	0.076 & 290 (4\%)  & -270 (4\%)~~  & 47  & 60~ & 0.23\\
Flame  & 1 & 		4  & 		0.12~ & 80 (4\%)  & -60 (3\%)~~  & 21  & 23~ & 0.78\\
W 40  & 1 &		20  & 	0.17~ & 38 (3\%)  & -26 (2\%)~~  & 24  & 17~ & 0.22\\
RCW 36 & 3 &		7  & 		0.11~ & 96 (3\%)  & -57 (2\%)~~  & 17  & 17~ & 0.82\\
NGC 2264 & 13 &	4  & 		0.28~ & 16 (3\%)  & -20 (4\%)~~  & 50  & 46~ & 0.62\\
Rosette & 15 &		0.7  & 	0.72~ & 3.8 (7\%)  & -2.5 (5\%)~~ & 126  & 62~ & $<$0.01\\
Lagoon & 11 &		0.2  & 	0.33~ & 14 (3\%)  & -9.1 (2\%)~~  & 61  & 57~ & 0.61\\
NGC 2362  & 2 & 	0.3  & 	0.58~ & 3.5 (7\%) & -2.0 (4\%)~~  & 9  & 9~ & 0.86\\
DR 21  & 9  & 		 3 & 		0.39~ & 13 (6\%)  & -11 (5\%)~~  & 72  & 31~ & $<$0.01\\
RCW 38 & 4 &		0.08  & 	0.30~ & 13 (1\%)  & -8.6 (1\%)~~  & 25  & 25~ & 0.88\\
NGC 6334 & 14 &	1  & 		0.50~ & 4.9 (4\%)  & 5.9 (5\%)~~  & 43  & 37~ & 0.41\\
NGC 6357  & 6 &	0.01  & 	0.38~ & 14 (3\%)  & -12 (3\%)~~  & 120  & 73~ & $<$0.01\\
Eagle  & 12 &		1  & 		0.41~ & 14 (6\%)  & -9.1 (4\%)~~  & 75  & 66~ & 0.42\\
M 17 & 15 &		 0.2 & 	0.24~ & 25 (3\%)  & -24 (3\%)~~  & 75  & 73~ & 0.80\\
Carina & 20 &		0.01  & 	0.59~ & 12 (4\%) & -3.5 (1\%)~~  &  200 & 145~ & $<$0.01\\
Trifid & 4 &		0.3  & 	0.8~~ & 1.5 (3\%)  & -1.2 (3\%)~~  & 11  & 12~ & 0.98\\
NGC 1893 & 10 &	0.06  & 	0.64~ & 3.0 (3\%)  & -7.0 (7\%)~~  & 51  & 34~ & 0.06\\
\enddata
\tablecomments{Properties of the best-AIC models. Column 1: Region name. Column 2: Number of subclusters in the model. Column 3: Observed surface density of the uniform model component. Column~4: Gaussian kernel size used to smooth residuals. Columns~5-6: Maximum and minimum residuals (as percent of peak observed surface density). Columns~7-9: Goodness of fit test results: Pearson's $\chi^2$ value, degrees of freedom, and the probability, respectively. \label{best_fit_models_tab}}
\end{deluxetable}
\clearpage\clearpage

\clearpage\clearpage

\begin{figure}
\centering
\includegraphics[angle=0.,width=5.0in]{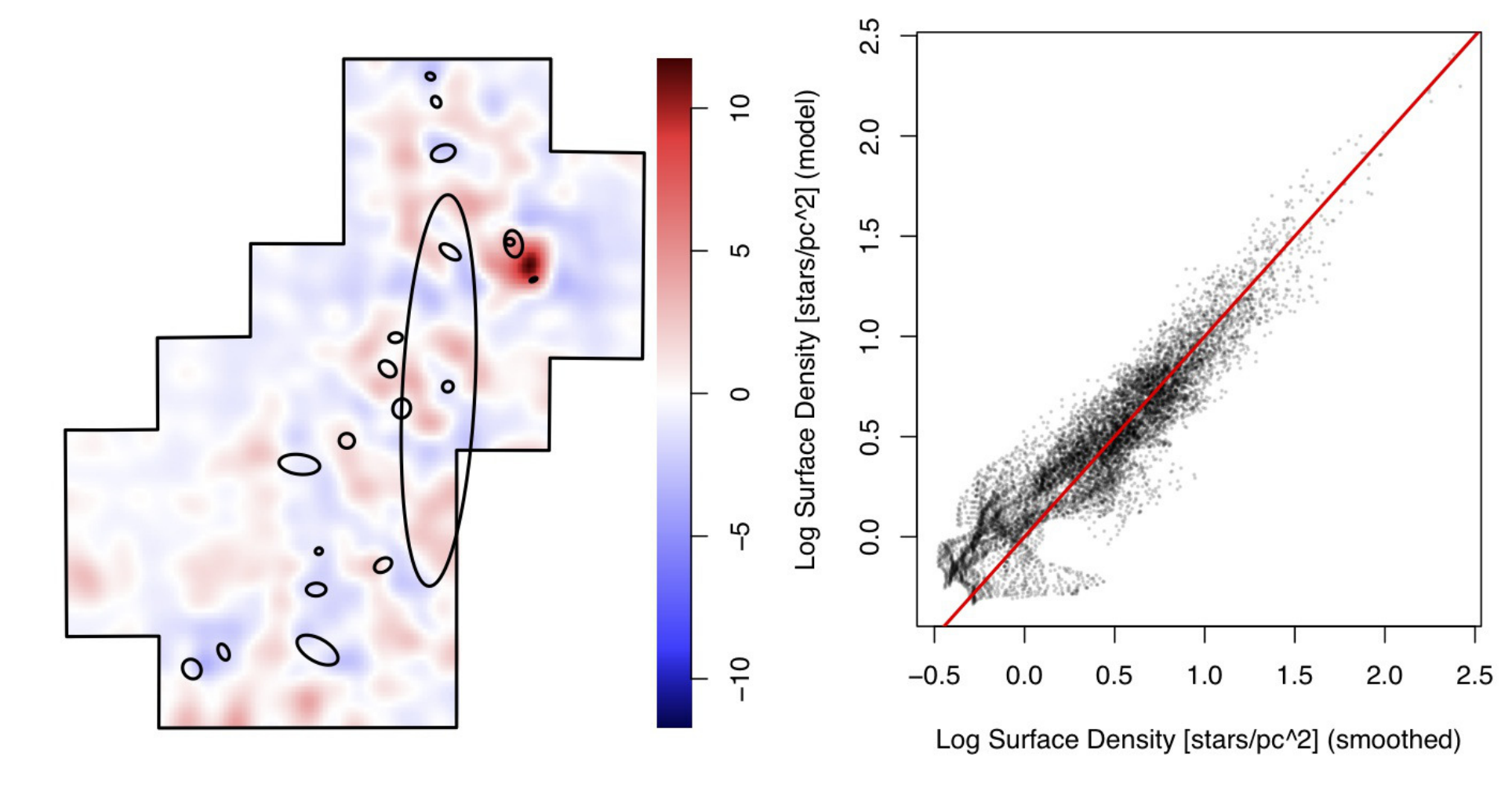}
\caption{Left: Residual surface density for the Carina Nebula. Negative residuals are shown in blue and positive residuals are shown in red as indicated by the color bar in units of stars~pc$^{-2}$. Right: Log surface density obtained from the finite mixture model is plotted against log surface density from the adaptively smoothed surface density maps. The $y=x$ lines is shown in red. Analogous plots for the other sixteen \mystix\ regions are provided by the electronic version of this article. 
}
\end{figure}

\setcounter{figure}{3}
\begin{figure}
\centering
\includegraphics[width=0.3\textwidth]{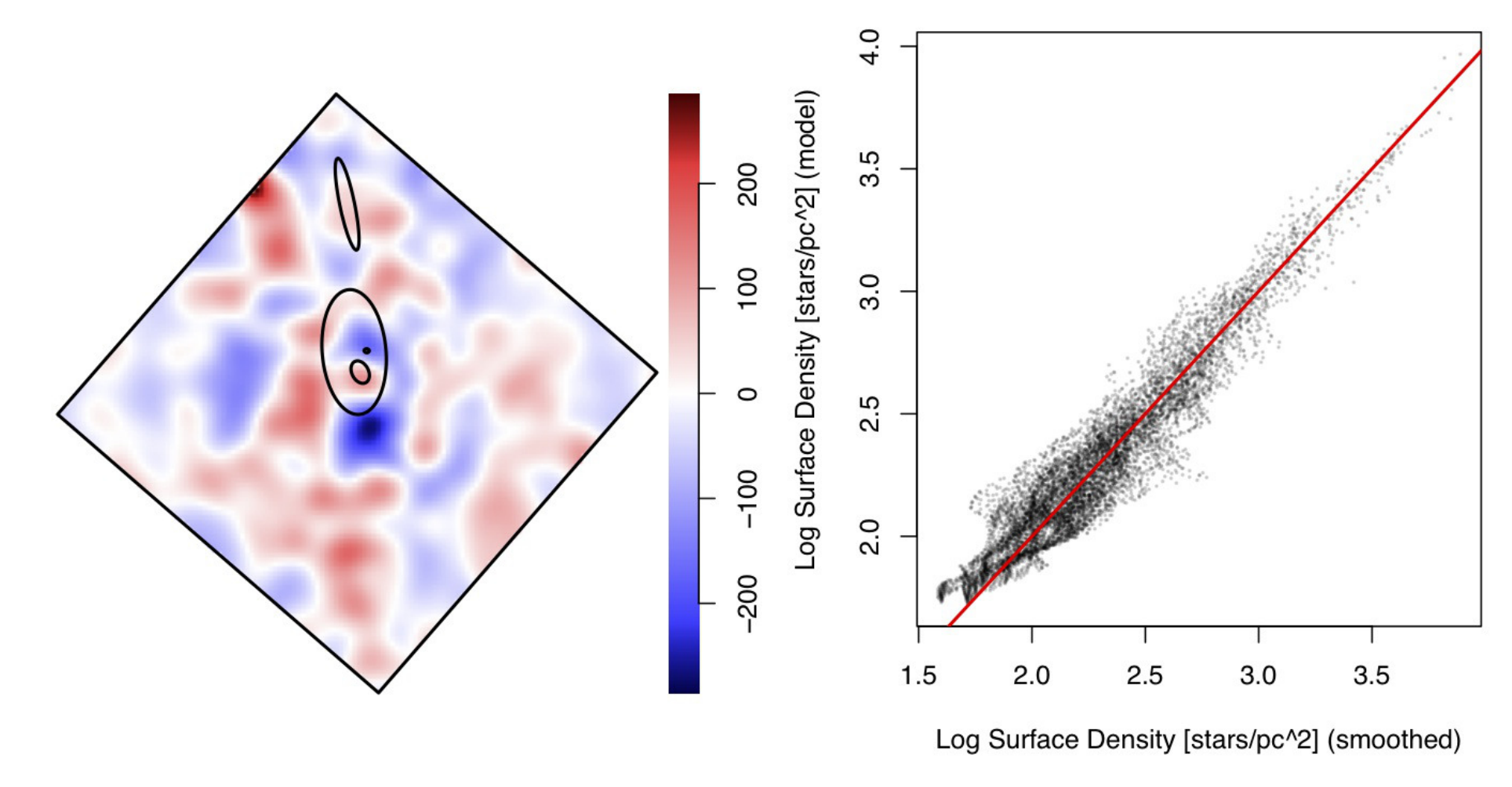}
\includegraphics[width=0.3\textwidth]{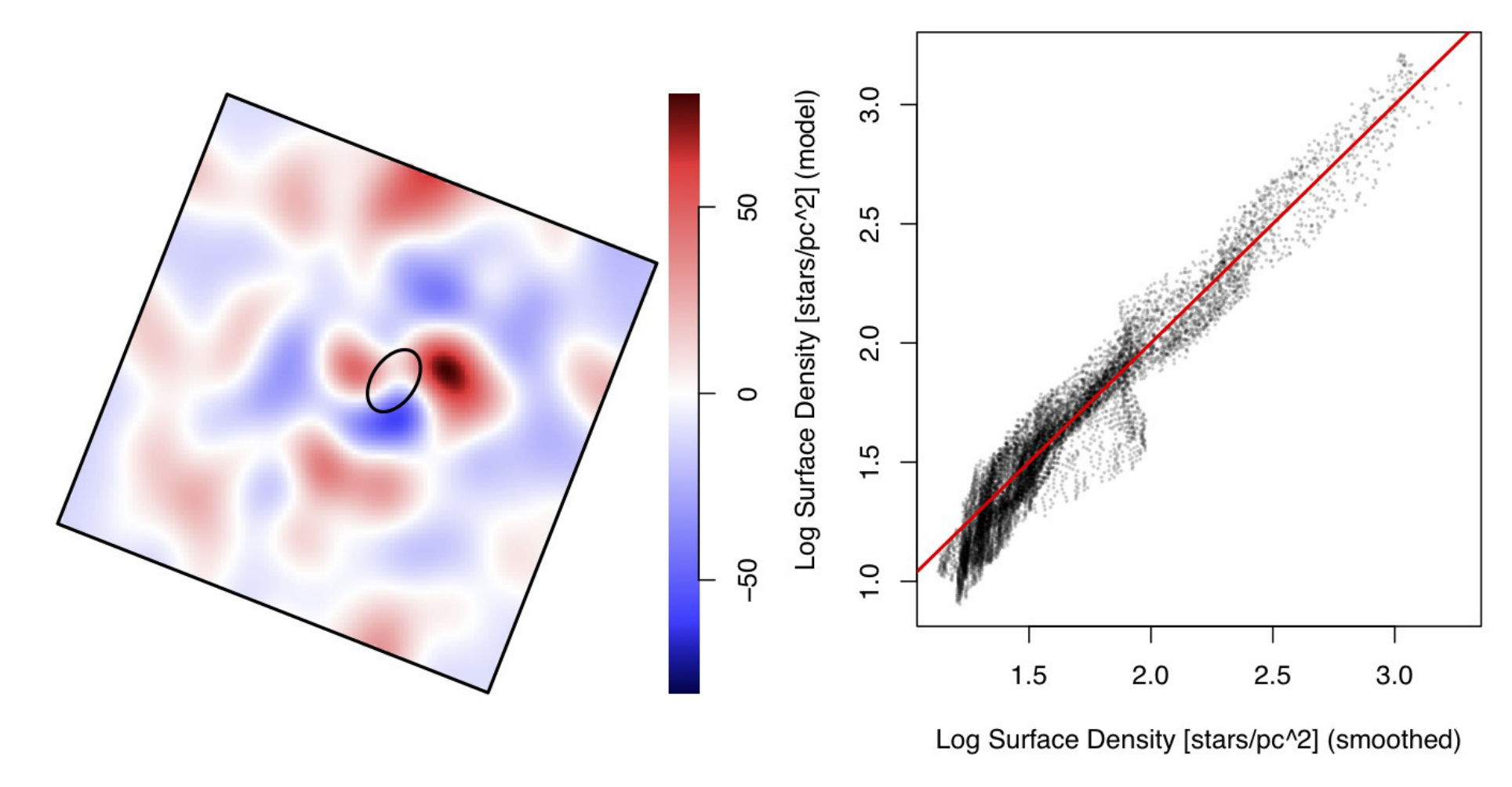}
\includegraphics[width=0.3\textwidth]{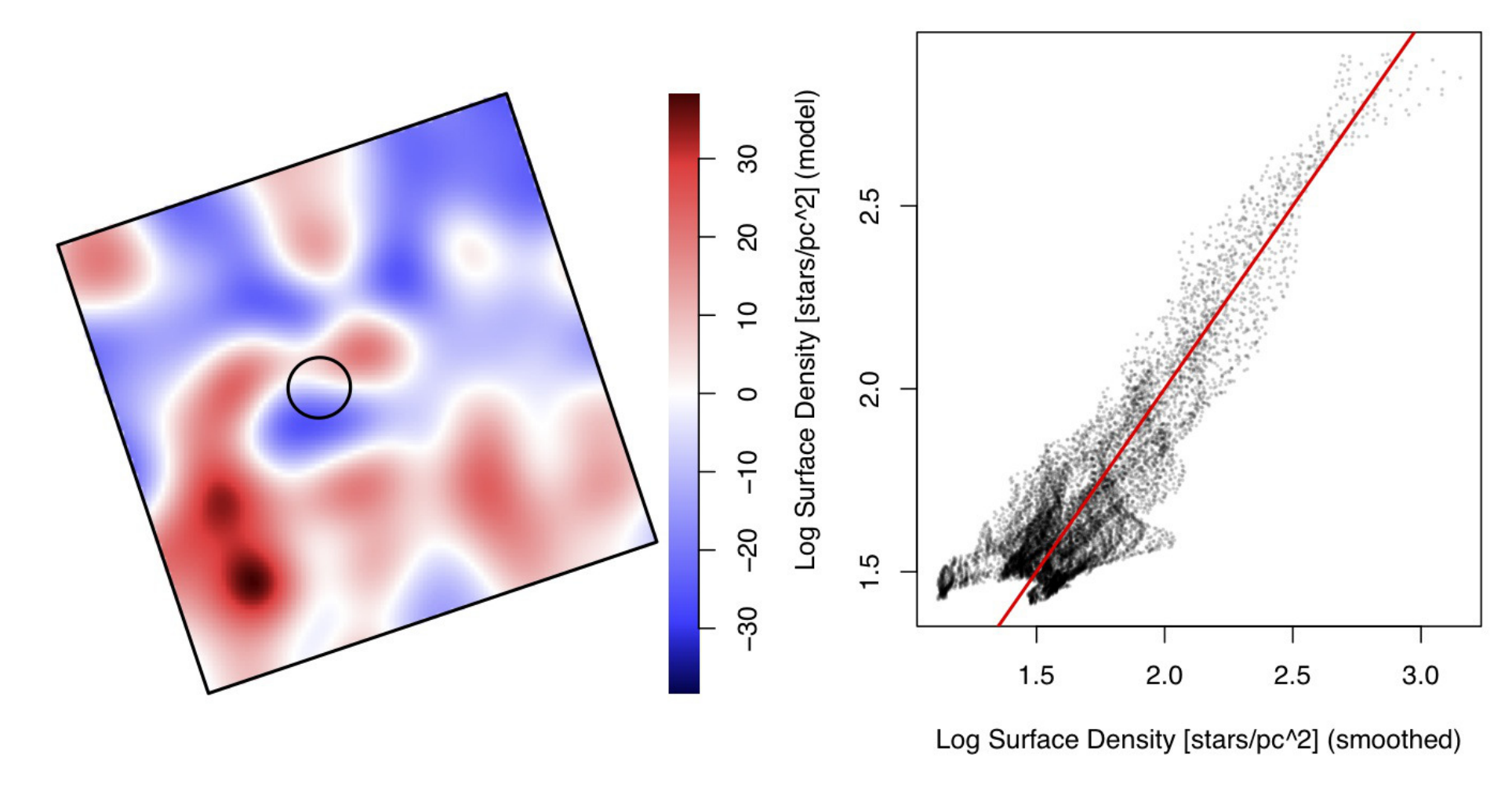}\\
\includegraphics[width=0.3\textwidth]{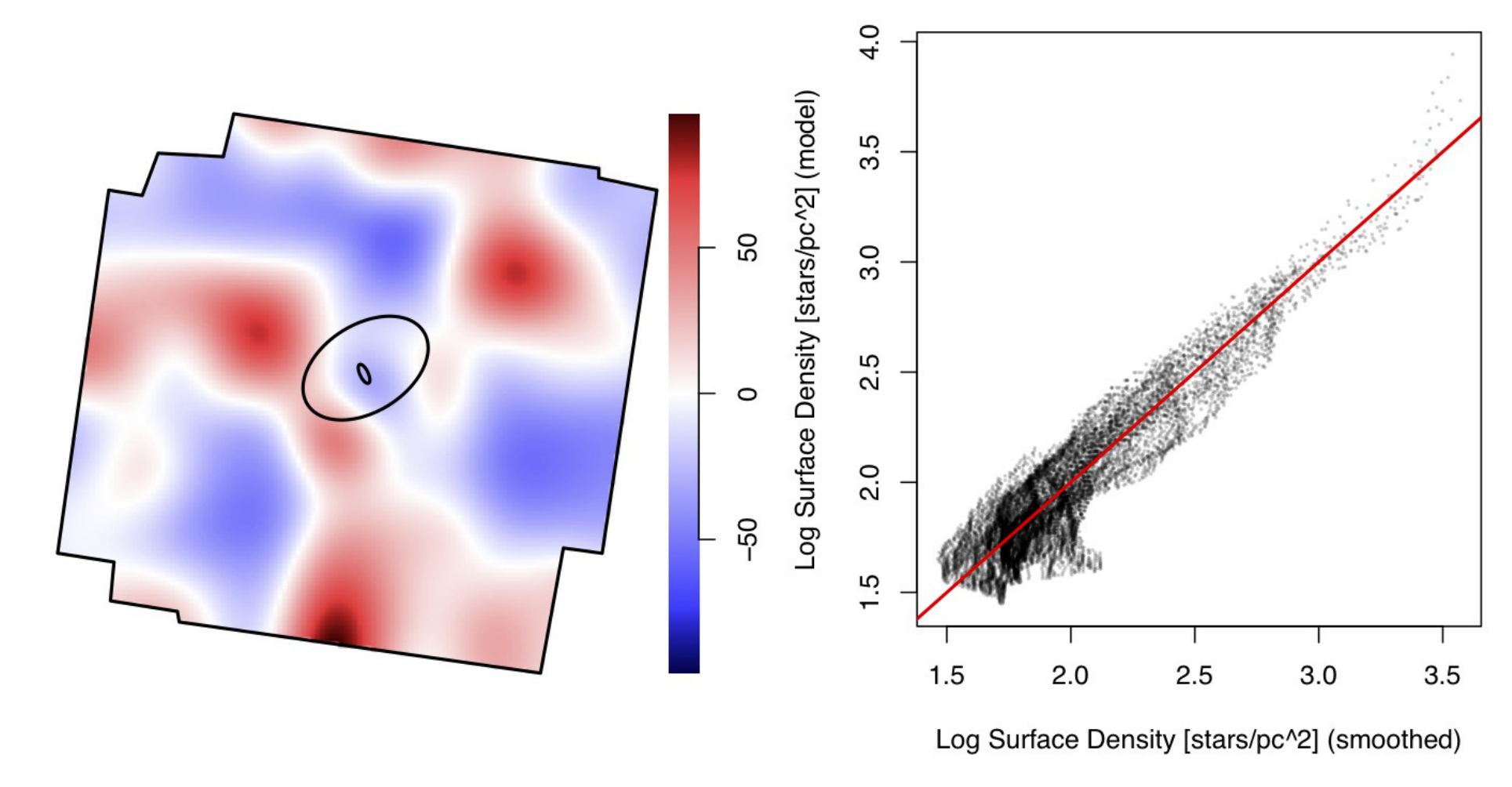}
\includegraphics[width=0.3\textwidth]{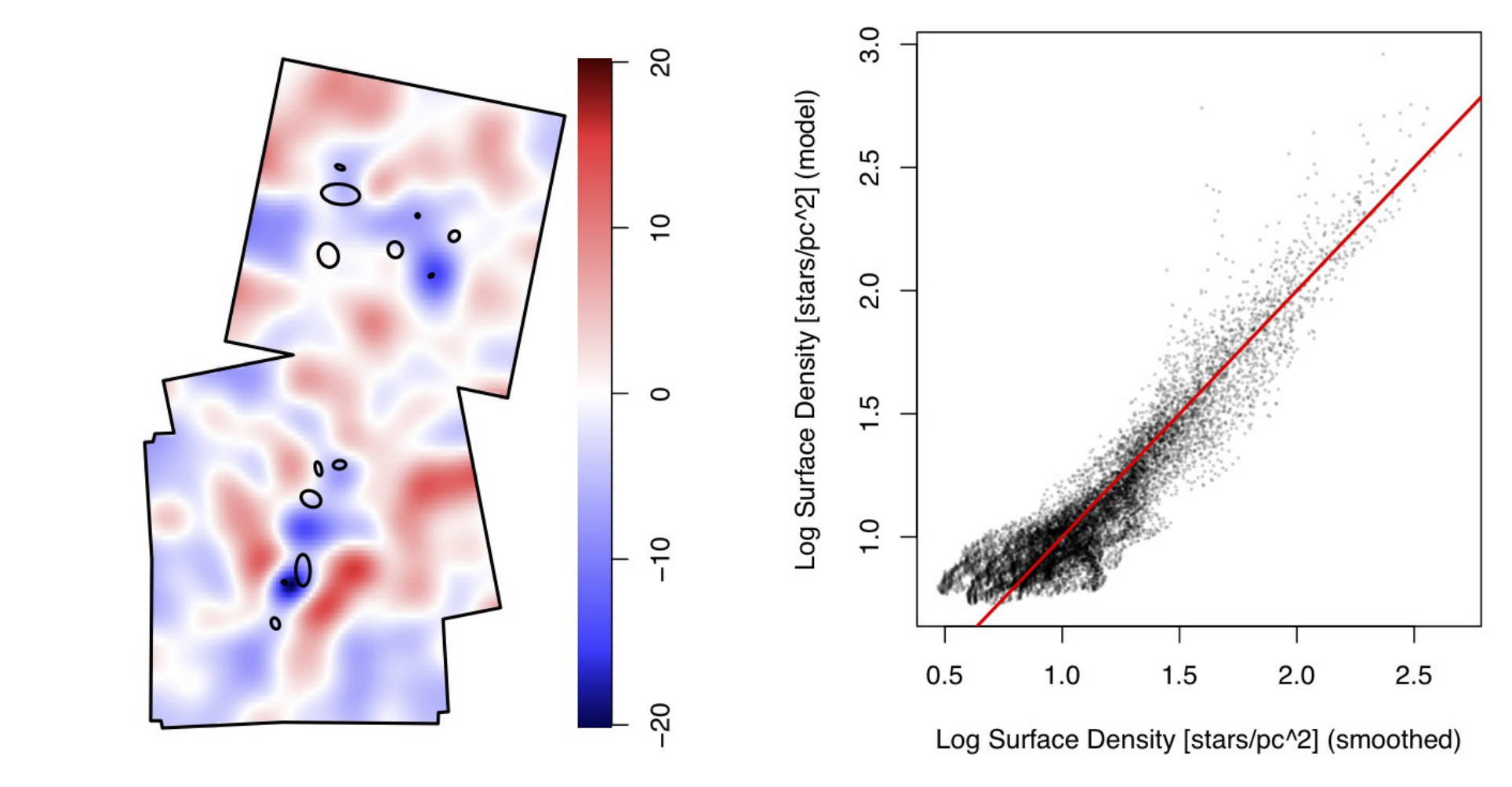}
\includegraphics[width=0.3\textwidth]{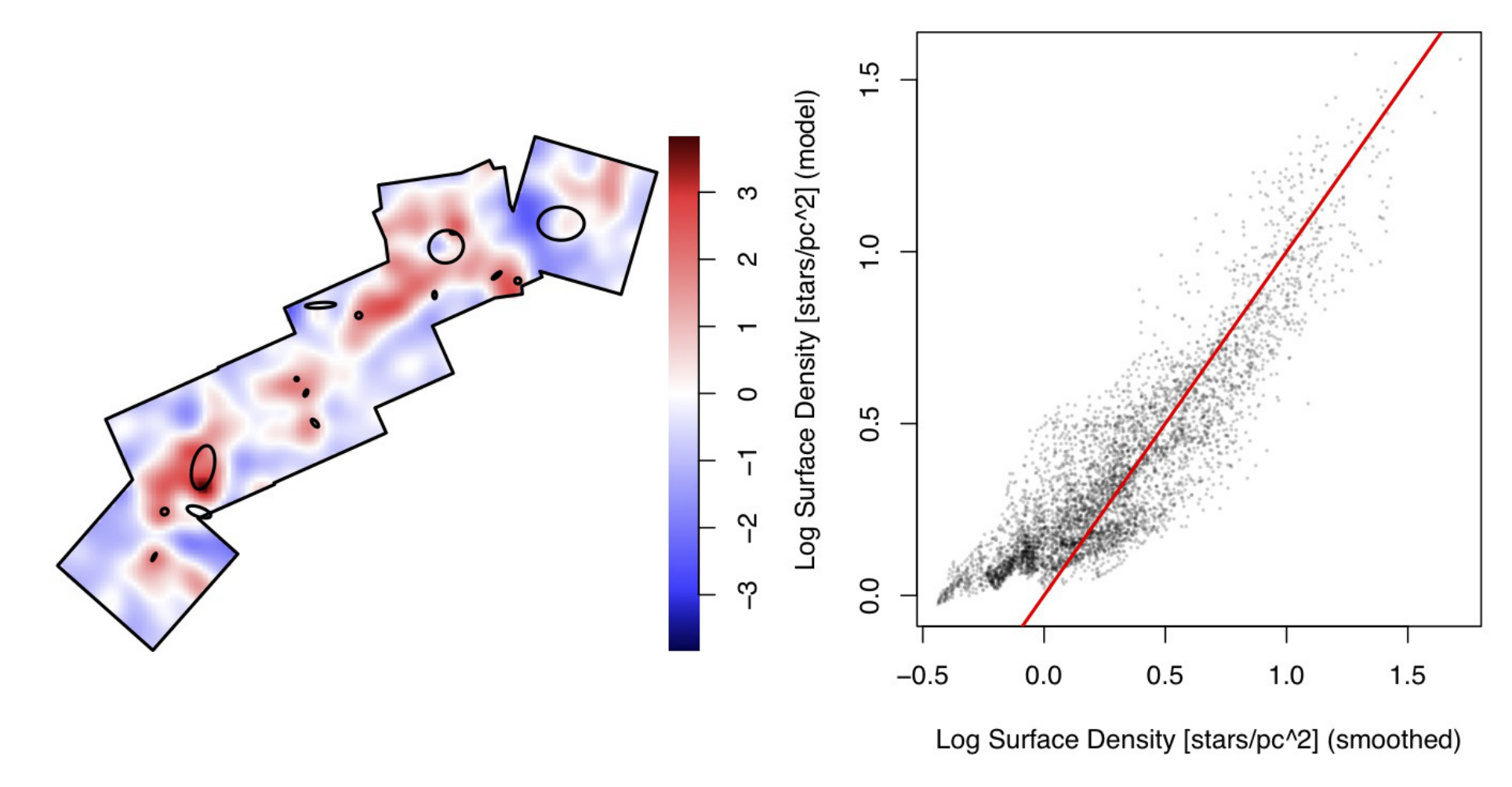}\\
\includegraphics[width=0.3\textwidth]{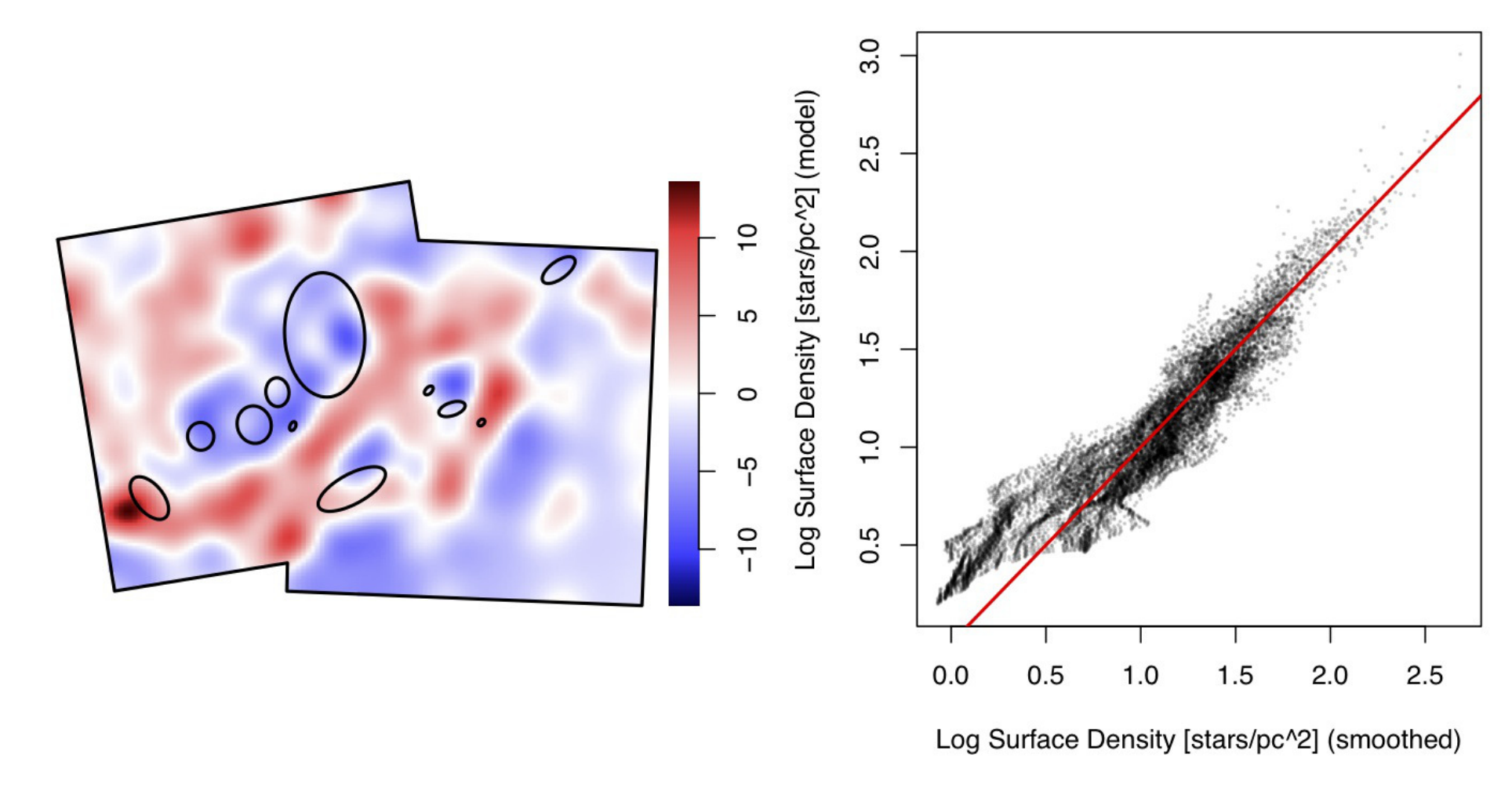}
\includegraphics[width=0.3\textwidth]{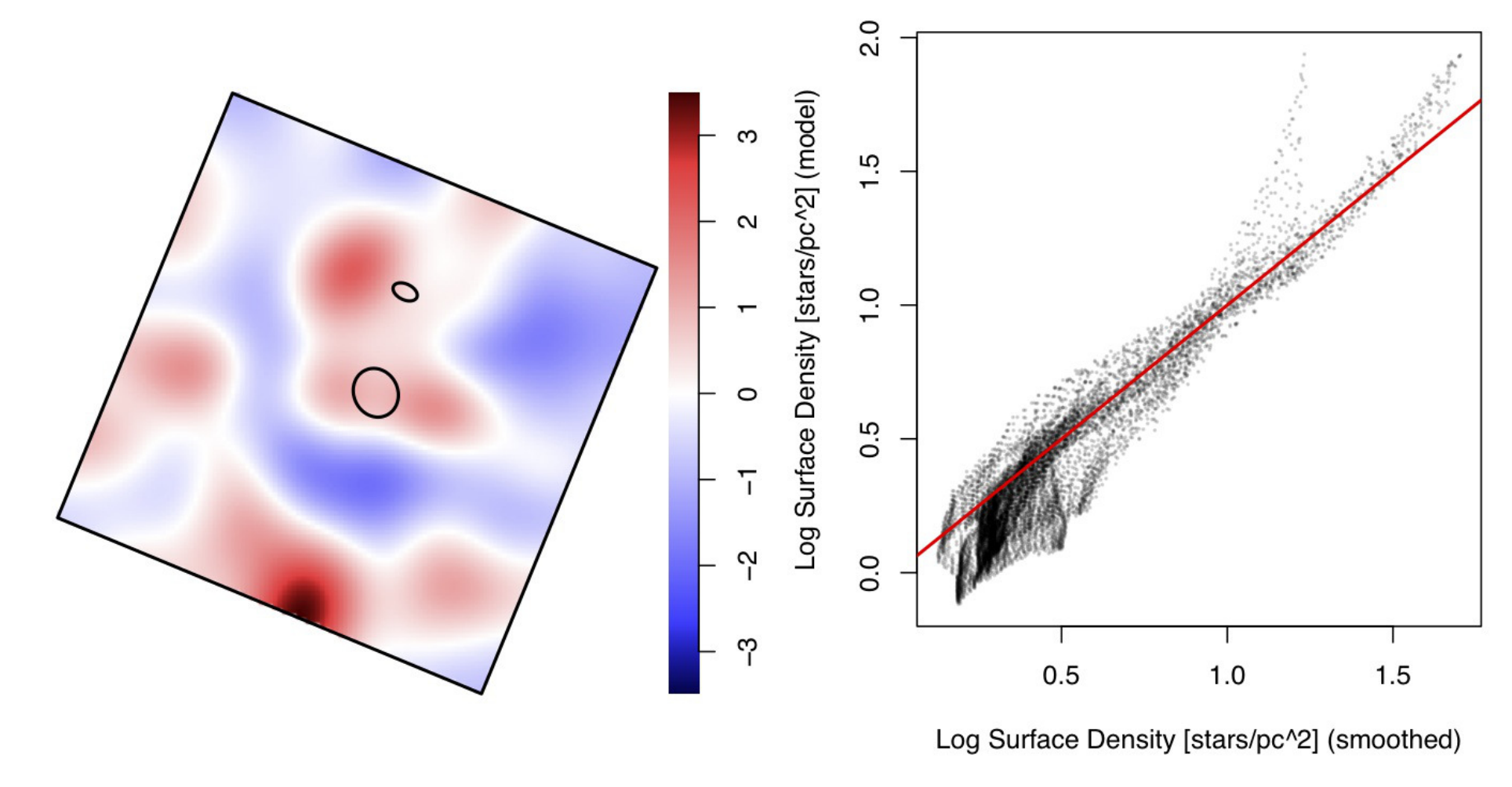}
\includegraphics[width=0.3\textwidth]{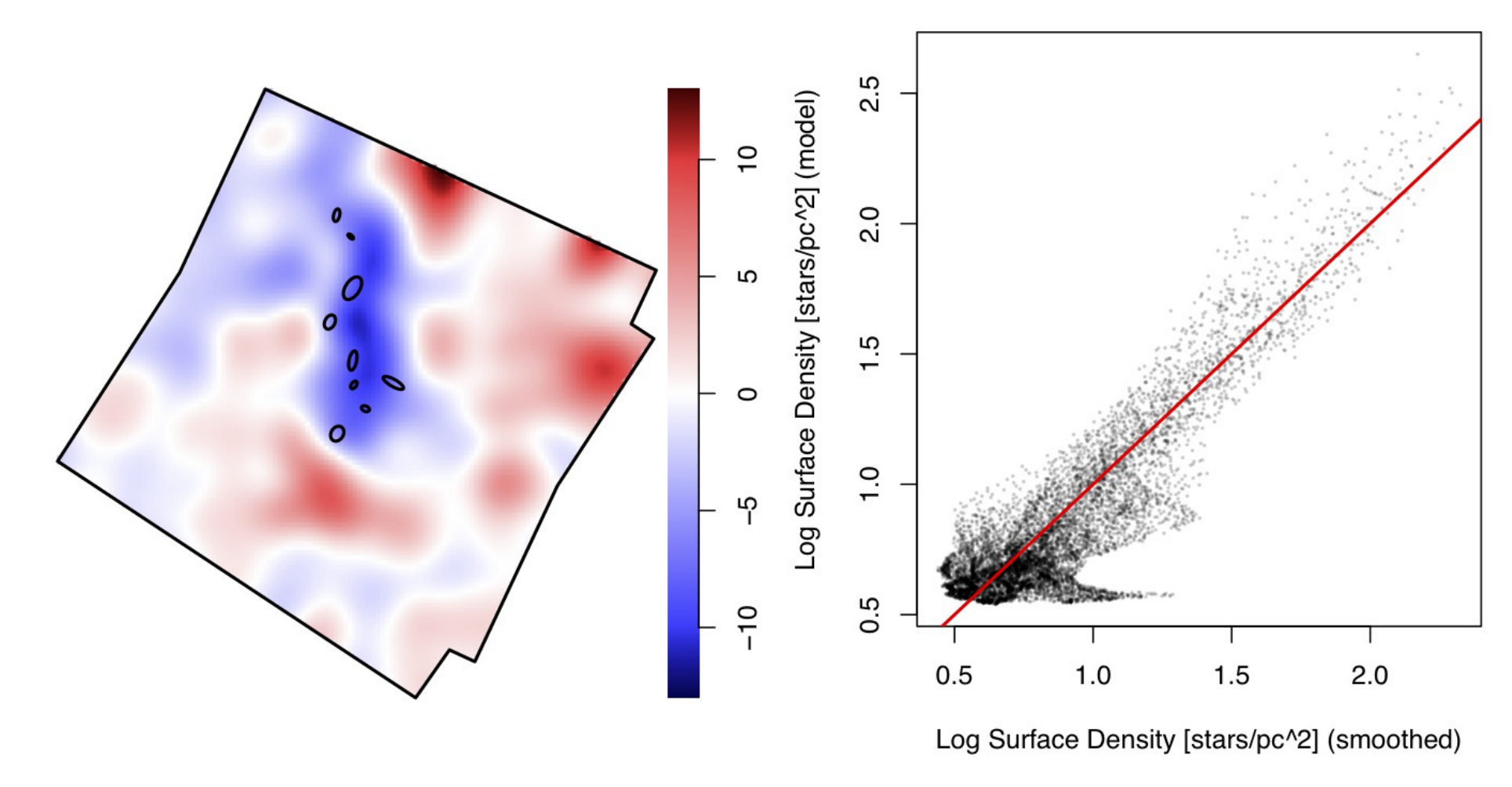}\\
\includegraphics[width=0.3\textwidth]{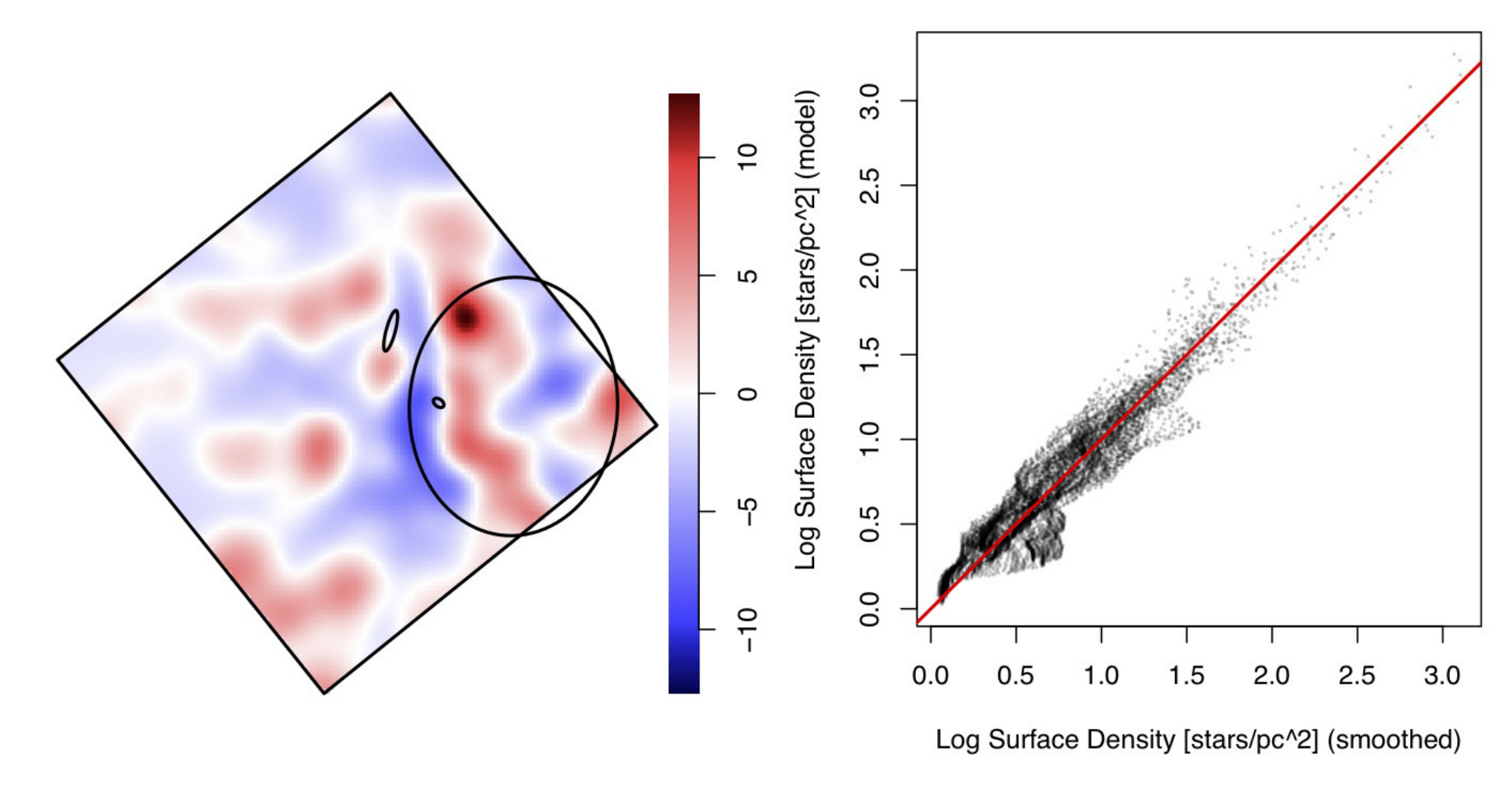}
\includegraphics[width=0.3\textwidth]{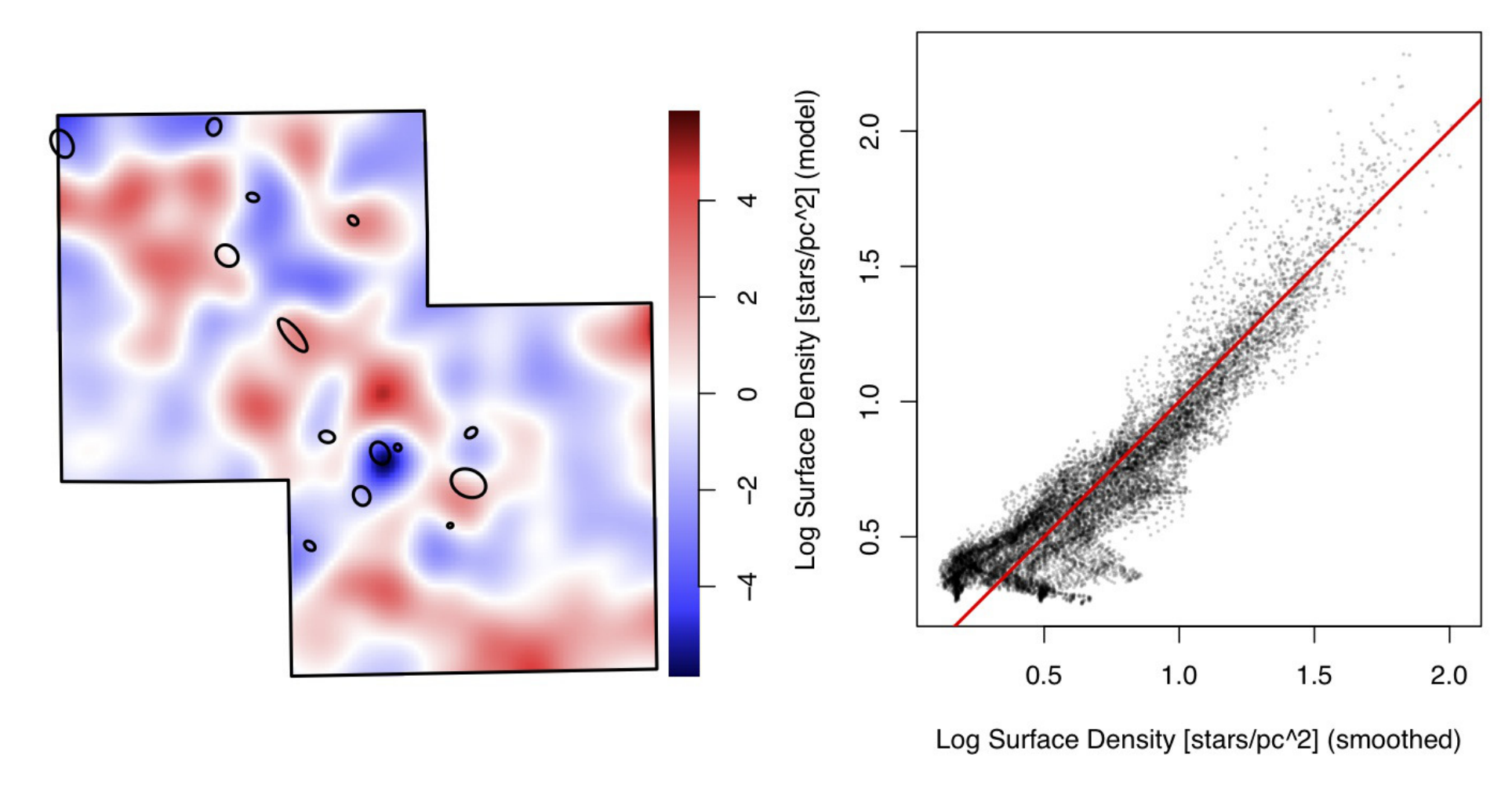}
\includegraphics[width=0.3\textwidth]{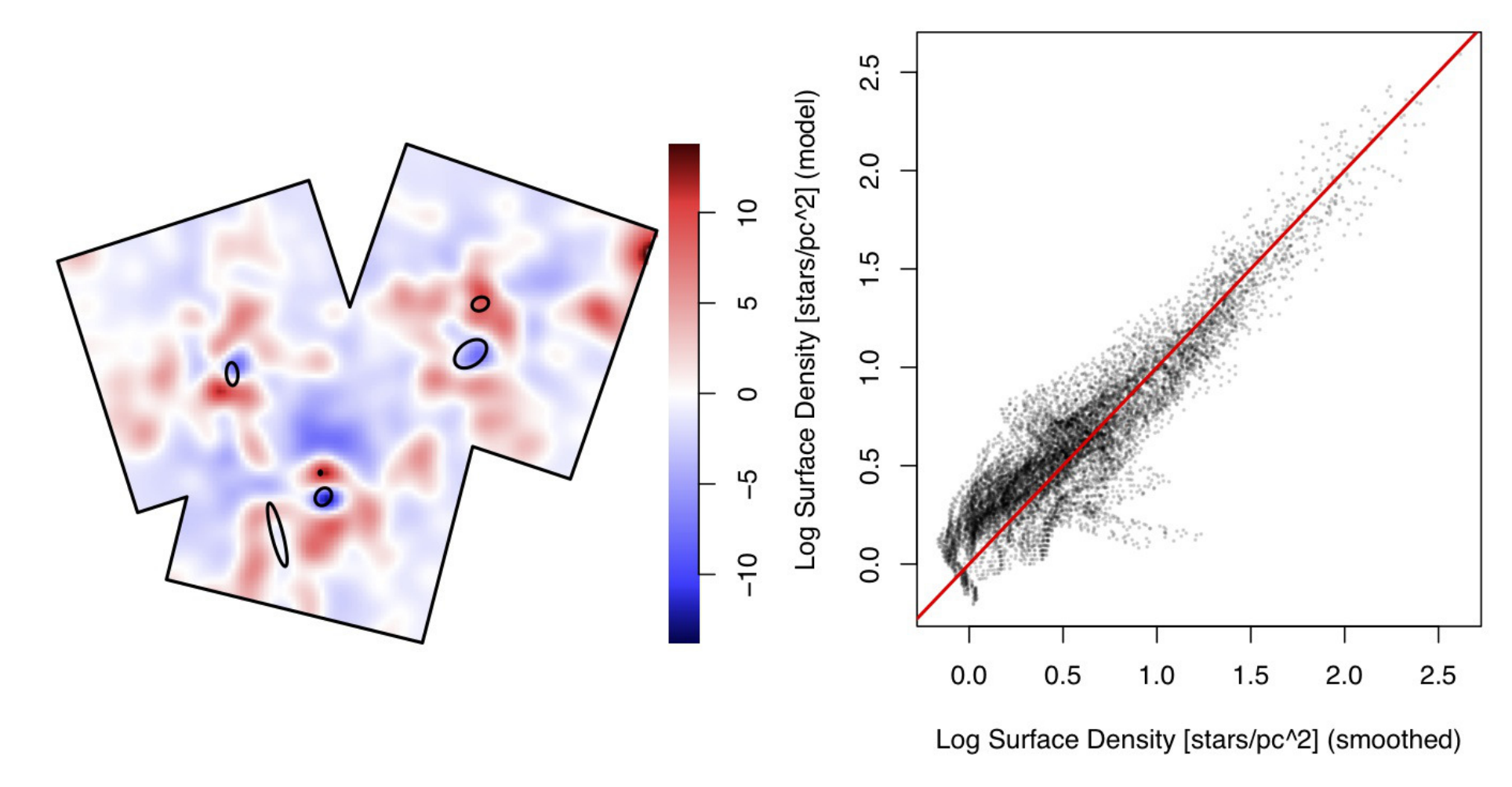}\\
\includegraphics[width=0.3\textwidth]{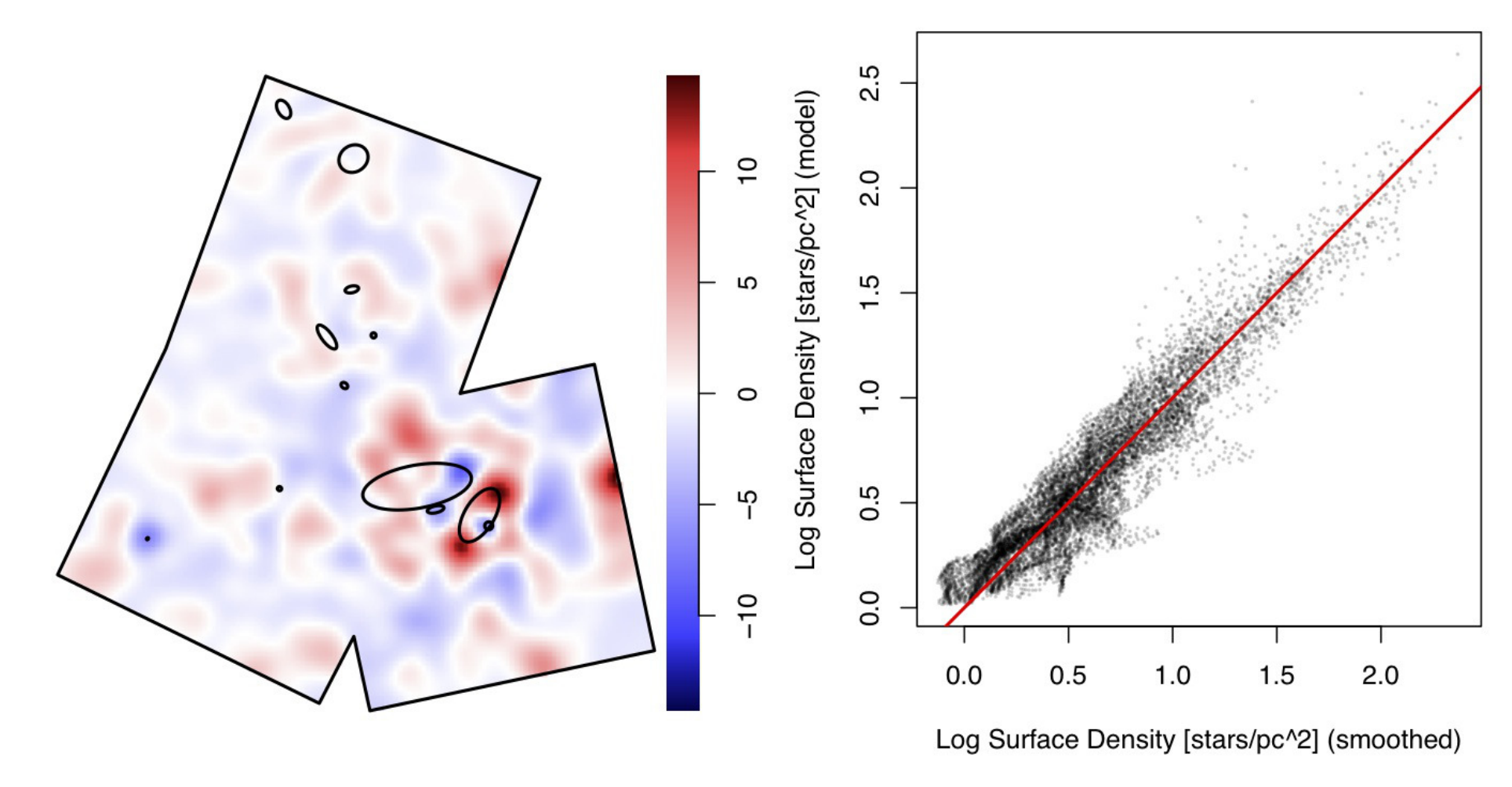}
\includegraphics[width=0.3\textwidth]{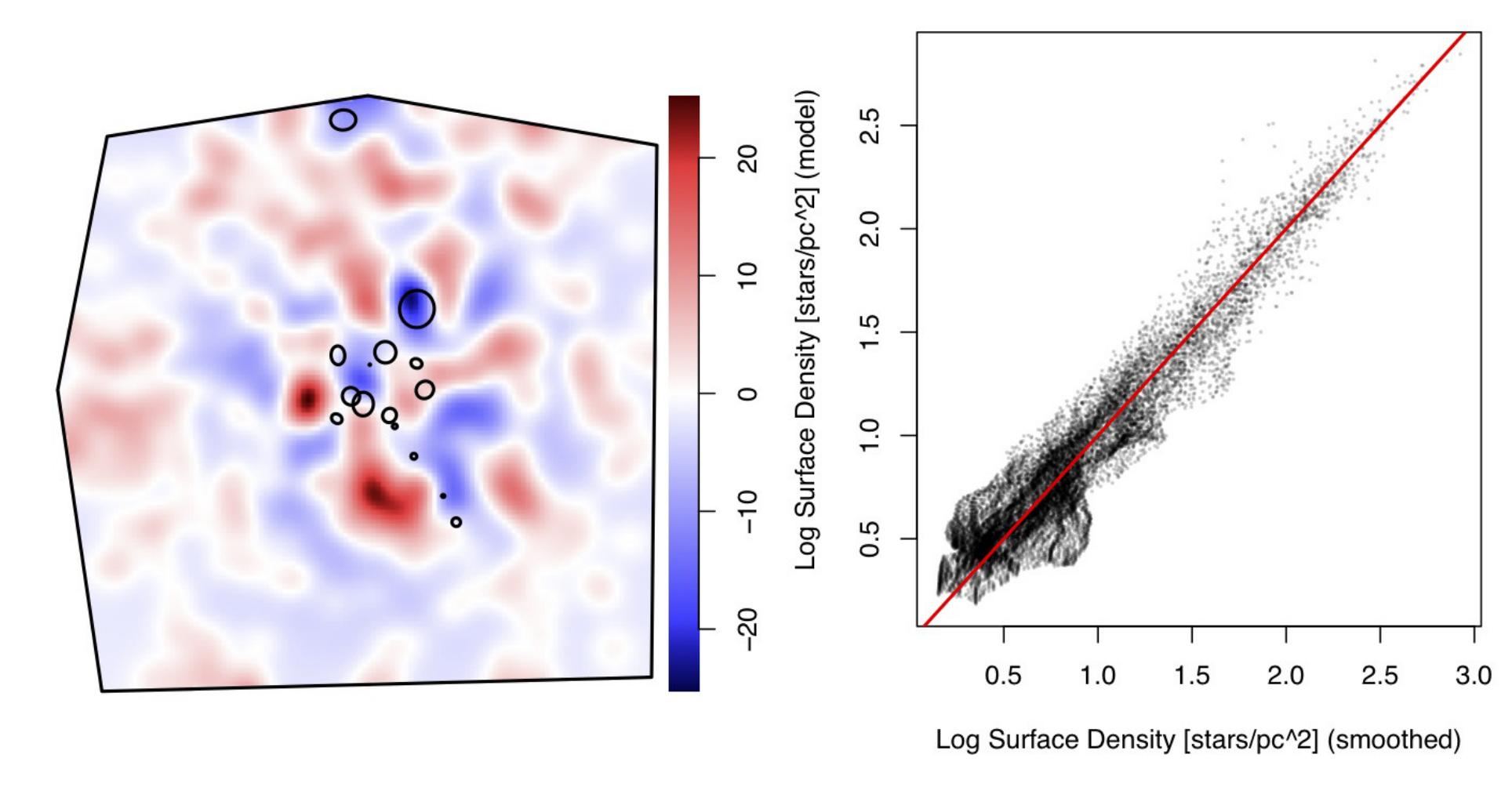}
\includegraphics[width=0.3\textwidth]{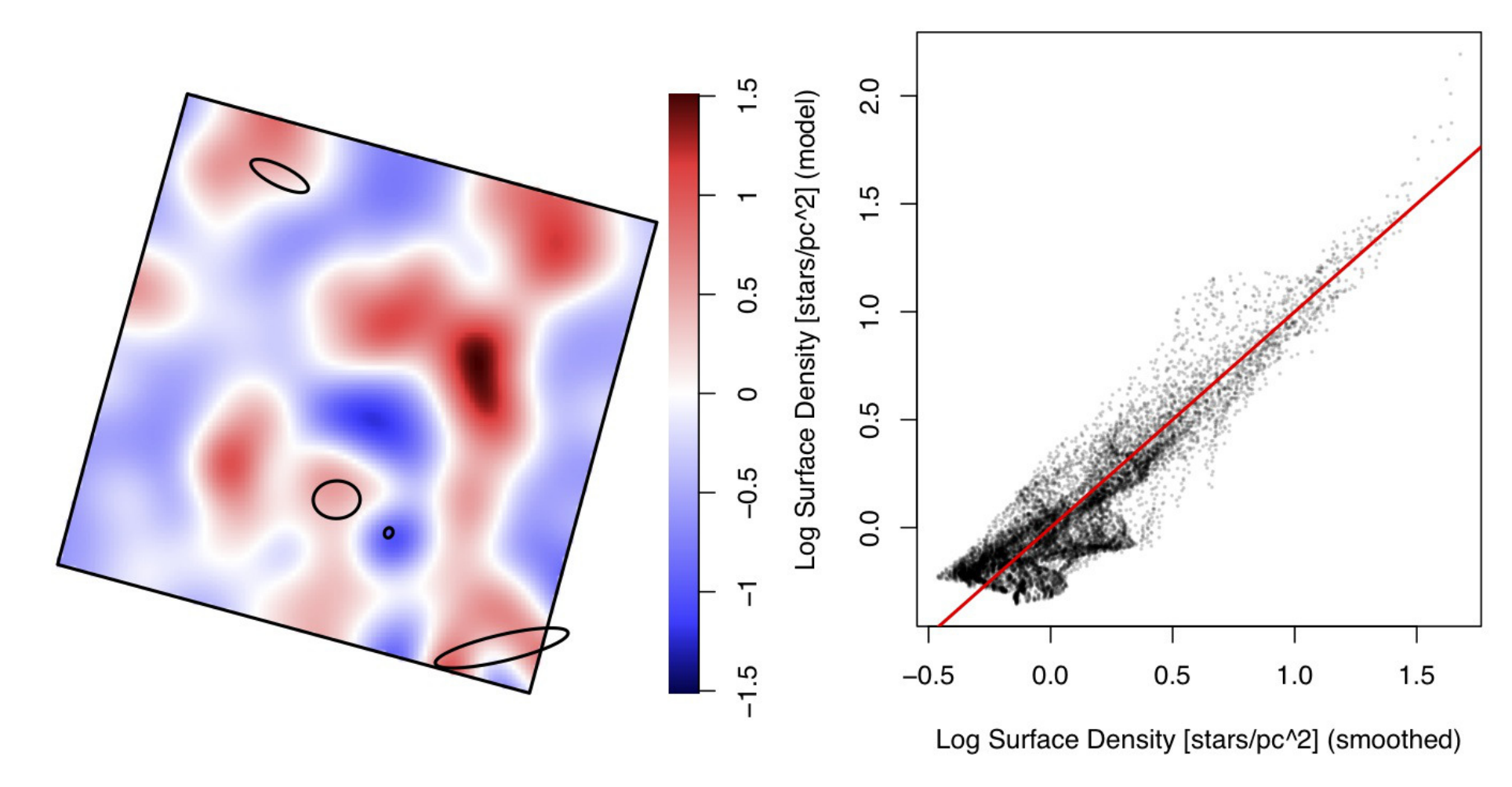}\\
\includegraphics[width=0.3\textwidth]{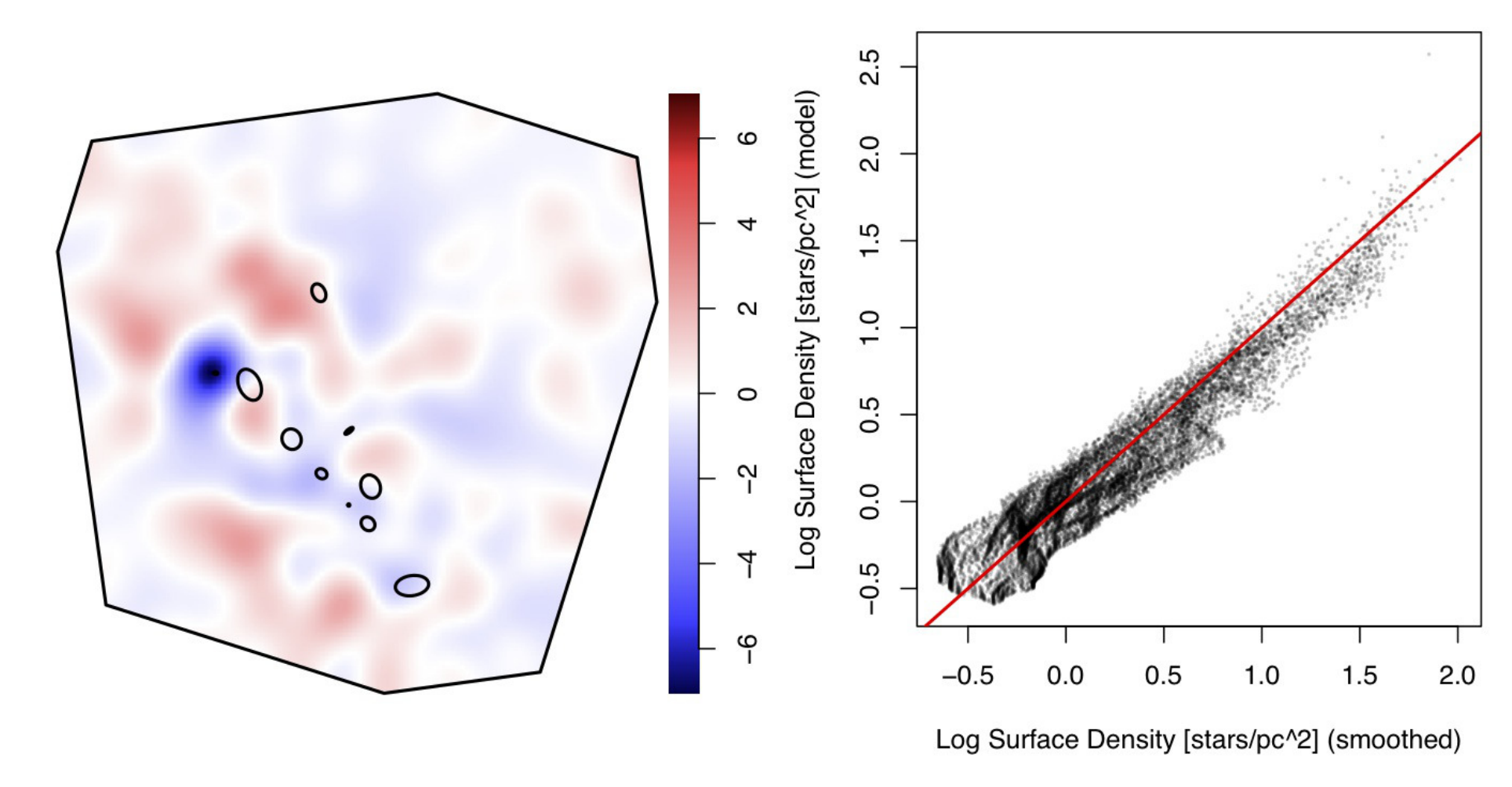}
\caption{Electronic Figure Set (left to right and top to bottom: Orion, Flame, W 40, RCW~36, NGC 2264, Rosette, Lagoon, NGC 2362, DR 21, RCW 38, NGC 6334, NGC 6357, Eagle, M 17, Carina, Trifid, NGC 1893)}
\end{figure}

\clearpage\clearpage

\section{Four Classes of Star-Forming Complexes} \label{classes_section}

\begin{figure}
\centering
\includegraphics[angle=0.,width=6.0in]{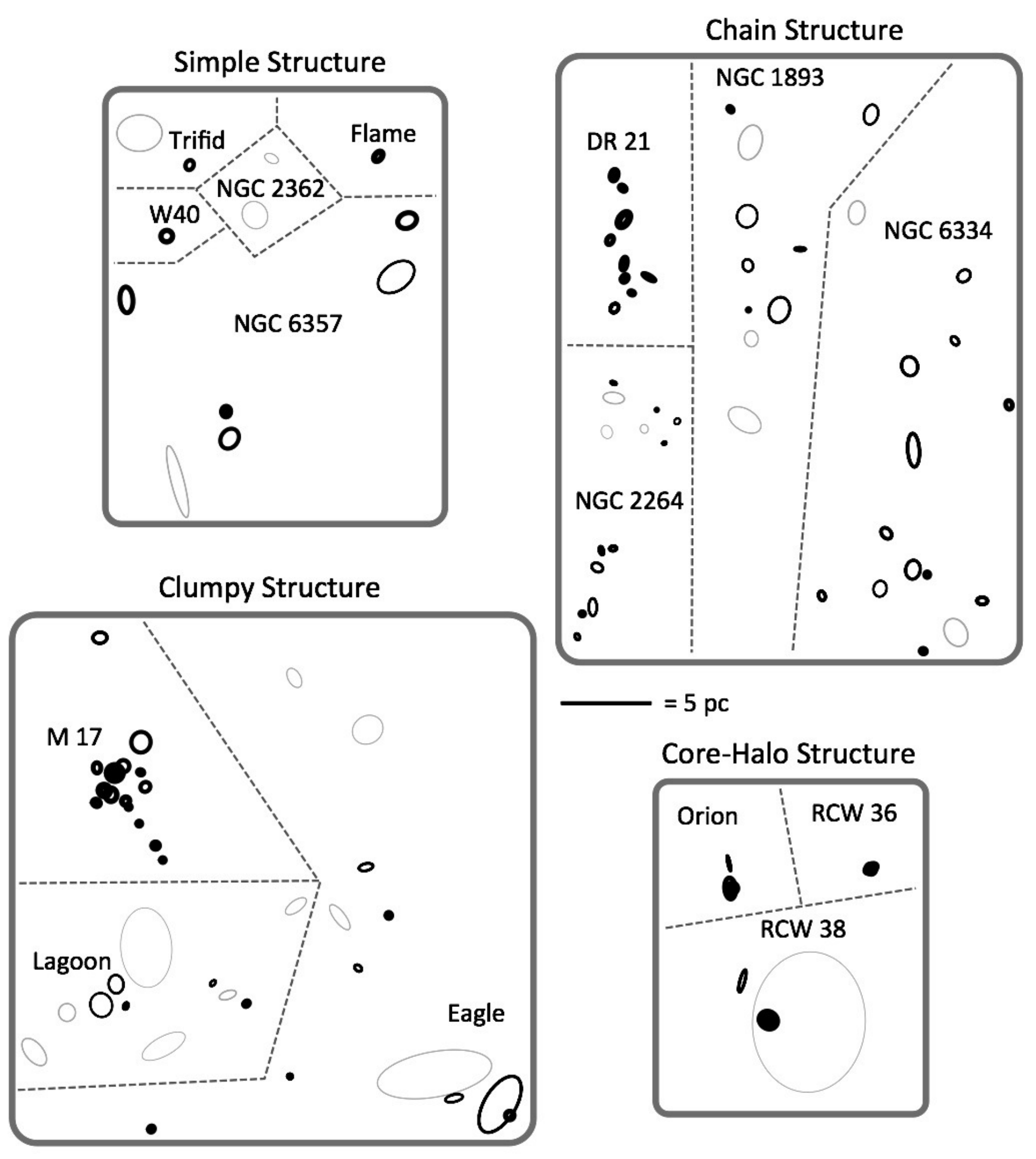}
\caption{
Diagrams of subcluster ellipsoids are shown on the same physical parsec scale, indicated by the 5~pc length scale in the center. The regions are grouped by morphological class --- simple, isolated isothermal ellipsoids (upper left); linear chain structure (upper right); clumpy structure (lower left); and core-halo structure (lower right). The Rosette Nebula and the Carina Nebula are omitted due to figure size constraints.  Orientations of the coordinate systems have been adjusted for easy comparison. Line thickness and shading indicate the relative stellar surface density for the subcluster, with thicker lines indicating higher surface density. 
\label{morph_class_fig}}
\end{figure}

The 17 MSFRs examined here have a wide diversity of structures, but they generally can be sorted into four classes. Figure \ref{morph_class_fig} compares the cluster structure (indicated by the patterns of core ellipses) for 15 of the regions on the same parsec scale. The Carina and Rosette Nebulae complexes are omitted because they can be viewed as composites of the four classes, as described below.

\begin{description}

\item[Simple Isolated Cluster Structure] Some regions are dominated by a simple, unimodal cluster that is well fit by a single ellipsoid model. Regions with this structure are the most likely to be dynamically relaxed.
This class includes the Flame Nebula, W~40, the principal clusters in the Trifid Nebula and NGC~2362, the two northern subclusters A and F in NGC~6357, and the main cluster NGC~2244 in the Rosette Nebula. The Flame Nebula is slightly elongated along the direction of the filamentary molecular cloud that passes through the region, while W~40 is round. NGC~2362 is composed of two overlapping subclusters, but both are well fit by ellipsoid models. Although the structure appears dynamically relaxed may suggest that these regions tend to be older, regions in this class are not always the oldest \mystix\ regions.  Some have molecular material and show signs of active star formation. The two-body relaxation time for many regions is probably longer than the age of the cluster. 

\item[Core-Halo Structure] Some clusters appear unimodal with little substructure around a main cluster, but still be poorly fit by a single isothermal ellipsoid model because of an excess of stars near the center of the cluster. The finite-mixture-model algorithm often breaks these up into two ellipsoids, one used to model the cluster core and the other to model the cluster halo. Examples include the ONC, RCW~36, RCW~38, Eagle Nebula subclusters A-B, and Tr~14 in the Carina Nebula.  These results confirm core-halo structures reported in the ONC and Tr~14 by \citet{Henney98} and  \citet{Ascenso07}, respectively.  In the cases of the Orion Nebula and RCW~36, the fitted core radius of the central cusp is extraordinarily small, $\sim0.01-0.02$~pc. The case of RCW~38 appears different because  the core subcluster is larger and off-center.  Although, the simple isolated clusters and core-halo clusters may be surrounded by peripheral subclusters or unclustered stars, more than half of individual MPCM stars from the flattened sample lie within 5 core radii of the principal clusters.

\item[Clumpy Structure] In some of the \mystix\ regions, it is difficult to model the surface density distribution into a few discrete isothermal ellipsoids because of clumpy structures within the main cluster. Regions with such clumpy structure include M~17, the Lagoon Nebula, and the Eagle Nebula. These three cases all have central dominant subclusters, but a substantial fraction of the stars lie in secondary subclusters that appear contiguous with the central concentration. The embedded subclusters in the South Pillars of the Carina Nebula or in the Rosette molecular cloud may also be considered clumpy, but these cases lie far from the dominant clusters.

\item[Linear chains of clusters] This structure consists of subclusters that are roughly arranged in a line. Five \mystix\ clusters exhibit such an arrangement: DR~21, NGC~2264, NGC~1893, NGC~6334, and the Carina Nebula. The length of clusters in the \mystix\ sample ranges from $\sim$10~pc to 30~pc, and in some cases can be as narrow as $\sim$1~pc perpendicular to the main axis. However, the stars are clearly broken up into distinct subclusters, rather than being continuously distributed along the length. The subclusters in these regions are often, but not always, elongated along the long axis of the filament. The young stars are more evenly distributed between the subclusters in these regions with no subcluster containing more than 20\% of the MPCMs (from the flattened sample) within 5 core radii.  Our data support arguments based on molecular and dust cloud studies that this structural class arises from star cluster formation along IR dark cloud filaments \citep{Rathborne06,Kauffmann10,Battersby11, Schneider12}. As expected from this model, the subclusters in DR~21, NGC~2264, and NGC~6334 are embedded in molecular clumps.  

\end{description}

\begin{figure}
\centering
\includegraphics[angle=0.,width=6.5in]{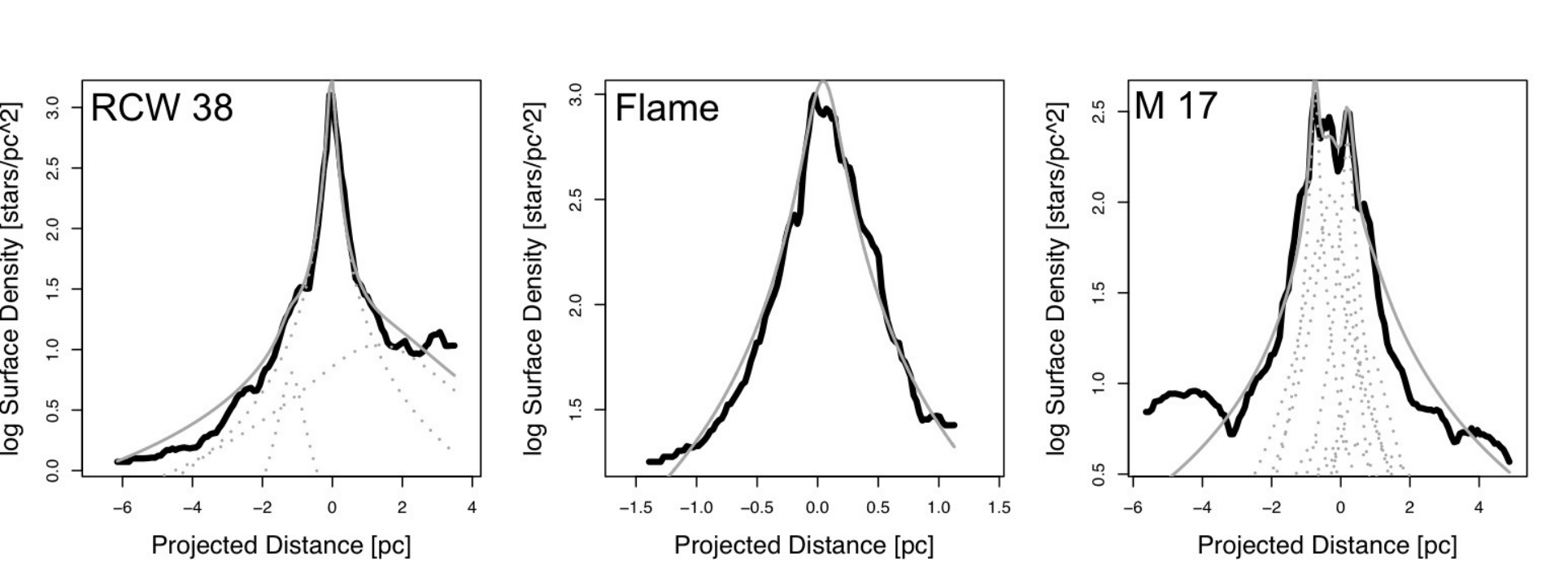}
\caption{Surface density profiles along lines of constant declination through RCW~38 (left; core-halo structure), the Flame Nebula (center; simple structure), and M~17 (right; clumpy structure). Surface density is shown on a logarithmic scale.  In each panel, the black line shows the adaptively smoothed data, the gray solid line shows the total model, and the gray dashed lines show the ellipsoidal model components. 
\label{slices_figure}}
\end{figure}

To exemplify the structure for three classes, Figure~\ref{slices_figure} shows surface density profiles along along one-dimensional cuts through the smoothed maps.  The Flame Nebula has a simple, single-ellipsoid morphology, RCW~38 has core-halo morphology, and M~17 has a clumpy morphology.   The core-halo structure in RCW~38 manifests as a shoulder to the right of the strong peak in surface density. The halo is difficult to see in the map of RCW~38 (shown in Figure~2) because it is overwhelmed by the extreme peak in surface density near the center;  however, the halo component creates a large asymmetry in the profile, and it is strongly favored by the AIC. In contrast, the Flame Nebula cluster is symmetric and both sides of the distribution match the single isothermal ellipsoid model well, except for a small discrepancy between the model and data near the peak of the distribution. The profile for M~17 contains multiple peaks modeled with different ellipsoid components.  The union of multiple subclusters produces a density profile that is much wider than the profile of any individual ellipsoid.

These qualitative morphological classes are intended to provide an intuitive description of regions, but they are not mathematically precise. The classifications can be ambiguous: for example, the Eagle Nebula and M~17 have some linear chain like structures although the preponderance of stars are concentrated in a few closely spaced subclusters.  In NGC~6334, most stars are spread along a linear chain, but some subclusters lie off the line.  Projection effects could also make a linear chain structure appear clumpy.
Further observations of proper motions and/or radial velocities will be helpful at clarifying ambiguities of the two-dimensional structure \citep[e.g.][]{Furesz08}.

\clearpage
\clearpage

\section{Properties of \mystix\ Subclusters} \label{ellipse_section}

\subsection{Subcluster Sizes} \label{sizes_section}

\begin{figure}
\centering
\includegraphics[angle=0.,width=4.0in]{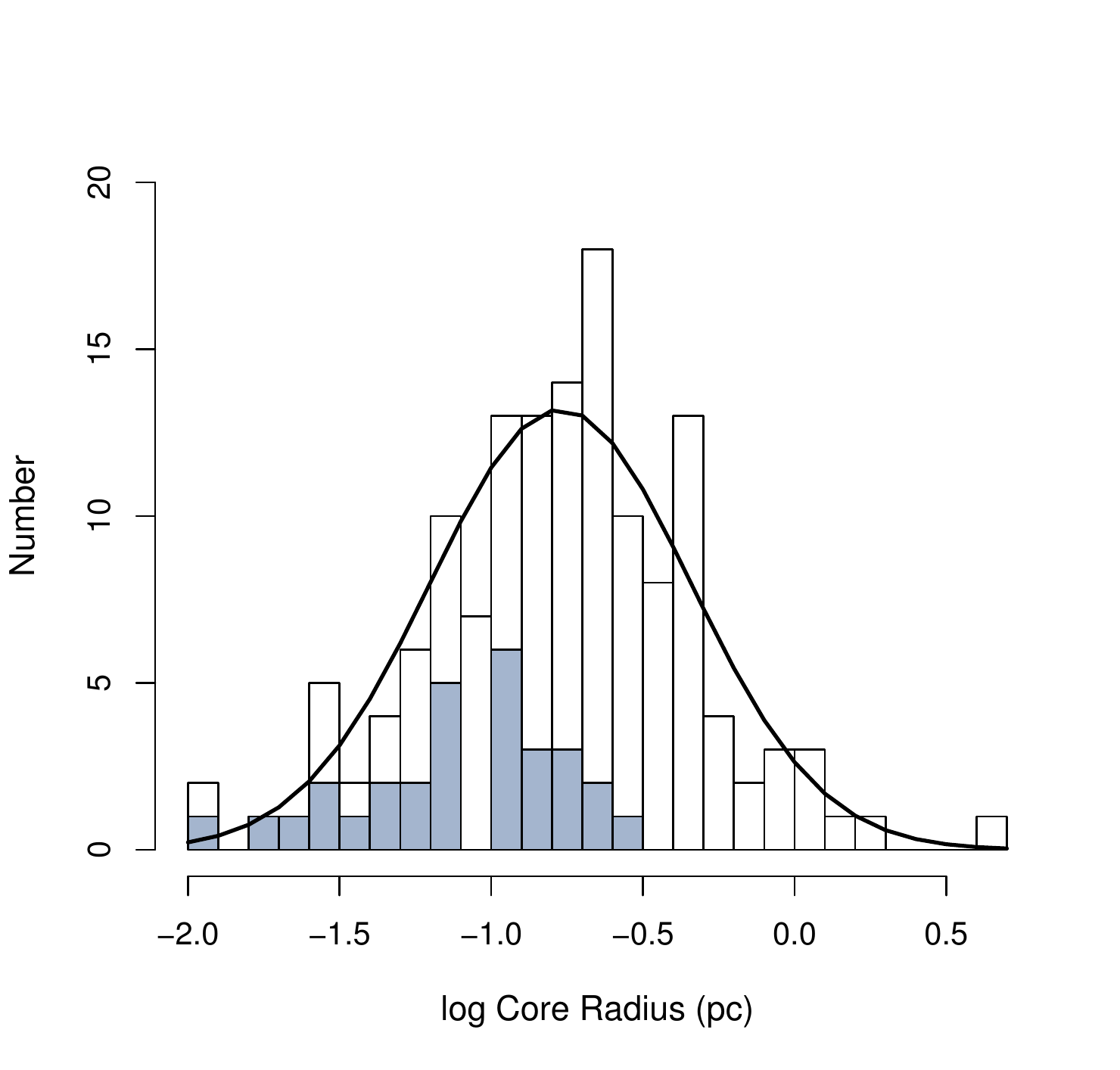}
\caption{Histograms of subcluster core radii, for all subclusters (white) and embedded subclusters (gray) with a lognormal fit. 
\label{size_distribution_figure}}
\end{figure}

The size of subcluster cores varies from 0.01~pc to $>$2~pc.  To investigate subcluster sizes, we use the geometric mean of semi-major and semi-minor axes to define a characteristic radius.  For the subclusters from Table \ref{ellipsoid_1_table}, the distribution of characteristic radii is shown in Figure \ref{size_distribution_figure}.  The distribution is fit well by a lognormal distribution with a mean  0.17~pc and standard deviation 0.43~dex; that is,  68\% of core radii lie between 0.06 and 0.45~pc.  The Anderson-Darling test for normality shows no statistically significant deviation from this distribution ($p>0.05$).  The peak of the distribution is similar to the ONC core radius of $\sim0.2$~pc found by \citet{Hillenbrand98}.  

The tails of this distribution are not well-defined because the smallest radii are seen in sparse subclusters without clear core regions while the largest radii arise from large-scale over-densities that do not appear to be distinct subclusters.   There may be a bias against the detection of subclusters with very large or small sizes. Detection of large subclusters is limited by the field of view and its stars may instead be assigned to the flat distributed star component in our statistical model. Very small subclusters will also be difficult to detect if they lack sufficient numbers of stars to significantly increase the likelihood of the model. A trend of subcluster size  increasing with MYStIX region distance suggests this bias is present.  On larger scales, in regions with where the young star surface density is low, distinct subclusters may be merged together to create larger artificial subclusters;  deeper X-ray and infrared surveys may result in more subclusters.   Nevertheless, the most common subcluster size ($\sim$0.2~pc) is consistent for \mystix\ regions at all distances.

\begin{figure}
\centering
\includegraphics[angle=0.,width=6.0in]{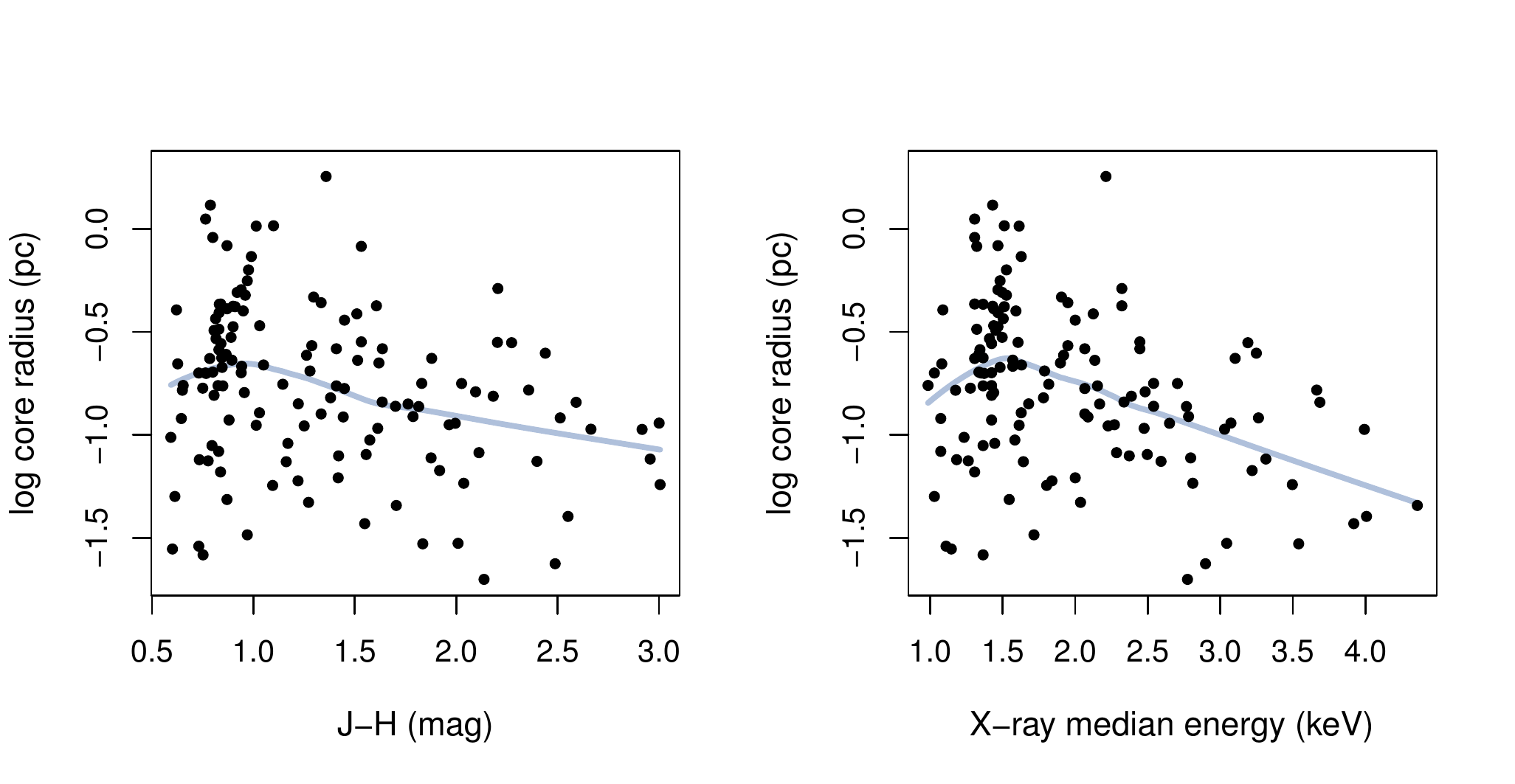}
\caption{Dependence of cluster size on absorption measures.  Left: Subcluster core radius vs.\ median $J-H$ mag. Right: Subcluster core radius vs.\ median X-ray median energy in the 0.5$-$8.0~keV band. Non-parametric LOWESS regression fits are shown in gray. 
\label{size_vs_me_figure}}
\end{figure}

Figure \ref{size_vs_me_figure} is a scatter plot of cluster core radius versus two measures of absorption along the line of sight, X-ray median energy $ME$ and $J-H$ for the cluster. A negative relation can be seen between these quantities, which is emphasized by the gray LOWESS non-parametric regression.   LOWESS is a well-established regression technique that combines robust least squares local regression technique with a $k$-nearest neighbor meta-model \citep{Cleveland79}.  The $J-H$ range corresponds approximately to $2 < A_V < 10$ mag, and the X-ray median energy range corresponds to $A_V<1$ to $A_V \simeq 50$ mag \citep{Getman10}.  Although there is much scatter in these relations, nonparametric correlation hypothesis tests using the Spearman rank coefficient (and confirmed using bootstrap resampling) give significant levels $p<0.001$ for the radius--$ME$ relation and $p<0.01$ for the radius--$J-H$ relation.

Some of the highly absorbed subclusters with small radius are well-known embedded star forming regions including the BN/KL cluster (subcluster A in Orion) and the Kleinmann-Wright Object cluster in M~17 (subcluster B in M~17). In particular, subclusters with median $ME>2.5$~keV (corresponding to $A_V>17$~mag) are rarely larger than 0.2~pc. This phenomenon is more evident in the histogram of core radii, which shows that clusters with median X-ray median energies greater than 2.5 (shaded gray region in Figure  \ref{size_vs_me_figure}) typically have smaller radii than less embedded clusters.  The peak of their lognormal distribution is at 0.08~pc, and the Anderson Darling test shows no statistically significant deviation from a lognormal distribution ($p>0.05$).  

Clusters lose their gas as they age, so the $J-H$ index for absorption can serve as a proxy for age --- this is directly demonstrated by \citet{Getman13a}. Thus, the negative size--absorption relation can be viewed as a positive size--age relationship suggesting that clusters expand as they age. \citet{Gieles12} find that the combination of multiple clusters at different stages of expansion can produce a distribution of stellar surface densities similar to what is seen in the \mystix\ regions.  

This effect may explain some of the differences in cluster sizes between different star formation regions. For example, DR~21 and NGC~1893 are both composed of linear chains of subclusters, but the subclusters in NGC~1893 are larger. This makes sense in light of the size--absorption relation because the younger DR~21 subclusters are deeply embedded in a molecular filament, while the older NGC~1893 subclusters are lying in a bubble and are lightly absorbed. \citet{Marks12} identify a typical half-mass radius for young embedded clusters of less than a few tenths of a parsec; for \mystix\ subclusters with $ME>3$ the mean core radius is considerably smaller than this around $0.04$~pc (Figure \ref{size_vs_me_figure}).

We note that the typical size of star clusters is similar to the width of filaments found by {\it Herschel} which appear to be ubiquitous in MSFRs and involved in the cluster formation process \citep{Andre10}. The variation in size and absorption of subclusters within a single star forming region can be interpreted as an indication of non-coeval clustered star formation.  

\clearpage\clearpage

\clearpage
\clearpage

\subsection{Subcluster Ellipticities} \label{ellipticities_section}

\begin{figure}
\centering
\includegraphics[angle=0.,width=6.0in]{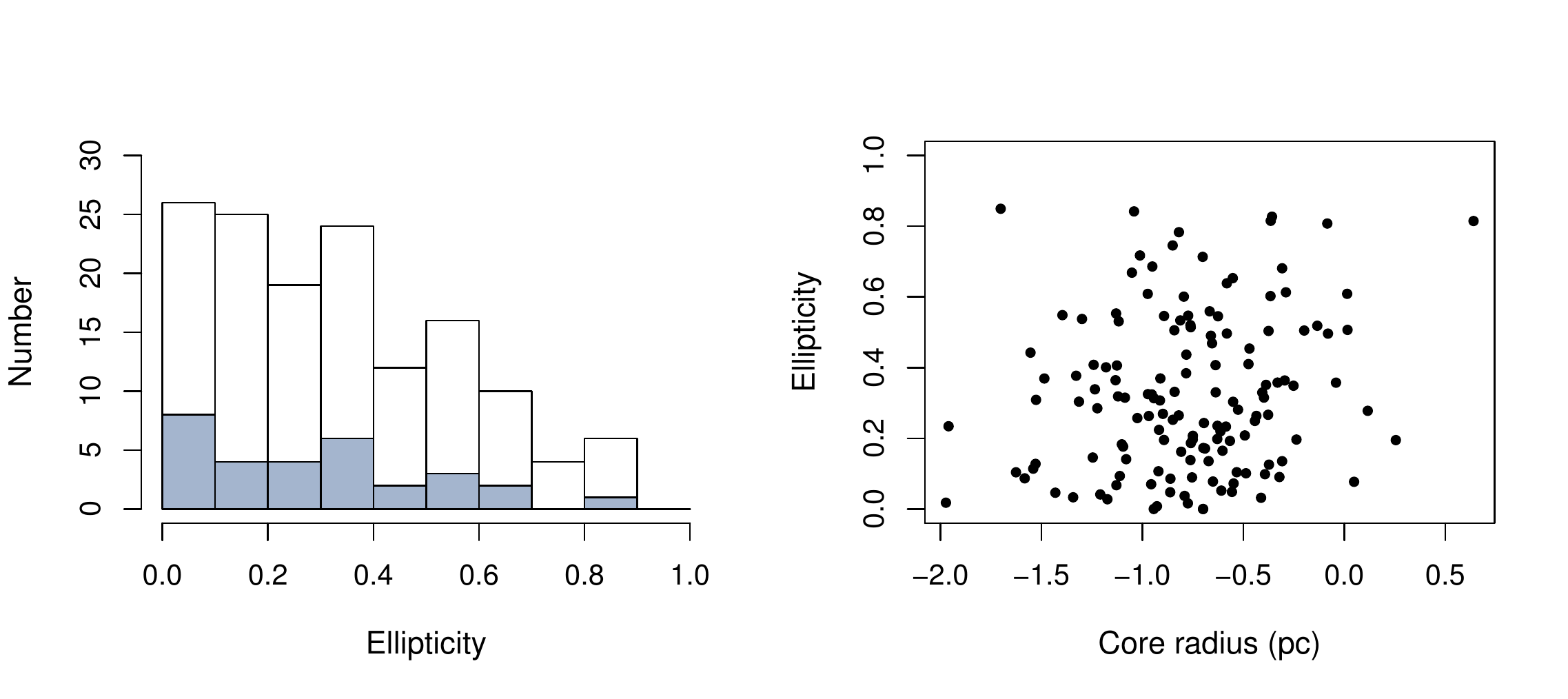}
\caption{Left: Histogram of all 142 subclusters' ellipticities is shown. The ellipticities range from $\sim$0, or nearly circular in projection, to 0.9, a 10 to 1 aspect ratio, with low ellipticity subclusters favored. 
Right: The scatter plot shows subcluster core size vs.\ core radius.
\label{ellipticity_figure}}
\end{figure}

Figure \ref{ellipticity_figure} (left) shows a histogram of \mystix\ cluster ellipticity. The measured ellipticities will be affected by projection, so they are lower limits on intrinsic ellipticity. Ellipticities range from 0 to $\sim$0.8, with a higher fraction having the lower ellipticities. Nevertheless, there are a significant number of subclusters with substantial ellipticities. Figure \ref{ellipticity_figure} (right) shows no relation between ellipticity and cluster core radius; this is confirmed using Spearman's rank correlation test. 

Dynamical studies have shown that highly elliptical cloud configurations are unstable to gravitational fragmentation \citep{Burkert04}. Note that the most extreme ellipticities with $\epsilon\simeq 0.8$ in our model fits may not distinct subclusters: for example, subcluster E in NGC 6357 appears to be part of a larger asymmetric complex, subcluster G in the Carina Nebula is a huge structure along the spine of the complex likely including unclustered stars and fitting residuals from other subclusters, and subcluster A in Trifid is one end of a larger filamentary structure that was truncated by the \Chandra\ field of view.

The ellipticity distribution for \mystix\ subclusters has some resemblance to the distribution of ellipticities in simulated star-forming regions found by \citet{Maschberger10}. There, ellipticites range from 0.0 to $0.82$, lower ellipticites are more common, and the peak ellipticity is $0.33$. In the simulated clusters, significant ellipticities are an effect of subcluster mergers, and this may also be the case for the \mystix\ clusters.  The frequent presence of high ellipticities in \mystix\ subclusters suggests that many have not reached dynamical equilibrium.  

\clearpage
\clearpage

\clearpage
\clearpage

\section{Discussion and Conclusions \label{discussion_section}}

The MPCM samples have been designed to improve the statistical characterization of stellar populations in MSFRs \citep{overview}.   Using a rich statistical sample of 16,608 MPCMs, ``flattened'' to remove X-ray spatial sensitivity variations,  and the finite-mixture-model technique to identify subclusters among these young stars, new details about the spatial structure of young stellar clusters emerge. In the diverse MSFRs surveyed by the \mystix\ project, we see a picture of star-forming regions with multiple non-coeval subclusters that probably formed with similar small sizes, but have expanded to a wide range of sizes.  A similar finding of cluster expansion in a different sample is studied by \citet{Pfalzner09}. The catalog of 142 \mystix\ subclusters derived here, combined with inferred intrinsic stellar populations (Paper~II) and age estimates for the subclusters \citep{Getman13a}, will help provide powerful links between observational and theoretical analysis star-cluster formation.

The finite mixture model, using isothermal ellipsoids, is a novel cluster analysis method for MSFRs that can reveal information that is hard to obtain from other cluster-finding methods.  \citet{Pfalzner11} finds that multimodal properties in regions with varying cluster surface density and richness are difficult to determine from surface density distributions alone.  But our method is an objective and mathematically founded procedure to separate multiply-clustered patterns into distinct isothermal ellipsoidal subclusters.  The procedure is flexible, finding statistically significant structures on all scales simultaneously, even treating cases such as small clusters reside inside large ones.  This is important because, even in relatively simple MSFRs like the Orion Nebula, layered subcluster structures are present and can affect inferred subcluster properties.  Furthermore, our algorithm does not require a single threshold in surface-density or separation to divide subclusters, which can help avoid bias when subclusters with very different properties exist in the same star-forming region. However, the finite-mixture-model method requires additional assumptions about region structure and subcluster shape. 

We validate the subcluster results using residual maps, $\chi^2$ tests of source counts, and comparison with results from previous studies of these regions. The models reproduce the observed surface densities reasonably well, although residual maps reveal some extra structure.  The improved stellar census in dense regions improves the accuracy of subcluster radius and peak surface density maps.  On larger scales, we caution that the MPCM sample coverage of widely distributed young stars is limited, and isothermal ellipsoid fitting is not sensitive to any truncation of cluster sizes that may be present. 

\subsection{The Structure of Stellar Populations in MSFRs}

The numbers of subclusters and their arrangements vary significantly from region to region. Heuristically, the clustering morphologies are divided into four classes: isolated clusters, core-halo clusters, clumpy clusters, and linear chains of clusters (Section~\ref{classes_section}).  The dominant subcluster in a region (i.e.\ the subcluster with most stars, $N_{4,\mathrm{obs}}$) often sits at the center of a group of subclusters (e.g.\ Subcluster L in M~17), but this is not always true (e.g.\ subcluster B in the Eagle Nebula).
In regions with linear chain structure there may be no subcluster that is particularly dominant. While, in contrast, regions with simple or core-halo structure may be much more centrally concentrated. 

In DR~21, NGC~2264, and NGC~6334, the linear chain cluster structure maps onto molecular filaments in the regions, and it is clear that the star cluster structure was at least partially inherited from the filamentary chain of molecular cloud clumps. More detail about correspondence between subcluster locations and molecular cloud clump locations will be given in a later \mystix\ study. Thus, regions with linear chain structure have not evolved much from their initial state at the time of the birth of stars. The more centrally concentrated clumpy, core-halo, and simple cluster structures are more likely to have dynamically evolved and, in cases where the subclusters are well fit by single isothermal ellipsoids, may have achieved some dynamical relaxation. 

The morphological classification suggests an evolutionary progression from linear chain structure to clumpy structure to core-halo structure to simple cluster structure.   Hierarchical cluster formation, often in molecular filaments, would evolve into dynamical relaxed unimodal structures.  However, the \mystix\ regions indicate that this progression is not always followed. For example, the linear chain region NGC~1893 is almost entirely unembedded, and the chain of subclusters in NGC~6334 exhibit a wide range of absorptions. 

The distribution of subcluster properties show trends common to all the regions. A lognormal distribution of subcluster size peaks at 0.17~pc, and many subclusters have ellipticities of $\epsilon \leq 0.3$, but few have ellipticities greater than 0.5.  No relationships between ellipticity and size are seen.   


An additional property of subclusters, their absorption, can help link these properties to cluster formation history and evolution. Line-of-sight absorption is measured both by the near-IR color index $J-H$ and the X-ray Median Energy indicator \citep{Getman10}. While absorption may be caused by overlaying molecular material, in most cases it represents local cloud cores within which the youngest clusters reside.  In complex \mystix\ star-forming regions, a wide range of absorption can be seen for the different subclusters, indicating these subclusters are not coeval.  A statistically significant anti-correlation between absorption and size of a subcluster is found, most easily explained if both absorption and size are related to age.   This issue is pursued in the accompanying study by \citep{Getman13a}  where spatial-age gradients in \mystix\ regions are found based on a new age indicator for individual pre-main sequence stars. A similar relationship was found for subclusters within the Rosette Molecular Cloud by \citep{Ybarra13} who infer a gas removal $e$-folding timescale of 0.4~Myr.

Cluster expansion is expected to produce a wide range of stellar surface densities seen in many star-forming regions \citep{Gieles12}.  Early cluster expansion arises principally from the loss of the gravitational potential of the molecular material, but stellar dynamics can also produce stellar halos around dense cores through three-body interactions with hard binaries and gravothermal core collapse associated with mass segregation \citep{Giersz96}.  This issue will be discussed further in Paper II when intrinsic populations, rather than observed population subject to different sensitivity effects, are derived.    

\subsection{Links to Astrophysical Theory of Cluster Formation}

The anti-correlation between radius and absorption provides evidence that subcluster expansion is important part of subcluster dynamical evolution. From numerical simulations, \citet{Moeckel10} find that even the very dense simulated cluster produced by \citet{Bate09a}, with a half-mass radius of $\sim$0.05~pc, grew to $>$2~pc over a period of $\sim$2~Myr. The typical small size that we find for the embedded clusters indicates that the high initial densities from the simulation in \citet{Bate09a} are realistic, which can make competitive accretion and dynamical interaction between protostars much more important than if clusters form $\sim$1~pc in size.

Simulation of cluster expansion after gas expulsion by \citet{Goodwin06} show a relationship between star-formation efficiency and the age-radius relationship. They find that, after 2.5~Myr, core radii will grow by a factor of $\sim$1.5 for a star-forming efficiency of 60\% and a factor of $\sim$3 for a star-forming efficiencies of 10--30\%.  Larger expansion factors are found in models of very rich clusters by \citet{Banerjee13}.   For comparison, \mystix\ results indicate that subcluster core radii start out at $\sim$0.08~pc and grow to the typical $\sim$0.2~pc for the typical unabsorbed \mystix\ subclusters (a factor of more than 2 expansion). The stars in \mystix\ regions are generally $<$5~Myr old, but more precise age determination is necessary to determine which age-radius relation provides the better fit. \citet{Pfalzner11} find a trend consistent with a star-formation efficiency of $\sim$30\% using the clusters in \citet{Lada03}.

The ellipticities of \mystix\ subclusters could be inherited from the structure of the parental molecular cloud or could arise from subcluster mergers.  The subcluster ellipticity distribution resembles those found by numerical simulations of merging subclusters \citep[][their Figure~10b]{Maschberger10} in which merger products have a distribution of ellipticities, including a tail at high ellipticities. 

The cluster morphological classes may correspond to structures seen in star formation simulations, such as those performed by \citet{Bate03,Bate09a} or \citet{Parker12}. Simulations often show multiple pockets of star formation occurring along gas filaments as the cloud collapses, which can end up merging and dynamically relaxing as the simulation progresses. Clumpy clusters may indicate early stages of partial merging, core-halo structures may be an intermediate stages of mergers, and isolated clusters with relaxed structure a final stage. \citet{Getman13b} add an important constraint to such a scenario: in at least two  \mystix\ regions, the dense core of the cluster has younger stars than the outer regions. This observational result requires some special behavior, such as continual feeding of gas towards the cluster center or dispersal of older core stars into the halo.

The simple clusters may represent those that have completed subcluster merging and dynamical relaxation.  Subcluster morphology can also influence these rates.  \citet{Smith11} find that the cluster formation rate is related to a filling factor measuring the fraction of volume containing subclusters. When this filling factor becomes high, tidal interactions accelerate subcluster mergers and formation of a simple unimodal cluster is promoted. This phenomenon may be occurring today in crowded, clumpy \mystix\ regions like M~17. 

The core-halo structures in 5 out of 17 \mystix\ regions may also be a result of subcluster mergers. \citet{Bate09a} report a hydrodynamical and N-body simulation of a MSFR resulting in a cluster with a half-mass radius of 0.05~pc and a halo of stars extending out $\sim$0.4~pc. \citet{Smith11} found that gravitational interactions between subclusters of young stars in simulated star-forming environments can scatter stars out of the clumps, which can lead to a dense core surrounded by a less dense halo of scattered stars. \citet{Maschberger10} also identify a halo of scattered low-mass stars around the dominant cluster, albeit with a surface-density--radius relation opposite of the effect we find with concentric ellipsoids.  However, \citet{Pfalzner13} report that 2-body encounters can produce a population of cluster stars with high ellipticities which will spend most of the time in a halo.  The accompanying paper by \citet{Getman13b} describe star formation scenarios that may account for core-halo structures. All of these issues relating to cluster formation and MSFR star formation histories will be pursued in \mystix\ papers by \citet{Getman13a} and Paper II where quantitative measures of subcluster intrinsic populations and cluster ages will be combined with the structural results obtained here.

\acknowledgements Acknowledgments:  

We thank the anonymous referee for taking the time and effort to provide useful feedback for this paper. The \mystix\ project is supported at Penn State by NASA grant NNX09AC74G, NSF grant AST-0908038, and the \Chandra\ ACIS Team contract SV4-74018 (G.~Garmire \& L.~Townsley, Principal Investigators), issued by the \Chandra\ X-ray Center, which is operated by the Smithsonian Astrophysical Observatory for and on behalf of NASA under contract NAS8-03060.  M.A.K. also received support from NSF SI2-SSE grant AST-1047586 (G. J. Babu, PI). We thank Leisa Townsley for providing advice on this paper and for the development of {\it Chandra} tools. This research made use of data products from the \Chandra\ Data Archive. This work is based on observations made with the {\it Spitzer Space Telescope}, obtained from the NASA/IPAC Infrared Science Archive, both of which are operated by the Jet Propulsion Laboratory, California Institute of Technology under a contract with the National Aeronautics and Space Administration. This research has also made use of SAOImage DS9 software developed by Smithsonian Astrophysical Observatory and NASA's Astrophysics Data System Bibliographic Services.

\appendix
\section{Finite Mixture Model with \Rlan\ Code}

The procedure for fitting finite mixture models is illustrated here for the \mystix\ region NGC~6357 with the code from the comprehensive \Rlan\ public domain statistical software system \citep{RCore13}.  The \Rlan\ code below makes use of online source code and machine readable tables. Two packages from the Comprehensive R Analysis Network (CRAN) are used: {\it plotrix} \citep{Lemon06} and {\it spatstat} \citep{Baddeley05}.  

The structure of the example is as follows: Step~1 loads the required \Rlan\ functions and data; Step~2 creates a first guess of the ellipsoid parameters; Step~3 conducts the first iteration of model fitting using the Nelder-Mead simplex algorithm; Step~4 displays the results of the first iteration; Step~5 refines the model with further fitting iterations; and Step~6 presents the results of the fit, producing figures analogous to Figures~2 and~4a in the paper.  These commands may be pasted into an \Rlan\ session following the instructions below.   

The input data consist of the polygonal boundaries of the fields of view and coordinates of the stars in the statistical sample in machine readable table format ({\it fov.mrt} and {\it stars.mrt}), which are available in the online version of this article in addition to the library of \Rlan\ functions {\it sourcecode.R}. The {\it library} commands below load the necessary libraries.\footnote{These libraries can be installed from the R session by \texttt{install.packages("spatstat",dependencies=T)} and \texttt{install.packages("plotrix",dependencies=T)}.} The function {\it star.ppp} creates a point pattern object (an internal format of the {\em spatstat} package) for the region NGC~6357 assuming a distance of 1.7~kpc. Finally, the function {\it unique} removes any duplicated positions in the catalog.


\begin{verbatim}
# Step 1 -- Setup

library(plotrix);
library(spatstat);
source("sourcecode.R");
clust <- star.ppp(target = "ngc6357", distance = 1.7);
clust <- unique(clust);
\end{verbatim}

To illustrate the analysis, we fit a model with $k=6$ subclusters and the uniform density component. The data structure ``param2'' specifies: the number of model components ($6+1=7$), the model that will be used for each component (ellipsoid or constant), and the number of additional parameters for each component (5 or 0 for ellipsoid or constant, respectively). The function {\it ell.model} is an implementation of Equation~4 for the isothermal ellipsoid in Section~3.1. The array ``param.init'' contains the initial guess parameters for all components of the finite mixture model --- we can inspect these parameters more easily by running the {\it param2ellipse} command, which returns a data-frame object with $k$ rows (subclusters) and 6 columns: ``x''~--~ellipsoid $x$ coordinate; ``y''~--~ellipsoid $y$ coordinate; ``core''~--~core radius in parsecs; ``theta''~--~ellipse orientation in radians; ``b''~--~axis ratio; and ``mix''~--~the log of the mixture coefficient corresponding to the relative peak surface densities for each subcluster (the log mixture coefficient for the first component is assumed to be 0.0 and not included in the ``param'' array.) The final entry in ``param'' is the log mixture coefficient for the unclustered component.

\begin{verbatim}
# Step 2 -- Initial Guess

k <- 6;
param2 <- c(k+1, ell.model, 5, ell.model, 5, ell.model, 5,
   ell.model, 5, ell.model, 5, ell.model, 5, const.model, 0);
param.init <- c(
   -1.00, -3.25, 0.3, 0, 1,  
    3.75,  1.25, 0.3, 0, 1, -0.4533811,
   -2.50, -4.50, 0.3, 0, 1, -0.7415462,
   -1.25, -2.50, 0.3, 0, 1,  0.8510107,
   -4.00,  0.75, 0.3, 0, 1,  0.1332489,
    4.00,  4.00, 0.3, 0, 1,  0.1667037,
   -2.0);
param2ellipse(param.init);
\end{verbatim}

The function {\it multi.model} is the finite mixture model, which is specified by the variables ``k,'' ``param,'' and ``param2.'' The function {\it model.lik} calculates log-likelihood of this model using Equation~6. Model fitting is done using the Nelder-Mead simplex algorithm via  {\em optim}. This \Rlan\ function takes the star cluster data, the model form and parameters, and the log-likelihood function. It is unlikely that the absolute maximum of the likelihood will be found on the first attempt, so this step must be performed iteratively --- with certain parameters held fixed in turn while other are free to vary --- until the likelihood converges on the maximum.

\begin{verbatim}
# Step 3 -- Likelihood Maximization Via Nelder-Mead

ocf <- optim(param.init, model.lik, model=multi.model,
   clust=clust, param2=param2, hessian=TRUE);
param <- ocf$par;
\end{verbatim}


The commands below plot the spatial point pattern object ({\it plot}) and draws ellipses ({\it draw.model}) indicating the initial guess parameters (red) and new parameters (green). The log-likelihood ``L'' and the AIC are calculated --- these values will improve with each iteration.

\begin{verbatim}
# Step 4 -- Model Inspection

plot(clust, pch=20, cex=0.5);
draw.model(param.init, param2, size=1);
draw.model(param, param2, size=1, border=3);
L <- -model.lik(param, model=multi.model, clust=clust, param2=param2);
AIC <- -2*L + length(param)*2.0; AIC;
\end{verbatim}

It is necessary to iteratively converge on the maximum likelihood. The {\it mask.freeze} function allows only certain parameters to be fit, while the other parameters are held constant. This can be used to concentrate on improving certain parameters without having to search the entire $(6k+1)$-dimensional parameter space. The commands below improve, sequentially, subcluster positions, subcluster sizes, subcluster orientations, subcluster shapes, subcluster positions again, and, finally, all parameters again. After each iteration, the choice of the next fitting function to run depends on inspection of the results of the previous iteration. Nevertheless, each iteration will always result in an AIC value that is greater or equal to the previous iteration's AIC values, so the order of these commands is not particularly important as long as there are enough iterations for the AIC to converge. (AIC is calculated again at the end of this fitting sequence.) The time for these steps to run depends on the number of points and the dimensionality of the parameter space being searched, so fits with frozen parameters will run more quickly than fits where all parameters are free to vary.

\begin{verbatim}       
# Step 5 -- Further Model Refinement Via Nelder-Mead

param <- mask.freeze(param, mask.adjust_positions(k), clust=clust);
param <- mask.freeze(param, mask.adjust_size(k), clust=clust);
param <- mask.freeze(param, mask.adjust_rotations(k), clust=clust);
param <- mask.freeze(param, mask.adjust_shape(k), clust=clust);
param <- mask.freeze(param, mask.adjust_positions(k), clust=clust);
ocf <- optim(param, model.lik, model=multi.model,
   clust=clust, param2=param2, hessian=TRUE);
param <- ocf$par;

L <- -model.lik(param, model=multi.model, clust=clust, param2=param2);
AIC <- -2*L + length(param)*2.0; AIC;
\end{verbatim}

Now that the model has been fit, we can visualize the results. The {\it adaptive.density} command produces the adaptively smoothed surface density maps described in Section~4.1. (Here, we only run 10 iterations rather than 100 used to produce Figure~2.) The function {\it make.fig2} produces an image of subcluster ellipses overlaid on the non-parametric surface density map similar to Figure~2. The function {\it make.fig2} produces an image of the spatial fit residuals similar to Figure~4. 


\begin{verbatim}
# Step 6 -- Visualizations of Subcluster Results

density <- adaptive.density(clust, f=0.1, nrep=10);
make.fig2(param, param2, clust=clust, image=density, 
   min.im=5.0, max.im=20000.0);

make.fig4(param, model=multi.model, clust=clust,
   param2=param2, bandwidth=0.38);
\end{verbatim}

A similar procedure is applied for different numbers of ellipsoids (with the initial guess parameters ``n'', ``param'', and ``param2'' adjusted accordingly) to find the number of subclusters that optimizes the AIC value. It is useful to try the fitting procedure starting with several different initial guesses in order to ensure that the results are independent of the initial parameters used.

\end{document}